UNIVERSITY OF CALIFORNIA, MERCED

**An Experimental Study of Distributed Quantile Estimation**

A thesis submitted in partial satisfaction of the
requirements for the degree
Master of Science

in

Electrical Engineering and Computer Science

by

Zixuan Zhuang

Committee in charge:

Professor Florin Rusu, Chair
Professor Mukesh Singhal
Professor Sungjin Im

2015

The thesis of Zixuan Zhuang is approved, and it is acceptable in quality and form for publication on microfilm and electronically:

---

---

---

Chair

University of California, Merced

2015

TABLE OF CONTENTS

LIST OF FIGURES

ACKNOWLEDGEMENTS

I would like to thank Professor Rusu for her expert advice and encouragement throughout, as well as Dr. Mattoon for his help on the organizations.

ABSTRACT OF THE THESIS

**An Experimental Study of Distributed Quantile Estimation**

by

Zixuan Zhuang

Master of Science in Electrical Engineering and Computer Science

University of California, Merced, 2015

Professor Florin Rusu, Chair

Quantiles are very important statistics information used to describe the distribution of datasets. Given the quantiles of a dataset, we can easily know the distribution of the dataset, which is a fundamental problem in data analysis. However, quite often, computing quantiles directly is inappropriate due to the memory limitations. Further, in many settings such as data streaming and sensor network model, even the data size is unpredictable. Although the quantiles computation has been widely studied, it was mostly in the sequential setting. In this paper, we study several quantile computation algorithms in the *distributed* setting and compare them in terms of space usage, running time, and accuracy. Moreover, we provide detailed experimental comparisons between several popular algorithms. Our work focuses on the approximate quantile algorithms which provide error bounds. Approximate quantiles have received more attentions than exact ones since they are often faster, can be more easily adapted to the distributed setting while giving sufficiently good statistical information on the data sets.

# Chapter 1

# Introduction

## 1.1 Motivating Examples

The Internet has grown exponentially since the late 1960s, and millions of people are using it every day. Therefore, websites, such as those used for online shopping, record huge amount of transactions from visitors, and these transactions logs are stored in a database system or plain files. These logs contain very important information for site owners, not any single record but the aggregate information. The aggregates are functions that summarize a series of data [11], such as *min*, *max*, *median*, and *average*. The owner of a online shopping site can use the aggregates information to analyze the impact of a sale or a promotion, or they could adjust sale strategies based on the aggregates information. In addition, retailers would like to analyze every customers' shopping preference by collecting their searching results, then based on the analyzation, retailers can push some goods suggestions to buyers, and those guidance will some how allure people to spend more money than they expect. For a large and famous online shopping site, there are millions of visits, even in a minutes, and the huge amount records are usually stored in multiple severs across the world. Then the problem arrises, collecting the aggregate information in this situation is not a easy task.

Another example is flight-tickets searching. The airline companies and travel agents typically have a website to allow travelers to search the flight tickets, compare the price, and book the tickets. The price of a flight ticket is some how determined by how many people search this flight and how many of them buy this flight ticket. A flight ticket always has a higher price if the flight is on a holiday or during an event, since more people are searching it and willing to buy it; a flight ticket is always cheaper if the flight departs at the midnight, because people are usually reluctant to fly overnight. Thus, the aggregate information is important for airline companies and travel agents to adjust the price of tickets.

## 1.2 Problem Definition

The $\phi$-quantile ($0 < \phi < 1$) of an $n$ element dataset is the element ($e$) in this dataset that has $\lfloor \phi n \rfloor$ number of elements that are no larger than $e$ [2]. For example, if we have a dataset that contains 10 numbers that are $1, 2, ..., 9, 10$, then the 0.2-quantile is 3, since 3 has 2 ($\lfloor 0.2 \times 10 \rfloor$) elements that are smaller than 3. The *min*, *median*, and *max* are special quantiles. The *min* is the smallest quantile, while the *max* is the largest quantile. The *median* is the 0.5-quantile.

## 1.3 Problem Analysis

The quantile problem can be solved in polynomial time. For local memory fit data, we can use QuickSort algorithm to sort the data in $O(nlogn)$, and then the exact quantiles are very easy to calculate. However, it is another story if the size of the data becomes too large to fit the limited memory or machine. As the size increases, we need much more time in loading data from disk to memory and writing from memory to disk, or communication between machines. Other situations, like data streaming and sensor networks, also face the similar problem; as an illustration, in data streaming, the unlimited data comes one by one into the limited memory, and in such situation, it is intractable to compute a $\phi$-quantile for a given period; for sensor networks, we have to consider the communication cost since the networks have a band-limit, and the networks will crash if you send "too" large of data sizes. In a database system with a very large data set (compared to the memory), we can only go through the data once to gather all quantile information, even one more process is too time consuming. Thus, approximate quantile is necessary, and an $\varepsilon$ is usually given to bound the error to $O(\varepsilon n)$. In most of cases, approximate quantile is enough to analyze the dataset, and exact quantile is inefficient and unnecessary. In this problem, we care more about communication cost or space usage than the running time, even though time-efficiency also plays a big role. This is because due to the limitation of the memory, using too much space will cause the system to crash. Moreover, we can generate the quantile summaries while loading the data, so if the increased time of the loading is acceptable, generating quantiles is feasible.

The simplest idea to solve the quantile problem would be sampling: given a probability $p$, we simply sample each coming data by probability $p$. By setting $p$ to be $\frac{1}{\varepsilon n}$, we can expect one data sampled every $\varepsilon n$ data, so that we can bound the error to $O(\varepsilon n)$. In this case, the size of sampled data is $O(1/\varepsilon)$. Z. Huang et al. [5] introduced their Sampling-Based algorithm based on this simple idea, we will discuss in detail in Section 3.2.

## 1.4 Existing Solutions and Their Limitations

There have been a large number of studies on quantile computation, which includes the algorithms we will study in this paper: GK [1], q-digest [4], Sampling Based [5], and Random Mergeable Summaries [6]. L. Wang et al. [2] have already done a survey on these algorithms: introduced and analyzed several quantile computation algorithms, devised some variants algorithms, and compared them on various measures; however, in this paper, we will focus on the distributed (parallel) setting instead of the centralized setting in Wang's paper. Since the size of data has been larger and larger in modern network, distributed computation is a very important method to process the large datasets. It is also very important to compare the quantile algorithms in distributed setting.

## 1.5 Contributions

Our main contribution in this paper is to compare the quantile computation algorithms in distributed setting. As we mentioned in Section 1.4, the survey of L. Wang et al. [2] is on a centralized setting. We express the quantile computation algorithms in a single formalism given by GLAs (see details in Chapter 4) and the experimental evaluation of the algorithms in a distributed setting. To summarize, we have the following contributions in this paper:

- Integrate the quantile computation algorithms to the *GLADE* database system.
- Explore possible implementations for some of the algorithms on the extension to a distributed setting.
- Compare the quantile computation algorithms in distributed setting.

- Execute the quantile computation algorithms with several datasets in the database; especially we tried 0, 0.5, and 1 zipfan-distributed datasets, which cover a much broader range of data distributions and are more common in databases.

- Measure the quantile algorithms in several aspects: ranking error, running time, space usage, and ratio time; the ratio time shows the scalability of an algorithm with the number of threads.

## 1.6 Organization

We start by introducing four quantile computation algorithms in Chapter 3. We introduce the *GLADE* framework and the interface GLA in Chapter 4. In Chapter 5, we evaluate the experimental results of all algorithms in detail, including a variant, FaseQDigest, introduced by L. Wang et al. [2]; while in Chapter 6 we conclude the paper.

# Chapter 2

# Formal Problem Statement

In this Chapter, we will formally state our problem, the quantile computation, and illustrate an example. In Chapter 3, we will show how each algorithm solves the same example.

We have two example datasets below.

$$\{3,4,0,7,1,0,0,2,6,0,2,1,0,4,2\} \tag{2.1}$$

$$\{3,6,2,3,2,1,7,3,3,5,2,3,3,7,2\} \tag{2.2}$$

To get the quantiles for these two datasets, we have to sort them first. The first example dataset has 15 elements, and it can be sorted as

$$\{0,0,0,0,0,1,1,2,2,2,3,4,4,6,7\} \tag{2.3}$$

Recall that the $\phi$-quantile of an $n$ elements dataset is the element ($e$) in this dataset that has $\lfloor \phi n \rfloor$ number of elements that are no larger than $e$. The 0.5-quantile of first dataset is 2, because there are 7 ($\lfloor 0.5 \times 15 \rfloor$) elements smaller than it. The 0.2-quantile is 0, since there are 3 ($\lfloor 0.2 \times 15 \rfloor$) elements are no larger than it.

The second example dataset also has 15 elements, and it can be sorted as

$$\{1,2,2,2,2,3,3,3,3,3,3,5,6,7,7\} \tag{2.4}$$

The 0.5-quantile of the second dataset is 3, because there are 7 elements that are smaller than it. The 0.7-quantile of the dataset is also 3, since there are 10 elements that are smaller than it.

To get the quantiles for the both data in the two example datasets, we need to merge the two datasets and sort them.

$$\{0,0,0,0,0,1,1,1,2,2,2,2,2,2,2,3,3,3,3,3,3,3,4,4,5,6,6,7,7,7\} \tag{2.5}$$

So the 0.5-quantile is 3, since there are 15 ($\lfloor 0.5 \times 30 \rfloor$) elements are smaller than it.

As stated above, the exact quantile is unnecessary for us, but the approximate quantile gives enough information, and it saves space and time. The $\varepsilon$-approximate $\phi$-quantile is the element that has $(\phi - \varepsilon)n$ to $(\phi + \varepsilon)n$ elements that are smaller than it.

Basically, the quantile computation algorithms process the data "one-pass" and store some of the elements along with additional information. "One-pass" means to read the data only once and get a small size summary that can answer a quantile query at any time; if there are new datasets, the summary can be merged with the new datasets, and can answer any quantile query for the whole datasets, including both the old and new datasets.

# Chapter 3

# Algorithms

## 3.1 GK

The GK algorithm [1] is a deterministic $\varepsilon$-approximation algorithm that provides a fancy way to compute quantiles with a summary size bounded in $O(\frac{1}{\varepsilon} log(\varepsilon N))$, where $N$ is the total number of data. We keep a list of tuples (using $std :: multimap$) $t = (v_i, g_i, \delta_i)$, where $g_i$ is the gap of lower bound of the rank between $v_i$ and $v_{i-1}$, and $\delta_i$ is the difference of the upper and lower bound of the rank of $v_i$. [2] shows that $g_i$ and $\delta_i$ follow the two restrictions:

$$(1)\ \sum_{j \le i} g_j \le r(v_i) + 1 \le \sum_{j \le i} g_j + \delta_i$$

$$(2)\ g_i + \delta_i \le \lfloor 2\varepsilon n \rfloor$$

$r(v_i)$ is the rank of of $v_i$.

To better illustrate, we define $r_{min}$ and $r_{max}$ as follows:

$$r_{min}(v_i) = \sum_{j \le i} g_j$$

$$r_{max}(v_i) = r_{min}(v_i) + \delta_i$$

The first restriction tells us that $r(v_i)$ is between $r_{min}(v_i)$ and $r_{max}(v_i)$, this restriction will also be used to answer the quantile queries. The second restriction bounded the error in $O(\varepsilon n)$.

### 3.1.1 Build

To add a new element $v$, we first find its successor $(v', g', \delta')$, which is the smallest $v_i$ that is greater than $v$, and insert tuple $(v, 1, g' + \delta' - 1)$ if not insert to either first or last in the list, or inset $(v, 1, 0)$ otherwise (algorithm 1). This insertion policy maintains the above two restrictions. The new element $v$ is inserted in between $v_{i-1}$ and $v_i$ ($v_i$ is $v$'s successor), so $v$'s rank can be as small as $r_{min}(v_{i-1}) + 1$ and as large as $r_{max}(v_i) - 1$. Thus, by making $g$ to be 1 and $\delta$ to be $g' + \delta' - 1$, we can get $r_{min}(v) \le r(v) + 1 \le r_{max}$. In the mean time, $g + \delta = g' + \delta' \le \lfloor 2\varepsilon n \rfloor$. For every $\lceil \frac{1}{2\varepsilon} \rceil$ insertion, we call COMPRESS [1] to reduce the size of the list.

**Bands and capacities.** To better explain the COMPRESS process, [1] introduced the *bands* and *capacities*. The basic idea of COMPRESS is to reduce the size of summary and keep the minimum number of tuples in the list. Thus, we need to remove the tuples with small capacities and keep the ones with large capacities. By partitioning the $\delta$s into *bands* which are (0, $\frac{1}{2}2\varepsilon n$, $\frac{3}{4}2\varepsilon n$, .., $\frac{2^i-1}{2^i}2\varepsilon n$, ..., $2\varepsilon n - 1$, $2\varepsilon n$)[1]; and the corresponding capacities are ($2\varepsilon n$, $\varepsilon n$, $\frac{1}{2}\varepsilon n$, ..., $\frac{1}{2^i}\varepsilon n$, ..., 4, 2, 1) [1]. To compute the *capacity*, we use algorithm 2.

**Algorithm 1** ADDITEM (GK)

**Input:** $v$, current synopses $\vec{t}$, number of elements we have seen $n$, $k$
**Output:** new synopses $\vec{t}$

**if** $n$ **mod** $k = 0$ **then**
  call COMPRESS
**end if**
find the position that $v$ should be inserted
**if** $v$ is inserted to the beginning or the end of $t$ **then**
  insert $(v, 1, 0)$ to $t$
**else**
  find its successor $(v', g', \delta')$ which is the smallest $v_i$ that is greater than $v$
  insert $(v, 1, g' + \delta' - 1)$ to $t$
**end if**

**Algorithm 2** CAPACITY

**Input:** $v$, $g$, $\delta$, $n$, $\varepsilon$
**Output:** *capacity*

$p = \lfloor 2\varepsilon n \rfloor$
$threshold = \lceil log(2\varepsilon n)/log2 \rceil$
**if** $\delta = 0$ **then**
  **return** $threshold + 1$
**end if**
**if** $\delta = p$ **then**
  **return** $0$
**end if**
**for** $\alpha = 1$ **to** $threshold$ **do**
  $lbound = p - 2^{\alpha} - (p \textbf{ mod } 2^{\alpha-1})$
  $ubound = p - 2^{\alpha-1} - (p \textbf{ mod } 2^{\alpha-1})$
  **if** $lbound < \delta \leq ubound$ **then**
    **break**
  **end if**
**end for**
**return** $\alpha$

The *descendants* of tuple $i$ is defined as contiguous segment of tuples next to tuple $i$ that have a smaller capacities than $i$.

To find the tuples that can be removed from the summary, we go through the list from the end to the begin, and if the tuple $i$ satisfies the following condition, then all the descendants of tuple $i$ can be deleted. We present the COMPRESS algorithm in Algorithm 3.

$$g* + g_i + \delta_i - 1 < 2\varepsilon n$$

$$g* = \sum_{\text{all the descendants of tuple } i} g$$

This condition ensures that the error bound restriction (mentioned above) is still satisfied. After deletion, the new gap of tuple $i$ is $g* + g_i$.

**Algorithm 3** COMPRESS (GK)

**Input:** $\vec{t}$, $s$, $n$, $\varepsilon$
**Output:** $\vec{t}$

```
for i = s − 2 downto 0 do
   g* = g_i
   j = i − 1
   while j > 0 and capacity_j < capacity_i do
      g* = g* + g_j
      j = j − 1
   end while
   if capacity_i < capacity_{i+1} and g* + g_{i+1} + δ_{i+1} − 1 ≤ 2εn then
      delete elements in t⃗ from j to i.
   end if
   g_{i+1} = g_{i+1} + g*
end for
```

### 3.1.2 Merge

To merge two summary lists from two nodes [3], for example:

- $t' =< (v'_1, g'_1, \delta'_1), ..., (v'_i, g'_i, \delta'_i), ..., (v'_{n'}, g'_{n'}, \delta'_{n'}) >$
- $t'' =< (v''_1, g''_1, \delta''_1), ..., (v''_i, g''_i, \delta''_i), ..., (v''_{n''}, g''_{n''}, \delta''_{n''}) >$

We combine them to a new list:

- $t =< (v_1, g_1, \delta_1), ..., (v_i, g_i, \delta_i), ..., (v_n, g_n, \delta_n) >$

where $n = n' + n''$, $g$ and $\delta$ is computed by the following process:

For an item in list $t$, $(v_i, g_i, \delta_i)$, if it was $(v'_j, g'_j, \delta'_j)$in $t'$, we find the last tuple that was in $t''$, which is in front of $v_i$, $(v''_k, g''_k, \delta''_k)$, and $(v''_{k+1}, g''_{k+1}, \delta''_{k+1})$ will be the first one in $t$ that is not smaller than $v_i$, vice-versa for item was in $t''$.

$$g_i = g'_j \tag{3.1}$$

$$\delta_i = \begin{cases} \delta'_j + \delta''_k & \text{if } (v''_{k+1}, g''_{k+1}, \delta''_{k+1}) \text{ not exists} \\ \delta'_j + g''_{k+1} + \delta''_{k+1} - 1 & \text{otherwise} \end{cases} \tag{3.2}$$

If $t'$ is a $\varepsilon'$-approximation quantile summary and $t''$ is a $\varepsilon''$-approximation quantile summary, it is very easy to figure out that $t$ is $\bar{\varepsilon} = max\{\varepsilon', \varepsilon''\}$-approximation quantile summary.

$$g_i + \delta_i \leq 2\varepsilon' n' + 2\varepsilon'' n'' \leq 2\bar{\varepsilon}(n' + n'') \leq 2\bar{\varepsilon} n \tag{3.3}$$

### 3.1.3 Estimation

To answer a $\phi$-quantile query, we develop the MERGE_QUANTILE, since the original QUANTILE [1] won't hold after merging due to the enlarged $\delta$s. We need to find a $v_i$ such that we have the least $|r - \sum_{k=1}^{i} g_k| + |\sum_{k=1}^{i} g_k + \delta_i - r|$, where $r$ is $\lceil \phi n \rceil$ (Algorithm 4).

**Algorithm 4** ESTIMAQTION (GK)

**Input:** $\vec{t}$, $\phi$, $n$
**Output:** $v$

$r = \lfloor \phi n \rfloor$
$r_{min} = 0$, $r_{max} = 0$, $min = \infty$
**for all** $(v, g, \delta)$ **in** $\vec{t}$ **do**
    $r_{min} += g$
    $r_{max} = r_{min} + \delta$
    $diff = |r - r_{min}| + |r_{max} - r|$
    **if** $diff < min$ **then**
        $min = diff$
        $min_v = v$
    **end if**
    **if** $r_{min} > r$ **then**
        **break**
    **end if**
**end for**
**return** $min_v$

In order to improve the running time, we use the variant GKMixed introduced by [2] to run our experiments. Instead of simply inserting a tuple into the list directly, we first see if this inserted tuple is removable. If it is removable, we remove this tuple immediately (the tuple $i$ is removable if $g_i + g_{i+1} + \delta_{t+1} \leq \lfloor 2\varepsilon n \rfloor$). And we run COMPRESS when the size of list doubles. The $O(\frac{1}{\varepsilon} log(\varepsilon N))$ bound is still satisfied in this variant [2].

### 3.1.4 Example

In this section, we will show the process that the GK algorithm uses to compute the 0.5-quantile for the first example dataset in Chapter 2 (Equation 2.1), and the merging process for both the example datasets (equation 2.1 and 2.2). Note that the algorithm will compress the synopses for every four ($\lceil \frac{1}{2\varepsilon} \rceil$) insertions. The order of the elements to be inserted is the same order as they are in example datasets. We choose 0.1414 as the $\varepsilon$.

The process of the insertion of the first example is as follows:

Insert 3. $\{< 3,1,0 >\}$. The $g$ is always 1 for every insertion. When we insert an element to the beginning or the end of the synopses, the $\delta$ is always 0.

Insert 4. $\{< 3,1,0 >< 4,1,0 >\}$.

Insert 0. $\{< 0,1,0 >< 3,1,0 >< 4,1,0 >\}$.

Insert 7. Since this is the fourth insertion, we compress the synopses before inserting 7 into the synopses. Since the band of $< 3,1,0 >$ is 2 which is not greater than $< 4,1,0 >$ (also 2), we remove $< 3,1,0 >$ from the synopses. We never remove the first element in the synopses. After removing $< 3,1,0 >$, we increase the $g$ of $< 4,1,0 >$ by 1, and we will get $\{< 0,1,0 >< 4,2,0 >\}$ after compressing. Then, we insert 7 into the sysnopses $\{< 0,1,0 >< 4,2,0 >< 7,1,0 >\}$.

Insert 1. $\{< 0,1,0 >< 1,1,1 >< 4,2,0 >< 7,1,0 >\}$. When inserting an element that is not inserted into either the beginning or the end of the synopses, the $\delta$ of the new element is $g' + \delta' - 1$, where $g'$ and $\delta'$ are in the successor $(v', g', \delta')$.

Insert 0. $\{< 0,1,0 >< 0,1,1 >< 1,1,1 >< 4,2,0 >< 7,1,0 >\}$.

Insert 0. $\{< 0,1,0 >< 0,1,1 >< 0,1,1 >< 1,1,1 >< 4,2,0 >< 7,1,0 >\}$

Insert 2. We compress the synopses before inserting 2 into the synopses. We will get $\{< 0,1,0 >< 0,2,1 >< 4,3,0 >< 7,1,0 >\}$ after compressing. Then, insert 2 into the synopses $\{< 0,1,0 >< 0,2,1 >< 2,1,2 >< 4,3,0 >< 7,1,0 >\}$.

Insert 6. $\{< 0,1,0 >< 0,2,1 >< 2,1,2 >< 4,3,0 >< 6,1,0 >< 7,1,0 >\}$.

Insert 0. $\{< 0,1,0 >< 0,2,1 >< 0,1,2 >< 2,1,2 >< 4,3,0 >< 6,1,0 >< 7,1,0 >\}$.

Insert 2. $\{< 0,1,0 >< 0,2,1 >< 0,1,2 >< 2,1,2 >< 2,1,2 >< 4,3,0 ><$

$6,1,0 >< 7,1,0 >\}$.

Insert 1. We compress the synopses before inserting 2 into the synopses. We will get $\{< 0,1,0 >< 0,2,1 >< 2,2,2 >< 4,4,0 >< 7,2,0 >\}$ after compressing. Then, insert 1, $\{< 0,1,0 >< 0,2,1 >< 1,1,3 >< 2,2,2 >< 4,4,0 >< 7,2,0 >\}$.

Insert 0. $\{< 0,1,0 >< 0,2,1 >< 0,1,3 >< 1,1,3 >< 2,2,2 >< 4,4,0 >< 7,2,0 >\}$.

Insert 4. $\{< 0,1,0 >< 0,2,1 >< 0,1,3 >< 1,1,3 >< 2,2,2 >< 4,4,0 >< 4,1,1 >< 7,2,0 >\}$.

Insert 2. $\{< 0,1,0 >< 0,2,1 >< 0,1,3 >< 1,1,3 >< 2,2,2 >< 2,1,3 >< 4,4,0 >< 4,1,1 >< 7,2,0 >\}$.

After finishing insertion of all elements in the dataset, we can query the algorithm for the 0.5-quantile. The algorithm will find the first element in the synopses that satisfies $\sum_{j\leq i} g_j \leq \lfloor \varepsilon n \rfloor + 1 \leq \sum_{j\leq i} g_j + \delta_i$, which is $< 1,1,3 >$.

The final synopses of the second example is $\{< 1,1,0 >< 2,1,0 >< 2,2,2 >< 2,1,3 >< 3,2,2 >< 3,1,3 >< 3,1,3 >< 6,4,0 >< 7,1,0 >< 7,1,0 >\}$, and its estimation of 0.50-quantile is $< 2,1,3 >$.

When merging the two synopses, the algorithm will combine them and calculate new $g$ and $\delta$ for every elements, then compress the synopses. The merged synopses is $\{< 0,1,3 >< 0,3,3 >< 1,1,4 >< 2,3,4 >< 2,2,10 >< 2,3,9 >< 3,2,10 >< 3,2,9 >< 4,4,5 >< 6,5,1 >< 7,2,1 >< 7,2,0 >\}$, and the estimation of 0.5-quantile of this synopses is $< 3,2,9 >$.

## 3.2 Sampling-Based

The Sampling-Based algorithm [5] is designed by Z. Huang et al. for quantile computation in sensor networks.

### 3.2.1 Build

The basic idea is simple: samples data in a certain probability, say $p$; when estimating, we can simply compute the rank of a sampled value by adding the previous one's (sampled data are sorted) rank with $1/p$, or 0 if it is the first one.

### 3.2.2 Merge

For multiple nodes, we need to compute the local rank $r(a,i)$, i.e., the rank of $a$ at node $i$, as above in each node, and send sampled data and their local ranks to the master node.

For any value of $x$ (may not in sampled data), we can compute its estimate rank in node $i$:

$$\hat{r}(x,i) = \begin{cases} r(pred(x,i),i) + 1/p & \text{if } pred(x,i)\ exists \\ 0 & \text{otherwise} \end{cases}$$

$pred(x,i)$ is the largest value in the sampled data in node $i$ that is not larger than $x$.

Then the global rank estimation of $x$ will be the sum of $\hat{r}(x,i)$ in every node:

$$\hat{r}(x) = \sum_i \hat{r}(x,i)$$

### 3.2.3 Estimation

To answer a $\phi$-quantile query, we first simply compute the global ranks of each sampled values from every node, then find the value whose rank is the closest to $\phi n$; this value is our answer.

To determine $p$, the following random variable is introduced:

$$X = \begin{cases} r(x,i) - r(pred(x,i),i) & \text{if } pred(x,i) \text{ } exists \\ r(x,i) + 1/p & \text{otherwise} \end{cases}$$

We can easily get:

$$E[X] = E'[X] + E''[X] = \sum_{i=1}^{r(x,i)} ip(1-p)^{i-1} + (1-p)^{r(x,i)}(r+1/p) = 1/p$$

$E'[X]$ is the expectation of X for the case $pred(x,i)$ exists, and $E''[X]$ is the expectation of X in the case $pred(x,i)$ does not exists. Then we can get variance

$$Var[X] = E[X^2] - E[X]^2 \leq 1/p^2$$

Since $\hat{r}(x,i) = r(x,i) - X + 1/p$, Var[$\hat{r}(x,i)$] = Var[X], and thus $\sum_i Var[\hat{r}(x,i)] \leq k/p^2$, where k is the number of nodes, by setting $p = \sqrt{k}/\varepsilon n$, the variance can be bounded in $O((\varepsilon n)^2)$.

The above algorithm has an $O(\sqrt{k}/\varepsilon)$ total communication cost, but to keep that every node has at most $O(1/\varepsilon)$ communication cost, $p_i$, probability for node $i$ would be determined as follows:

$$p_i = \begin{cases} p & \text{if } s_i \leq n/\sqrt{k} \\ 1/\varepsilon s_i & \text{otherwise} \end{cases}$$

where $s_i$ is the size of data in node $i$.

### 3.2.4 Improved Merging for Tree Model

The above algorithm is used in a flat model since all nodes send data directly to the master node. But a general system would be in a tree model. Thus, [5] introduced a merging algorithm for the routing tree-distributed networks.

Let $d_i$ denote the sampled data in node $i$, $s_i$ denote the size of data in node $i$, and $p_i$ denote the sample probability in node $i$. We can classify the sample $d_i$ is a small sample if $s_i < n/\sqrt{k}$, and large sample otherwise. $p_i$ is $\sqrt{k}/\varepsilon n$ for small samples, and $1/\varepsilon s_i$ for large ones. [5] also defines a class number for large samples

$$c_i = \lfloor log(s_i\sqrt{k}/n) \rfloor$$

We can say $c_i$ does not exist if $d_i$ is a small sample.

The merging idea is the following:

When a node receives the data from its children, merge all the small samples (including its own) if the total size of these small samples is no less than $n/\sqrt{k}$, otherwise just keep these small samples as they were. For the new sample from the merging, we need to compute an estimate rank for each value in the new sample as its "local rank" for this new sample. Since the new sample will be a large sample, the probability $p'$ is $1/\varepsilon s'$, where $s'$ is the new size (sum of all merged small samples). Subsampling each value in sample $d_i$ with probability $p'/p_i$ we can get the new sample.

For the large samples, we merge based on the class number. Every two samples in the same class $c$ will be merged using same method as above, subsampling with probability $p'/p_i$, and estimating the rank as new "local rank," and then the class number for a new sample would be $c+1$. We start from $c = 0$, and if there is no more than one sample in this class, we move on to the next $c+1$.

It has been shown that the total communication cost is bounded by $O(h\sqrt{k}/\varepsilon)$ in the original paper. To further improve the communication cost, they partition the routing tree into $k/h$ ($h$ is the height of the tree) connected components, so that the total communication cost will reduce to $O(\sqrt{kh}/\varepsilon)$.

**Algorithm 5** Merge for Tree Model (Sampling-Based)

**Input:** a vector of samples $\vec{d}$, probabilities $\vec{p}$ for each sample, class numbers $\vec{c}$
**Output:** new samples $\vec{d'}$, new probabilities $\vec{p'}$, new class numbers $\vec{c'}$

//We say that the class numbers of small samples are -1, so $\vec{c}$ stores all class numbers for samples in $\vec{d}$.
say all small samples in $\vec{d}$ is $\vec{d_s}$
**if** $size_of(\vec{d_s}) \leq n/\sqrt{k}$ **then**
    $p'$ = $1/(\varepsilon \times size_of(\vec{d_s}))$
    **for all** $d_i$ in $\vec{d_s}$ **do**
        samples $d_i$ with probability $p'/p_i$
        put sampled data into $d'$
    **end for**
    push $d'$ into $\vec{d'}$, $p'$ into $\vec{p'}$, $c' = \lfloor log(size_o f(d')\sqrt{k}/n) \rfloor$ into $\vec{c'}$
**end if**
**for** $c_i$ **from** 0 **to** $max(\vec{c})$ **do**
    **while** there are two or more samples in $\vec{d}$ whose class number are $c_i$ **do**
        select two samples $d_a$ and $d_b$
        $s' = size_of(d_a) + size_of(d_b)$, $p' = 1/\varepsilon s'$
        samples $d_a$ with probability $p'/p_a$
        samples $d_b$ with probability $p'/p_b$
        push all samples into $\vec{d'}$
        push $p'$ into $\vec{p'}$, $c' = \lfloor log(size_o f(thesampleddata)\sqrt{k}/n) \rfloor$ into $\vec{c'}$
    **end while**
**end for**

### 3.2.5 Example

In this section, we will show the process that the Sampling-Based algorithm uses to compute the 0.5-quantile for the first example dataset in Chapter 2 (Equation 2.1), and the merging process for both the example datasets (Equation 2.1 and 2.2). Since there are only two synopses to be merged, it is a flat mode and the estimation can be computed easily.

The data are sampled with a probability $p = \sqrt{k}/\varepsilon n = \frac{1}{3}$, where $k = 2$ is the number of synopses to be merged, $n = 30$ is the total number of items and $\varepsilon = \frac{\sqrt{2}}{10}$. The sampled items for the two example datasets are shown below.

$$\{4,7,6,2,0\} \tag{3.4}$$

$$\{6,7,3,2,3\} \tag{3.5}$$

After sorting the samples, we can compute the local ranks for each item, $r_i = r_{i-1} + 1/p$. The local ranks for the first sample $\{0,2,4,6,7\}$ are $\{0,3,6,9,12\}$; the local ranks for the second sample $\{2,3,3,6,7\}$ are $\{0,3,3,6,9\}$. To estimate the 0.5-quantile, we get the global ranks for each item by summing all the local ranks. Since the global rank for 6 is 15 (9 + 6) and is the closest to $\phi n$.

## 3.3 Q-Digest

The q-digest [4] introduced by Shrivastaca et al. is another deterministic quantile computation algorithm. It is only for integers and assumes a fixed universe $[u]$. The q-digest uses a virtual complete binary tree with $u$ leaves in which each leaf represents an integer in $[u]$ and each non-leaf node represents a dyadic interval. For instance, the root is the interval of $[0,u)$, and its children are $[0,u/2)$ $[u/2,u)$. Every node keeps a corresponding counter, say $c_v$. Initially all $c_v = 0$. To save the memory, a node will not be allocated unless its $c_v$ is not 0, and will not delete when $c_v$ becomes 0 again. The basic ideal of inserting and compressing is simple: It first reads all the data and counts the frequency, getting a distribution in the leaves; pushes up the values whose counter is small (condition will be provide later) through the tree. In such ways, we still can keep relatively precise information of high frequency values but make some error for low frequency ones.

For every internal node $v$ with $c_v > 0$ except the root, the following two conditions must be satisfied [4, 2]:

$$(1)\ c_v \leq \lfloor \varepsilon n/logu \rfloor$$

$$(2)\ c_v + c_{v_p} + c_{v_s} > \lfloor \varepsilon n/logu \rfloor$$

where $v_p$ is the parent of $v$, and $v_s$ is the sibling of $v$.

### 3.3.1 Build

To build a q-digest tree, we read all the data; for each value, we increase the corresponding leaf's counter by 1. After reading all the data, we start to compress the tree from bottom to top, push up the nodes that violate the conditions (2), since (1) will never be violated when we go from the bottom up [4].

The compress algorithm simply goes from the leaves to the root, scans every node in each level, if node $v$ is not satisfied condition (2), we add $c_v$ and $c_{v_s}$ to $c_{v_p}$, and delete them (see Algorithm 6).

**Algorithm 6** COMPRESS (Q-Digest)

**Input:** the binary tree $T$, ε, $n$, $u$

**Output:** new binary tree $T$

1. **for all** level $l$ in $T$ **from** bottom **to** top **do**
2. **for all** node $v$ in level $l$ **do**
3. **if** $c_v + c_{v_p} + c_{v_s} \leq \lfloor \varepsilon n / logu \rfloor$ **then**
4. $c_{v_p} = c_v + c_{v_p} + c_{v_s}$
5. $c_v = 0$, $c_{v_s} = 0$
6. **end if**
7. **end for**
8. **end for**

The condition (2) also helps us to bound the size of structure to $O(logu/\varepsilon)$ nodes, since for a q-digest Q

$$\sum (c_v + c_{v_p} + c_{v_s}) > |Q| \lfloor \varepsilon n / logu \rfloor$$

and

$$\sum (c_v + c_{v_p} + c_{v_s}) \leq 3n$$

### 3.3.2 Merge

To merge q-digests from two different datasets with same ε and $u$ is easy, simply add two q-digest together and do a compress.

### 3.3.3 Estimation

To answer a ϕ-quantile query, accumulate the counters of nodes in post-order till the sum is greater than $\phi n$, and return the right end point of the interval.

This q-digest algorithm needs to read all the data before doing anything, and it may cause problems when $u$ and $n$ are large, so we also implement the variant called FastQDigest introduced by [2] for our experiment as a comparison. In FastQDigest we have a “real” tree-structure, initially an empty tree T. When inserting a new element $e$, we first need to find its lowest ancestor $v$ (ancestor of $e$ means $e$ is in the range of $e$'s dyadic interval). We will choose the root if T is empty. Then, increase $c_v$ by 1 if it won't violate condition (1); otherwise we add the child of $v$ that is also ancestor of $e$ to T and set its counter to 1. We can call the compressing algorithm as above when $n$ doubles. Since this is a top down process, condition (2) will still hold.

### 3.3.4 Example

In this section, we will show the process that the q-digest algorithm uses to compute the 0.5-quantile for the first example dataset in Chapter 2 (Equation 2.1) and the merging process for both the example datasets (Equation 2.1 and 2.2). For our example, we choose 0.4 as the $\varepsilon$ so that $\lfloor \varepsilon n/logu \rfloor = 2$.

Figure 3.1 shows the q-digest processes the first example dataset. To get the estimation of the first example, we get the post-order of the tree in Figure 3.1b, which is 0, 1, [0,2), 2, 3, [2,4), [0,4), 4, 5, [4,6), 6, 7, [6, 8), [4,8), [0,8). Then, we sum the counters of each node in such order till the sum is larger than $\phi n$ (7.5). Thus, we can get 2 as the estimation.

Figure 3.2 and 3.3 show the q-digest processes in the second example dataset. To get the estimation, we sum the counters of each node in the post-order of the tree of Figure 3.3b till the sum is larger than $\phi n$ (7.5). Thus, we can get 3 as the estimation.

Figure 3.4 shows the processes merging the two example datasets. To get the estimation of the merged tree, we sum the counters of each node in post-order till the sum is larger than $\phi n$ (15). Thus, we can get 3 is the estimation.

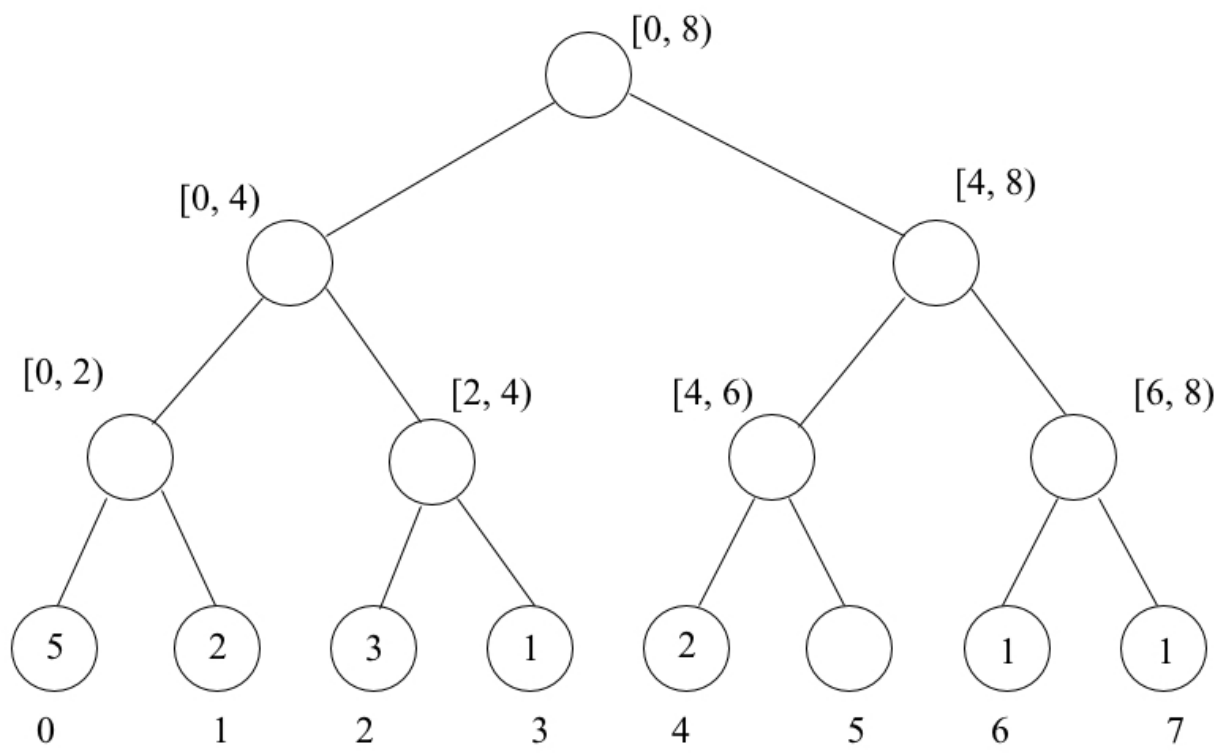

(a) Step 1, pass all the data and count each value

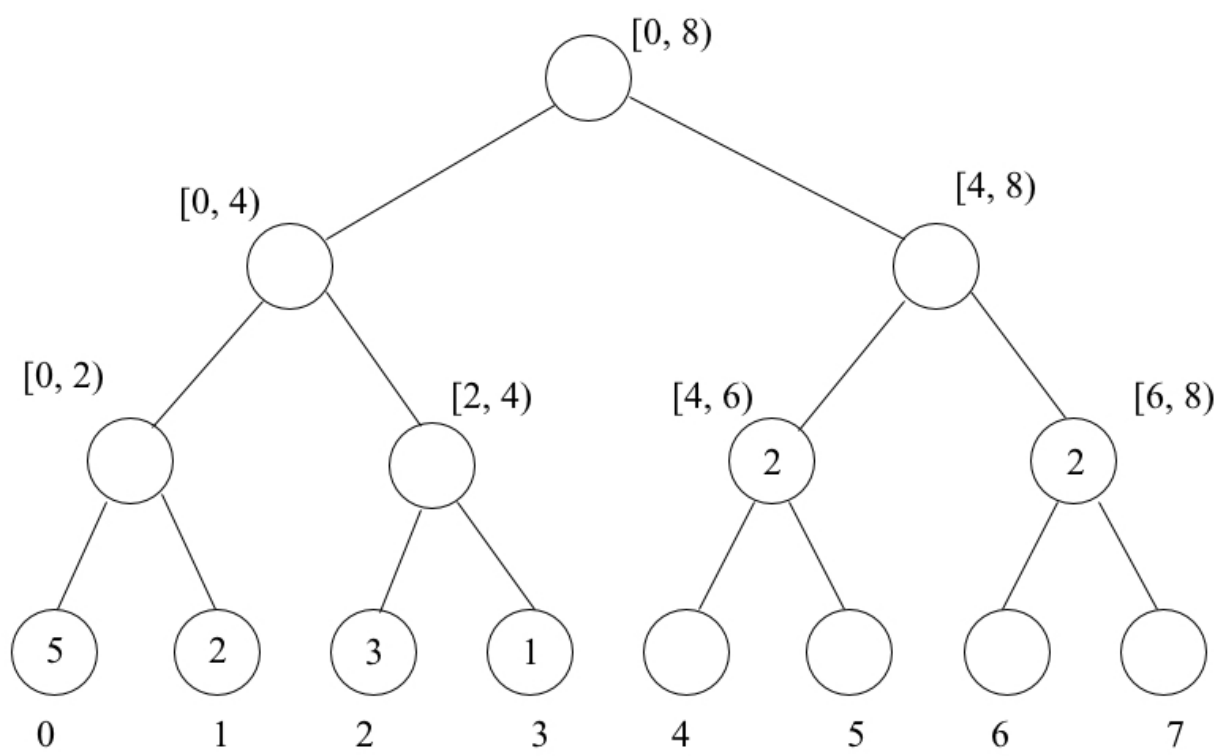

(b) Step 2, from the bottom to top, if a node violates the condition ($c_v + c_{v_p} + c_{v_s} \leq 2$), push it up

**Figure 3.1**: Example 1.

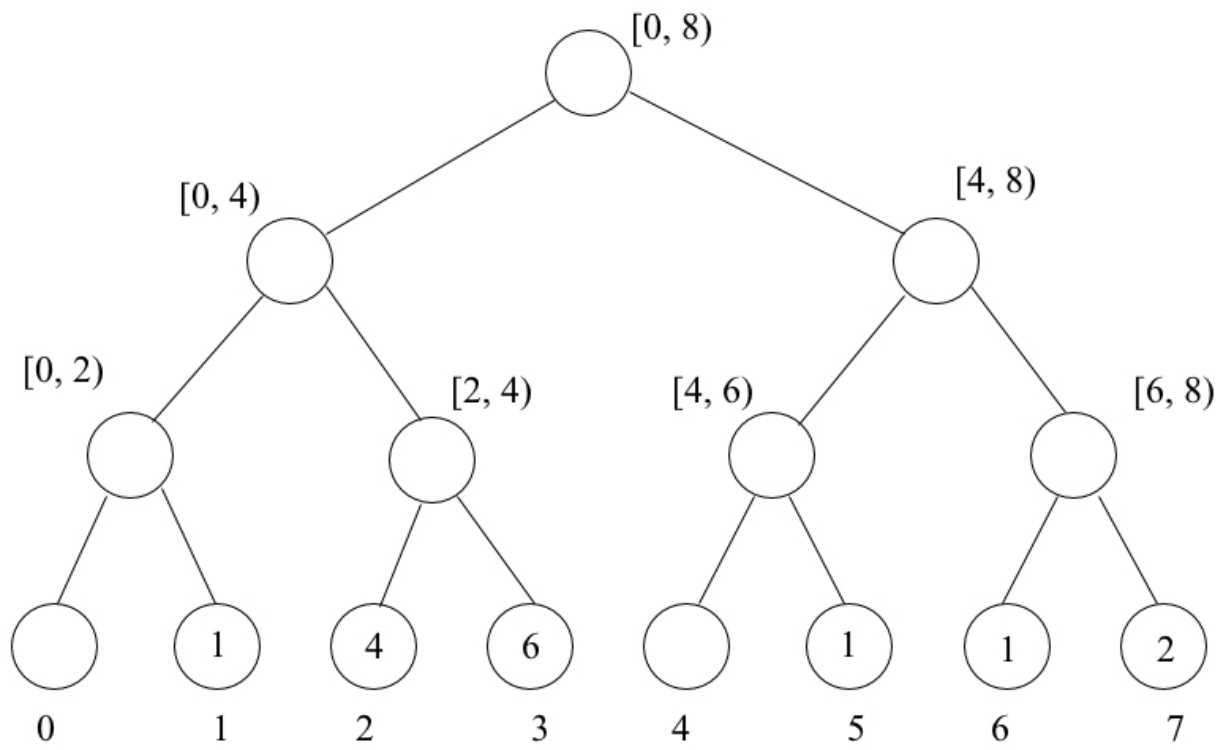

(a) Step 1, pass all the data and count each value

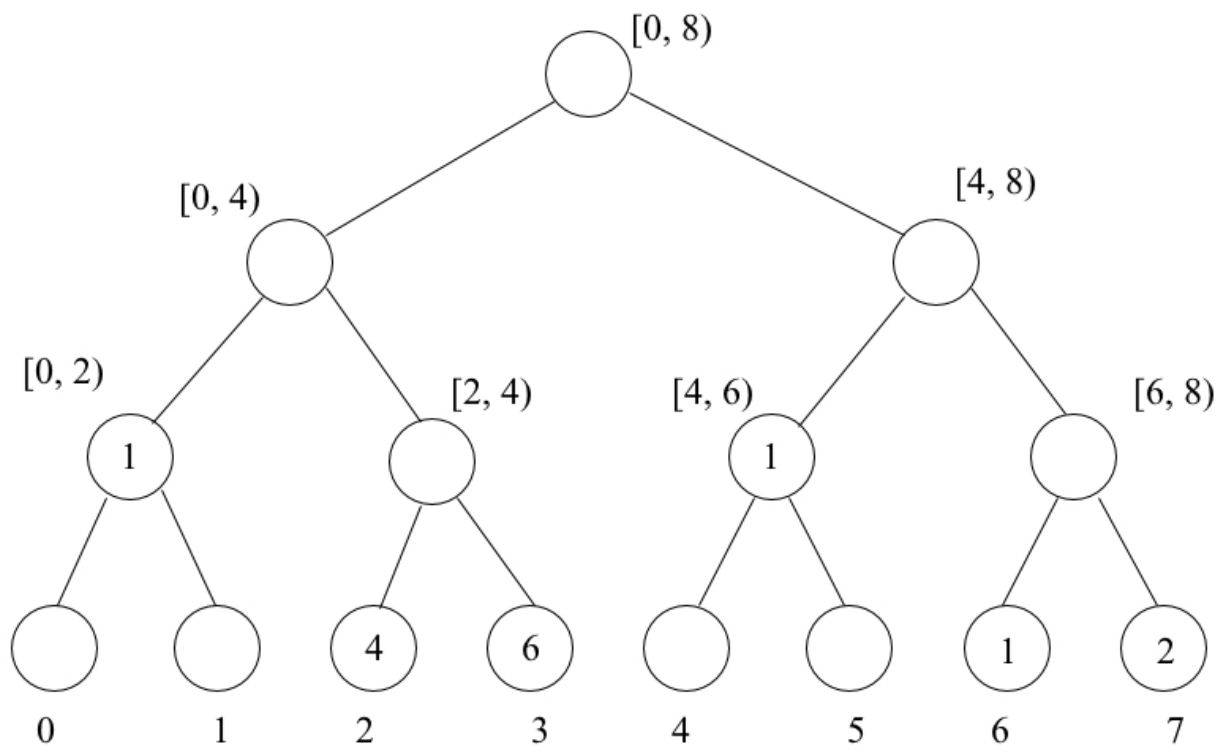

(b) Step 2, from the bottom to top, if a node violates the condition ($c_v + c_{v_p} + c_{v_s} \leq 2$), push it up

**Figure 3.2**: Example 2.

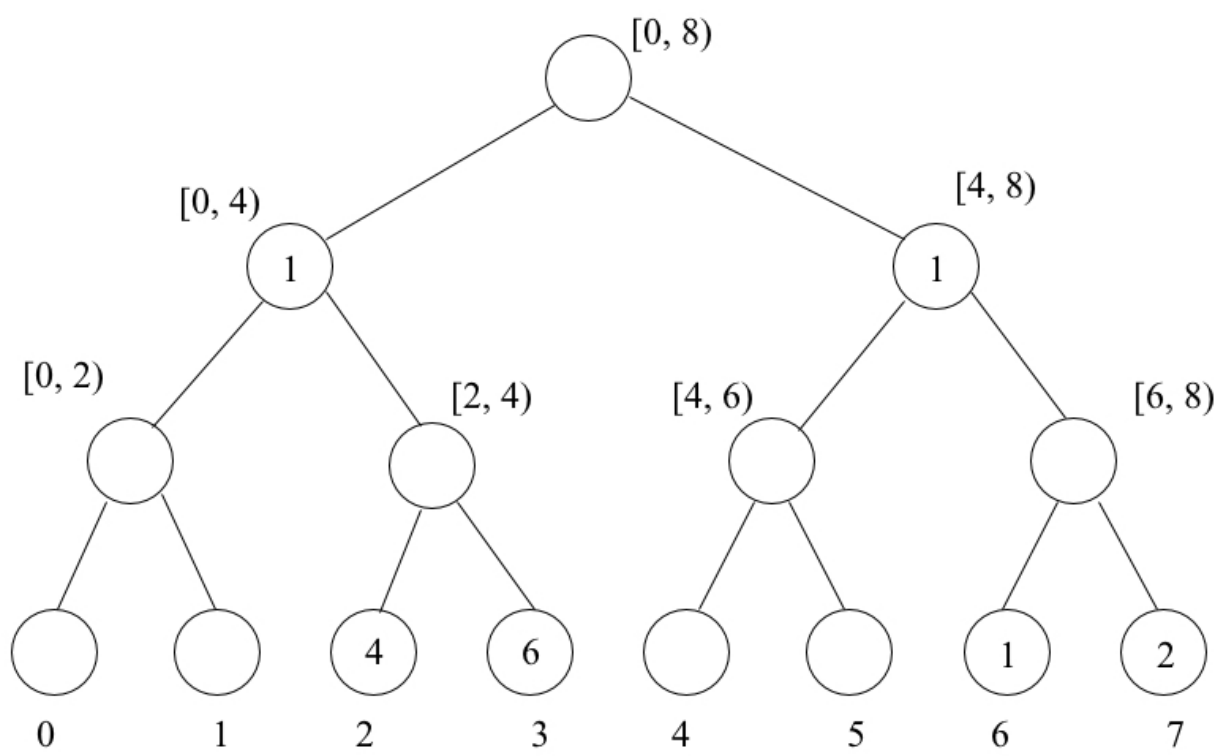

(a) Step 2, from the bottom to top, if a node violate the condition ($c_v + c_{v_p} + c_{v_s} \leq 2$), push it up

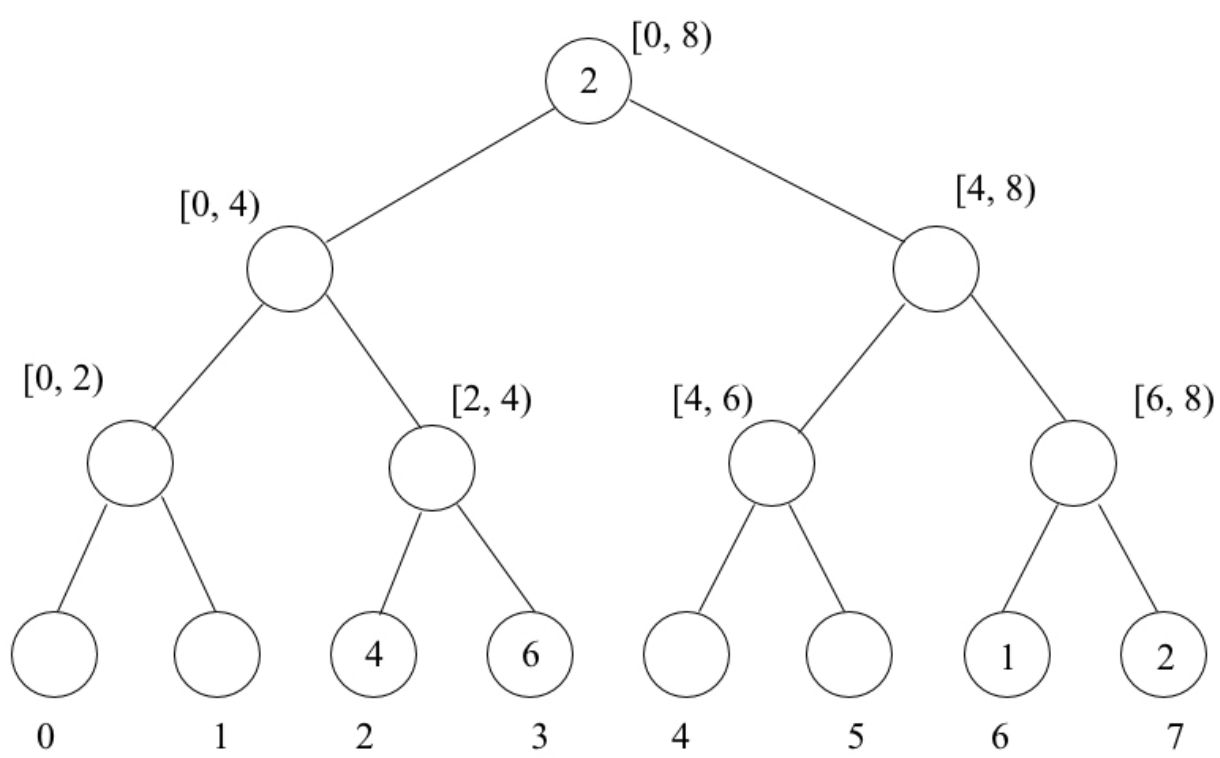

(b) Step 2, from the bottom to top, if a node violates the condition ($c_v + c_{v_p} + c_{v_s} \leq 2$), push it up

**Figure 3.3**: Example 2 (continued).

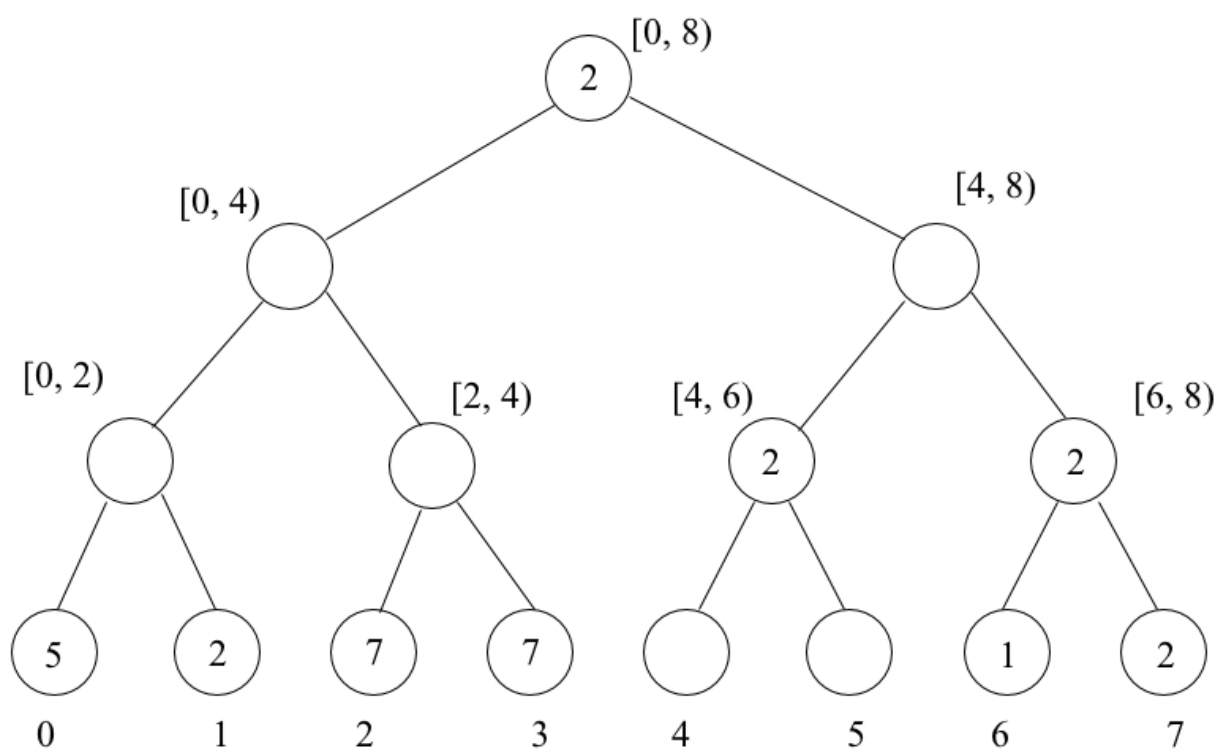

(a) Step 1, merge two trees by summing up counters of each nodes in two trees.

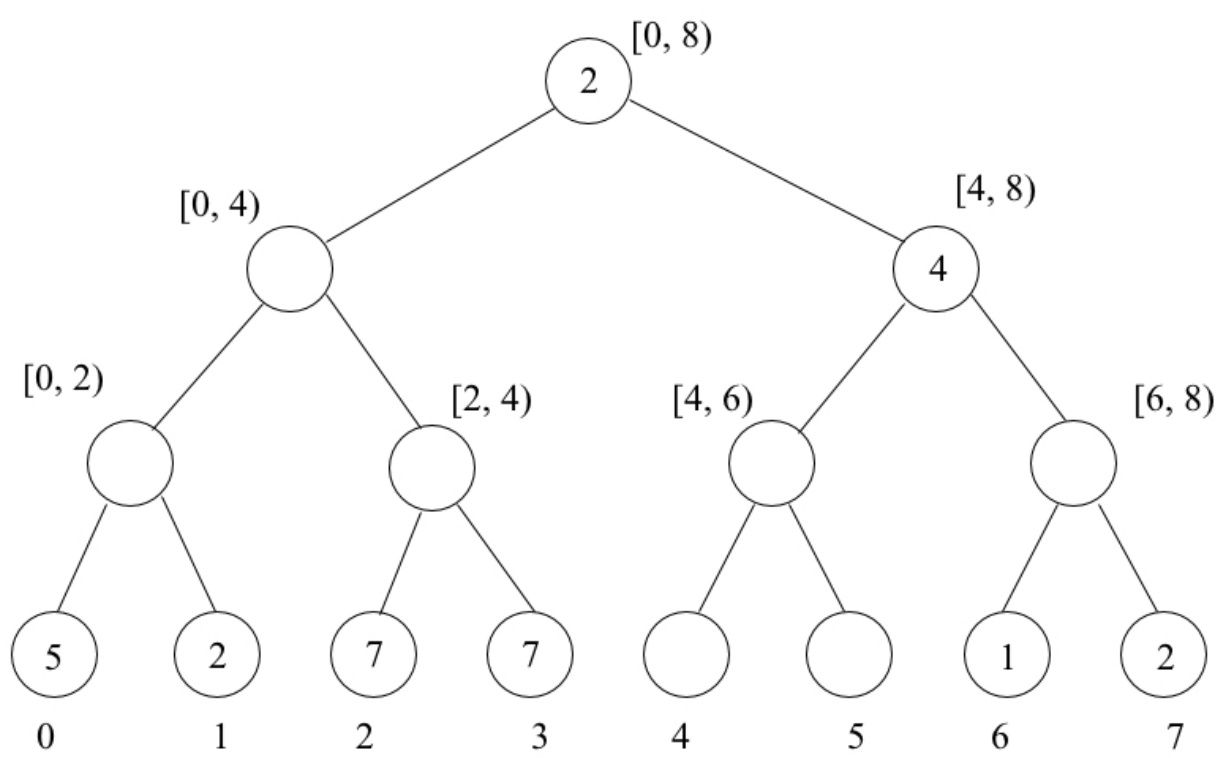

(b) Step 2, from the bottom to top, if a node violates the condition ($c_v + c_{v_p} + c_{v_s} \leq 4$), push it up

**Figure 3.4**: Merge Example 1 and 2.

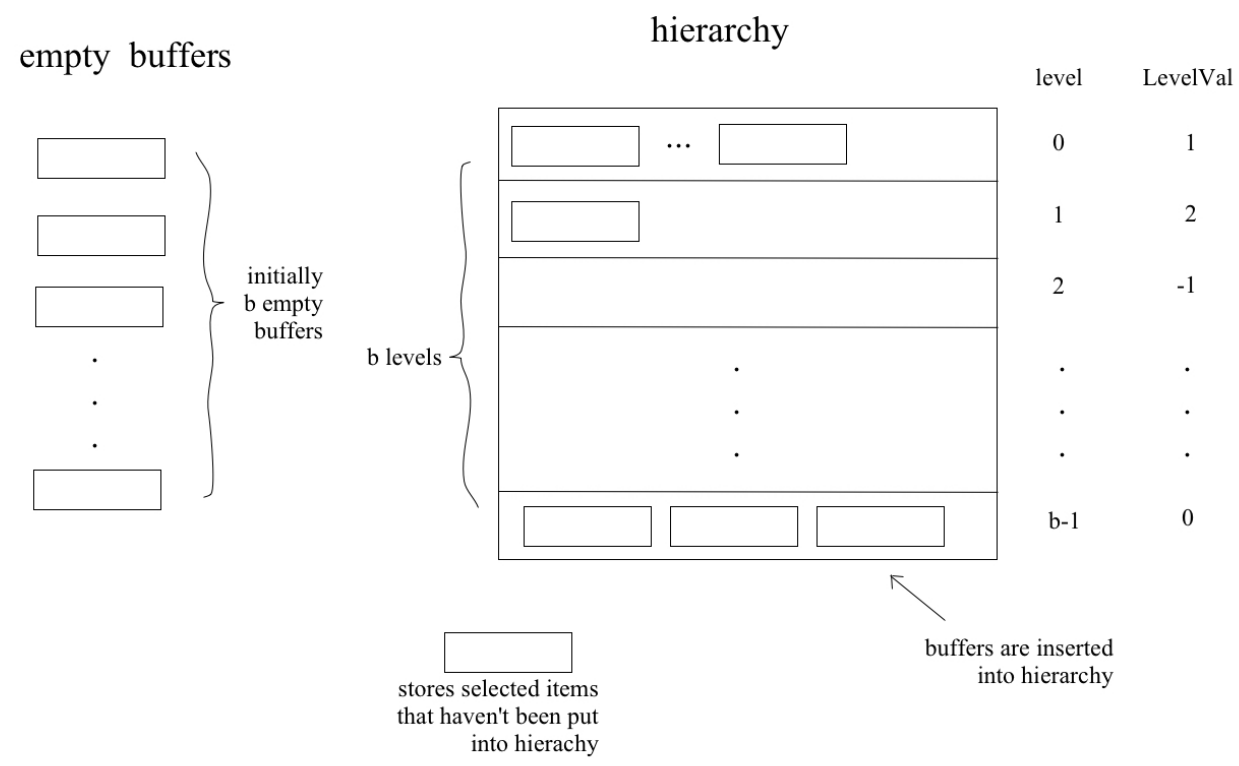

**Figure 3.5**: The structure of the Random Mergeable Summaries.

## 3.4 Random Mergeable Summaries

The Random Mergeable Summaries [6] is a ε-approximation non-deterministic quantiles algorithm introduced by P. Agarwal et al. Similar to the Sampling-Based algorithm [5], the Mergeable Summaries are also based on random sampling. It has some empty buffers at the beginning, when the data comes one-by-one we randomly choose some of them and put them to an empty buffer (if no empty buffer is available, we find two buffers in the same level and merge them), when this buffer becomes full, we put it on a buffers list with a level based on the number of elements we have seen. Note that, the elements in a buffer are always sorted. Figure 3.5 shows the synopses of the Random Mergeable Summaries.

We first define the following variables:

$$h = \lfloor \log \frac{1}{\varepsilon} \rfloor$$

$$b = h + 1$$

$$s = \lfloor \frac{1}{\varepsilon} \times \sqrt{\log \frac{1}{\varepsilon}} \rfloor$$

$$constLevel = 1 - 2 \times log\frac{1}{\varepsilon} - \frac{1}{2} \times log(log\frac{1}{\varepsilon})$$

We totally have $b$ buffers, each buffer is an array with size s. There is a $b$ levels hierarchy storing full buffers and each level is associated with a level value *LevelVal*, which initially is $-1$. In addition, there is a *curLevel*, which initially is 0, and will increase when total number of the items in the summaries increases.

### 3.4.1 Build

When new items are coming, randomly sample from the data until we get $s$ items. Then, we put these items into an empty buffer, if there, into the hierarchy, either on the first level we find that the level value is $-1$ or that the level value equals *curLevel*. If no empty buffer is available, we find the lowest *LevelVal* in hierarchy with at least two buffers and merge them randomly, and the resulting buffer goes into the next level and we will get an empty buffer (Algoritm 8). After putting the buffer into the hierarchy, we can update *curLevel* to either 0 or $\lceil constLevel + log(n+1) \rceil$ whichever is larger.

### 3.4.2 Merge

To merge two summaries $s_1$ and $s_2$, we first create a random buffer $B$ in $s * 2$ size. The new *curLevel* (*newLevel*) is $max(0, \lceil constLevel + log2(n_1 + n_2) \rceil)$. In summary $s_1$, there are some items that haven't been put into the hierarchy; we sample these items with

**Algorithm 7** ADDITEM

**Input:** $h$ is the hierarchy, $v$, $s$, $curLevel$, $n$, a list $\vec{t}$, an empty buffers list $\vec{b}$, $constLevel$
**Output:** new hierarchy $h$

randomly decide if we keep the item $v$.
**if** we don't keep this item **then**
  **return** $h$
**end if**
push $v$ to $\vec{t}$.
**if** $\vec{t}$ has $s$ items **then**
  **if** there is an empty buffer in $\vec{b}$ **then**
    get a buffer $b$ from $\vec{b}$
  **else**
    find the level $h_i$ with the lowest $LevelVal$ in $h$ with at least two buffers
    call MERGE_TWO_BUFFERS to get an empty buffer $b$
  **end if**
  put all items in $\vec{t}$ to buffer $b$ and clear $\vec{t}$.
  **for** level $l$ in $h$ **do**
    **if** $LevelVal = -1$ or $LevelVal = curLevel$ **then**
      push $b$ into $l$
      **break**
    **end if**
  **end for**
  $LevelVal = max(0, \lceil constLevel + log(n+1) \rceil)$
  **return** $h$
**end if**

**Algorithm 8** MERGE_TWO_BUFFERS

**Input:** $h$ is the hierarchy, $l$ is the level of hierarchy that have the lowest $LevelVal$, $s$
**Output:** an empty buffer

**for all** levels in $h$ starting from $l$ **do**
  **if** there are at least two buffers in this level $i$ **then**
    Pop out two buffers, say $b1$ and $b2$.
    Randomly choose $s$ elements in $b1$ and $b2$; put these elements into $b3$.
    Push $b3$ into level $i+1$ of $h$; $LevelVal_{i+1} = LevelVal_i + 1$.
    **break**
  **end if**
**end for**
Erase $b1$
**return** $b1$

a $min(1, \frac{1}{2^{newLevel-curLevel_1}})$ factor. For summary $s_2$, we do the same thing, sampling the items that haven't been put into the hierarchy with a $min(1, \frac{1}{2^{newLevel-curLevel_2}})$ factor. Then, we can put those sampled data into the random buffer $B$. If $B$ has more than $s$ items, put the first $s$ items into the hierarchy of $s_1$.

We divide both hierarchies of $s_1$ and $s_2$ into two parts, the levels of buffers in a hierarchy that are less than *newLevel* and the levels greater than *newLevel*. For the first part we take out one by one from the hierarchies, sample and put into $B$, then put the first $s$ items into the hierarchy of $s_1$ with *newLevel*. After all buffers whose level is less than *newLevel* have been processed, put the rest items in the $B$ into the hierarchy. For the second part, the levels of buffers in the hierarchies, we take out every buffer (say $b$) in the hierarchy of $s_2$ and merge it with a buffer in the same level (say level $l$) in the hierarchy of $s_1$. If no more buffers are left in the level $l$, we find an empty buffer in $s_1$ (Algorithm 8) and put items from $b$ to this empty buffer to push to level $l$ of $s_1$'s hierarchy.

### 3.4.3 Estimation

Before querying a quantile, we need to finalize the summary, this is because of the items that haven't been put into the hierarchy. Recall that, when we add items, we first sample from incoming items and wait until we get the $s$ sampled items we put them in an empty buffer and plug this buffer into the hierarchy. If there is no empty buffer available at that time, use Algorithm 8 to get an empty buffer. Thus at the time we want to query, there may be some items haven't been plugin hierarchy, we need to find an empty buffer, put these items into this buffer and then put this buffer into the hierarchy. Besides, to easier query, we also merge all the buffers together to a single array.

To compute a quantile, $\varepsilon$, we find from the smallest item to the largest, and sum the $2^{LevelVal}$ of each item, the first item we find when the sum of $2^{LevelVal}$ (previous items' *LevelVal*) is no less than $\varepsilon n$ or the last item is the final estimation. The size of the

**Algorithm 9** MERGE_HIERARCHIES

**Input:** hierarchies $h_1$ $h_2$, $newLevel$, $B$, $s$

**Output:** $h_1$

```
for all levels l in h1 and h2 whose LevelVal < newLevel do
    samples the elements in all buffers in l and put the sampled data into B
    push first s items in B into h1 with newLevel
end for
push the rest items in B into h1
for all levels l in h2 whose LevelVal > newLevel do
    find the level l' in h1 that has same level with l
    for all buffers b in l do
        if l' has a buffers b' then
            merge two buffers b and b'
        else
            get an empty buffer b' in h1
            put all items in b into b'
            push b' into l'
        end if
    end for
end for
return h1
```

structure can be bounded in $O(s \times b)$.

### 3.4.4 Example

In this section, we will show the process that the Random Mergeable Summaries algorithm uses to compute the 0.5-quantile for the first example dataset in Chapter 2 (Equation 2.1), and the merging process for both the example datasets (Equation 2.1 and 2.2). Figures 3.6, 3.7, and 3.8 show the process for the first example; Figure 3.9 shows the merged result of the two examples and the final estimation.

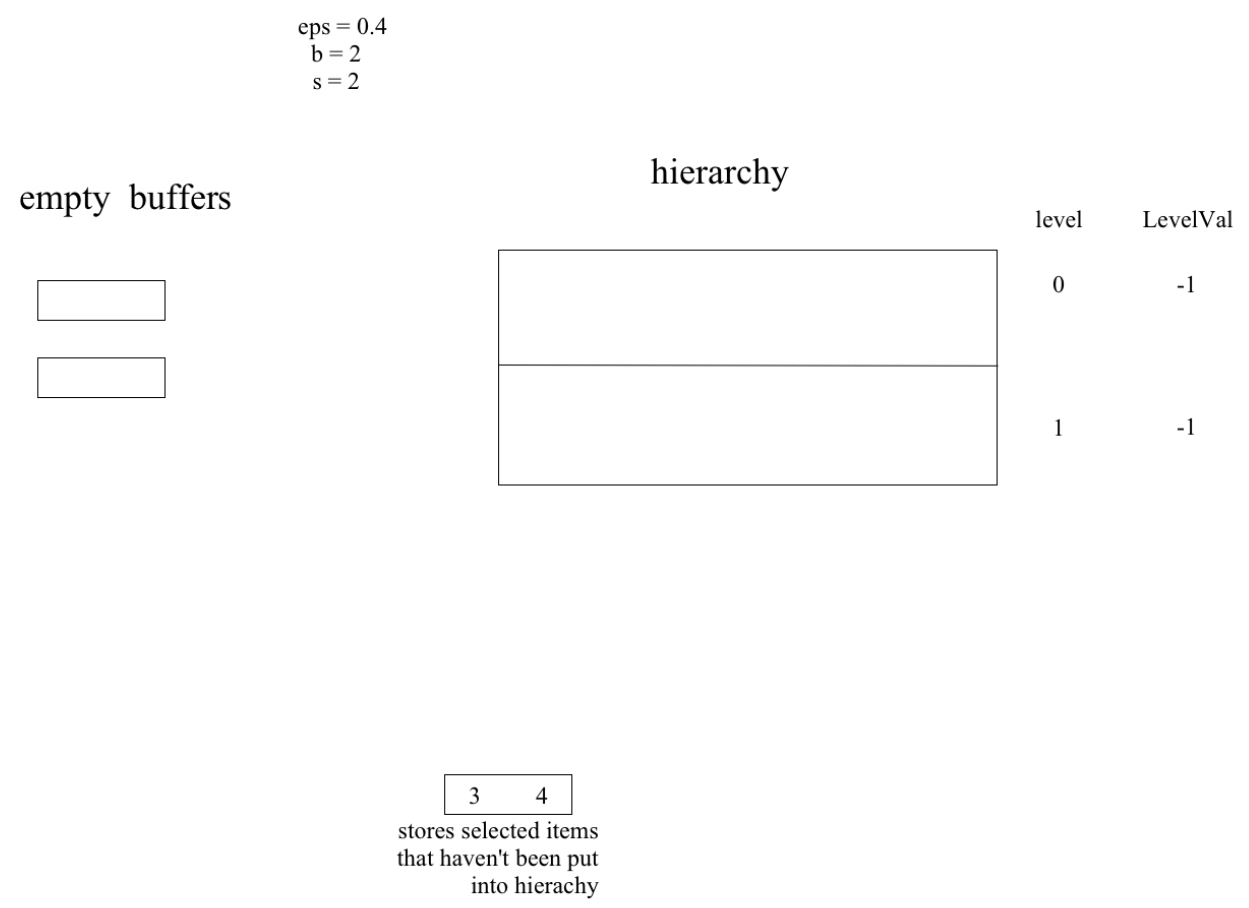

(a) At the beginning, there are two empty buffers, and a two-level hierarchy. The first two selected items, 3 and 4 are stored in an additional stage before put into hierarchy.

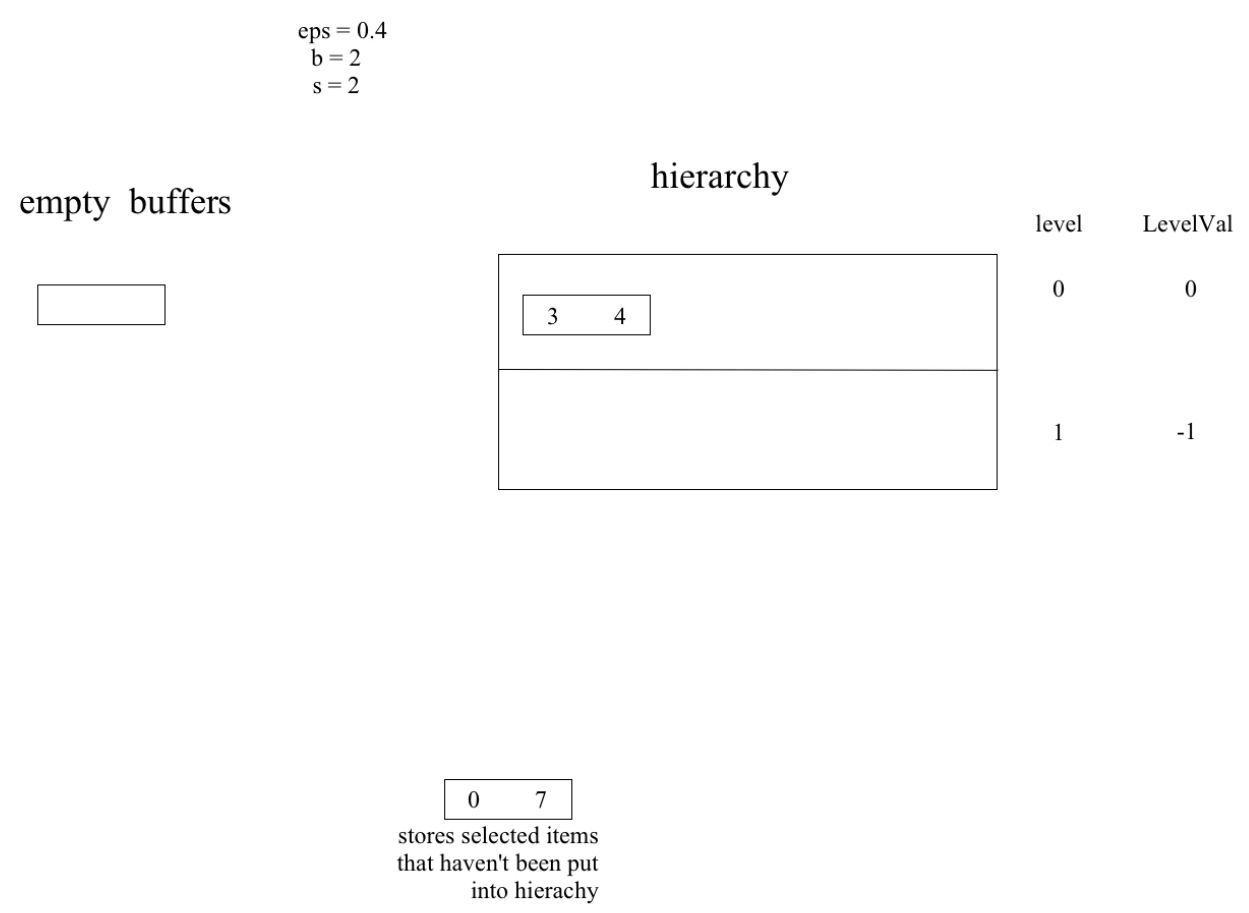

(b) Use an empty buffer to take the items in the stage and put them into level 0 of the hierarchy. The LevelVal becomes 0. Then, the next selected items, 0 and 7, are stored in the stage.

**Figure 3.6**: Example 1.

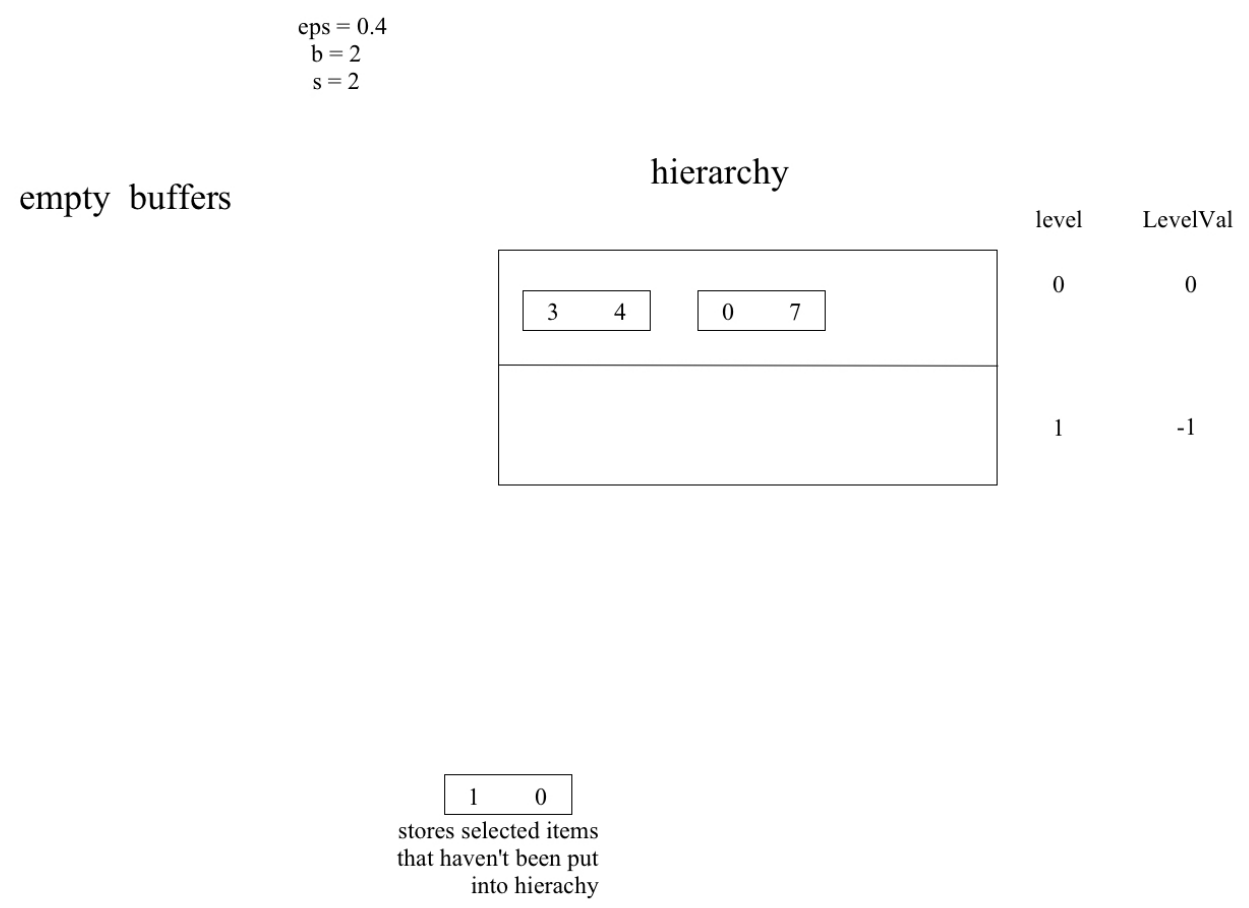

(a) Put 0 and 7 into level 0. Then, select the 1 and 0; note that the 0 is not token.

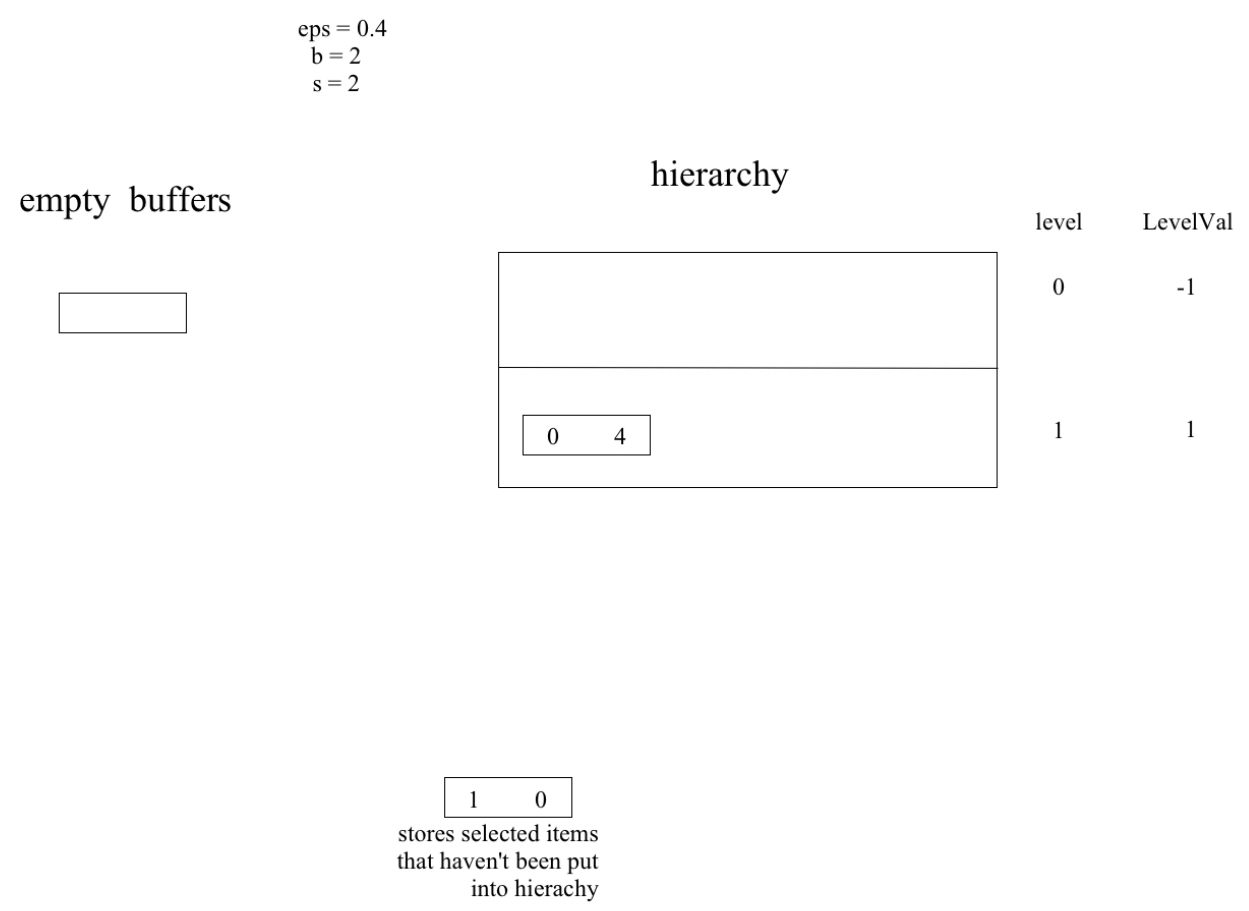

(b) Before taking the items in the stage, we first need to find an empty buffer. We merge the two buffers in level 0 of the hierarchy into one buffer, and then put it into level 1 (now LevelVal of level 1 is 1).

**Figure 3.7**: Example 1 (continued).

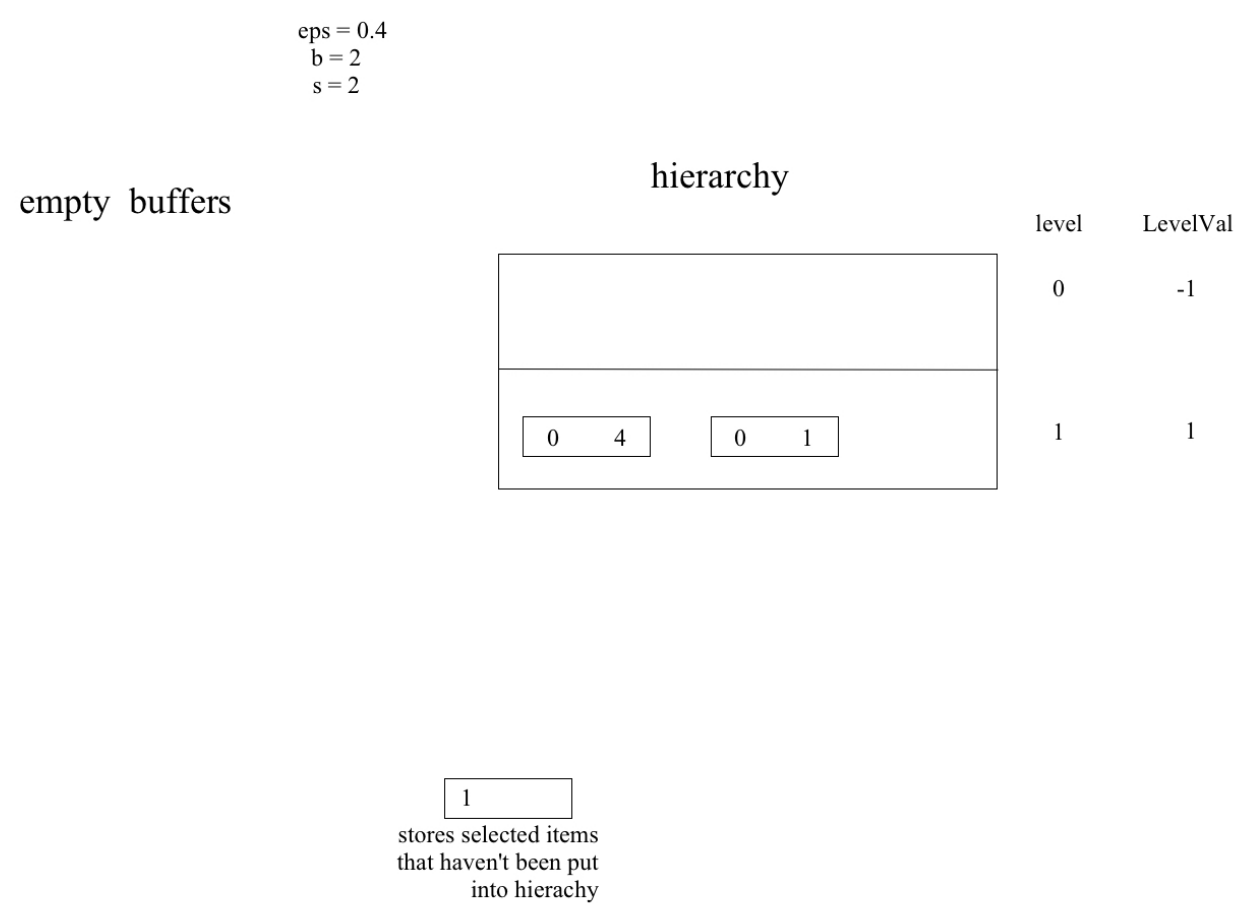

(a) Then, we can suck 1 and 0 into the hierarchy with the empty buffer. In the rest of data, only a 1 is chosen due to the random process.

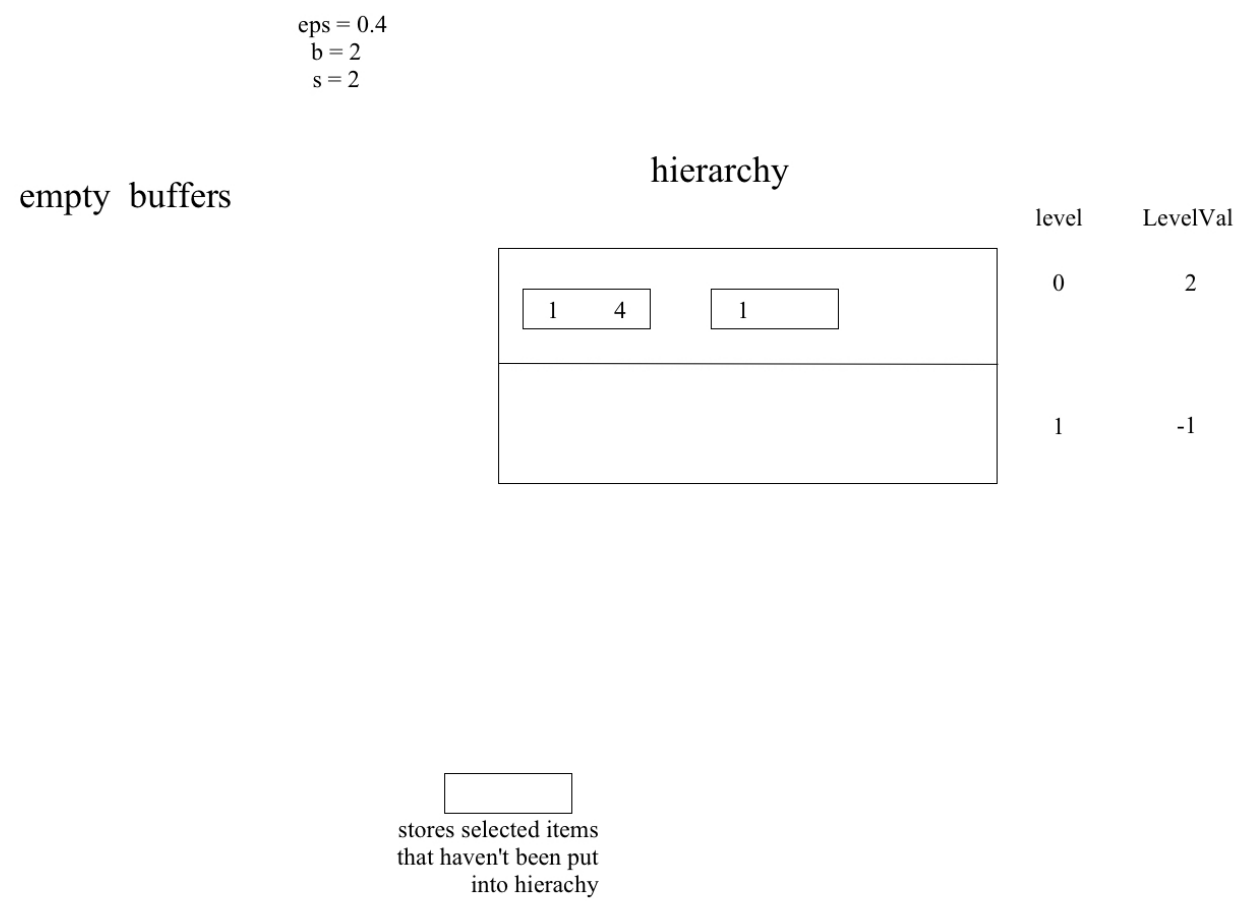

(b) The final step before estimation is to suck the item that is in the stage into the hierarchy. Since no more empty buffers are available, we first merge the two buffers in level 1. The estimation is 4, since we sum up $2^{LevelVal}$ until 4 to get more than 7.5.

**Figure 3.8**: Example 1 (continued).

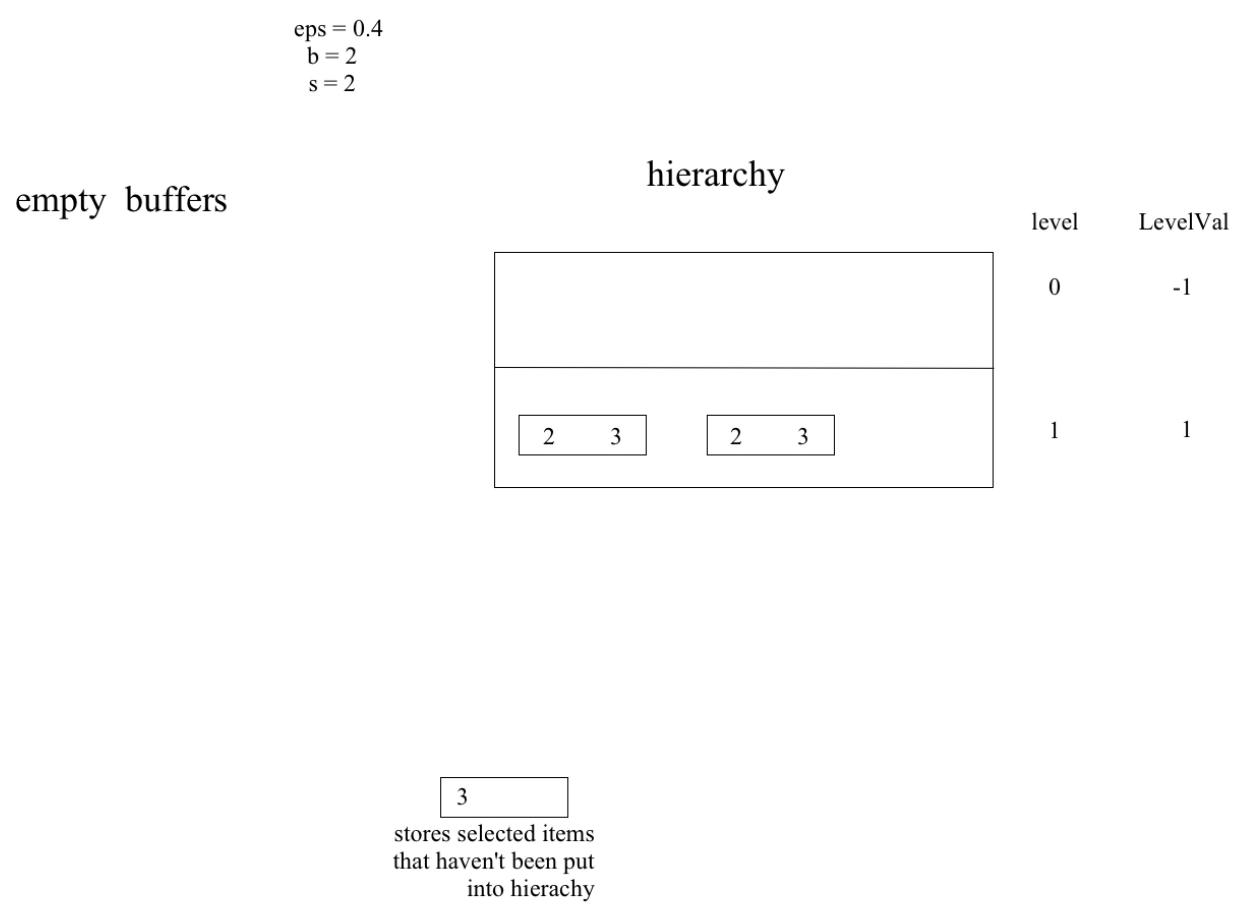

(a) This is the result of the second example dataset (before the final step).

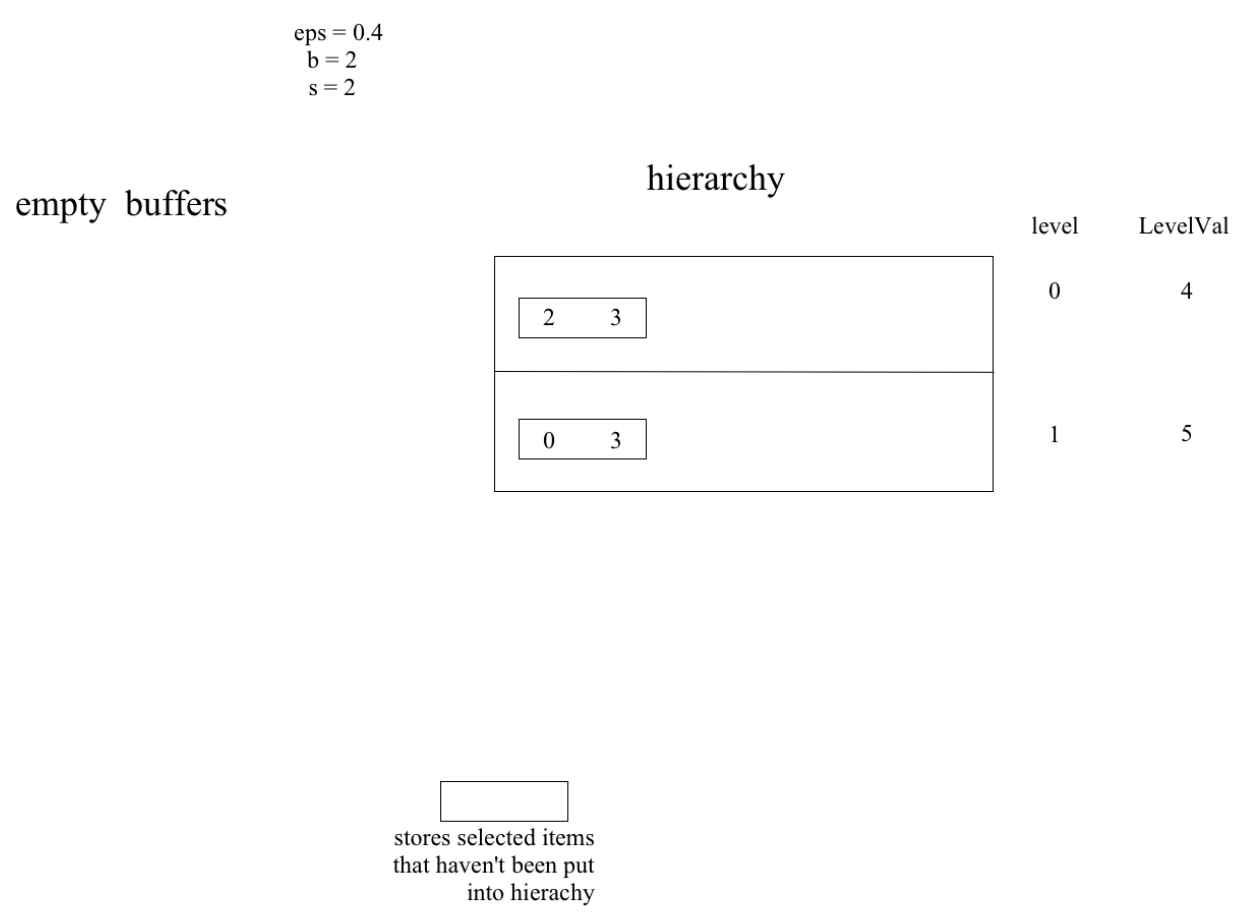

(b) This is the merged result of summaries in 3.8a and 3.9a. The estimation is 2 since we sum up $2^{LevelVal}$ until 2 to get more than 15.

**Figure 3.9**: Merge Examples and estimation.

# Chapter 4

# Interface of GLADE

In this chapter, we will introduce the *GLADE* [9] system, in which we implemented and executed all the experiments, present more details about the user interface of *GLADE*, Generalized Linear Aggregates (GLA) [9], and show how we implement the parallel quantile computation for the five algorithms in *GLADE*. Figure 4.1 shows the system architecture of **GLADE**.

## 4.1 Introduction to GLADE

*GLADE* is short for Generalized Linear Aggregate Distributed Engine and is a scalable distributed system for large scale data analytics [9]. *GLADE* provides an engine to optimize execute user-defined aggregates functions. With well-organized architecture, *GLADE* has highly efficiency and very good performance on multi-query processing.

The storage system of *GLADE* is a relational multi-query database system, DataPath [9] [10]. After the initial loading, the dataset will be stored in the database system, and will be passed to GLAs [9] when *GLADE* is executing. In addition, the data in the database system will be partitioned into several chunks, and user can configure the number of chunks. While *GLADE* is executing, the chunks will be loaded into memory

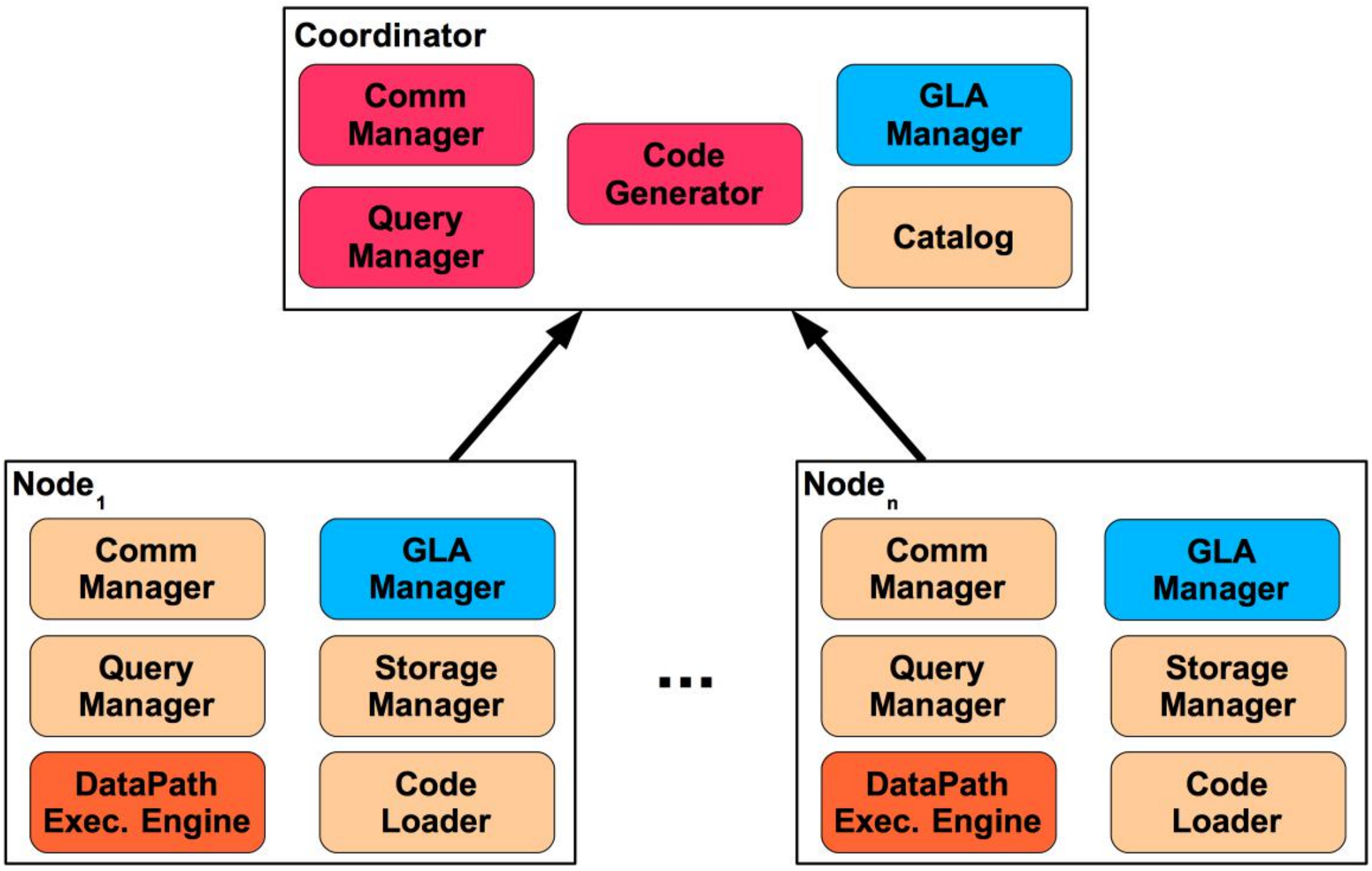

**Figure 4.1**: GLADE system architecture.

one by one.

Our *GLADE* framework has nine clusters total, one of them is a master node, and other eight nodes are organized as a binary tree. Each node will send its result GLA to its parent node, and the parent node will merge its own GLA and incoming GLAs then send to its parent node; finally, the root node will send the final result to the master node. Furthermore, for every single node, *GLADE* can start multiple threads to run the GLA. The number of threads is also configurable. If there is a thread available, *GLADE* will start and execute a GLA. Once there is no data chunk to be process, this GLA is finished and is ready to be merged.

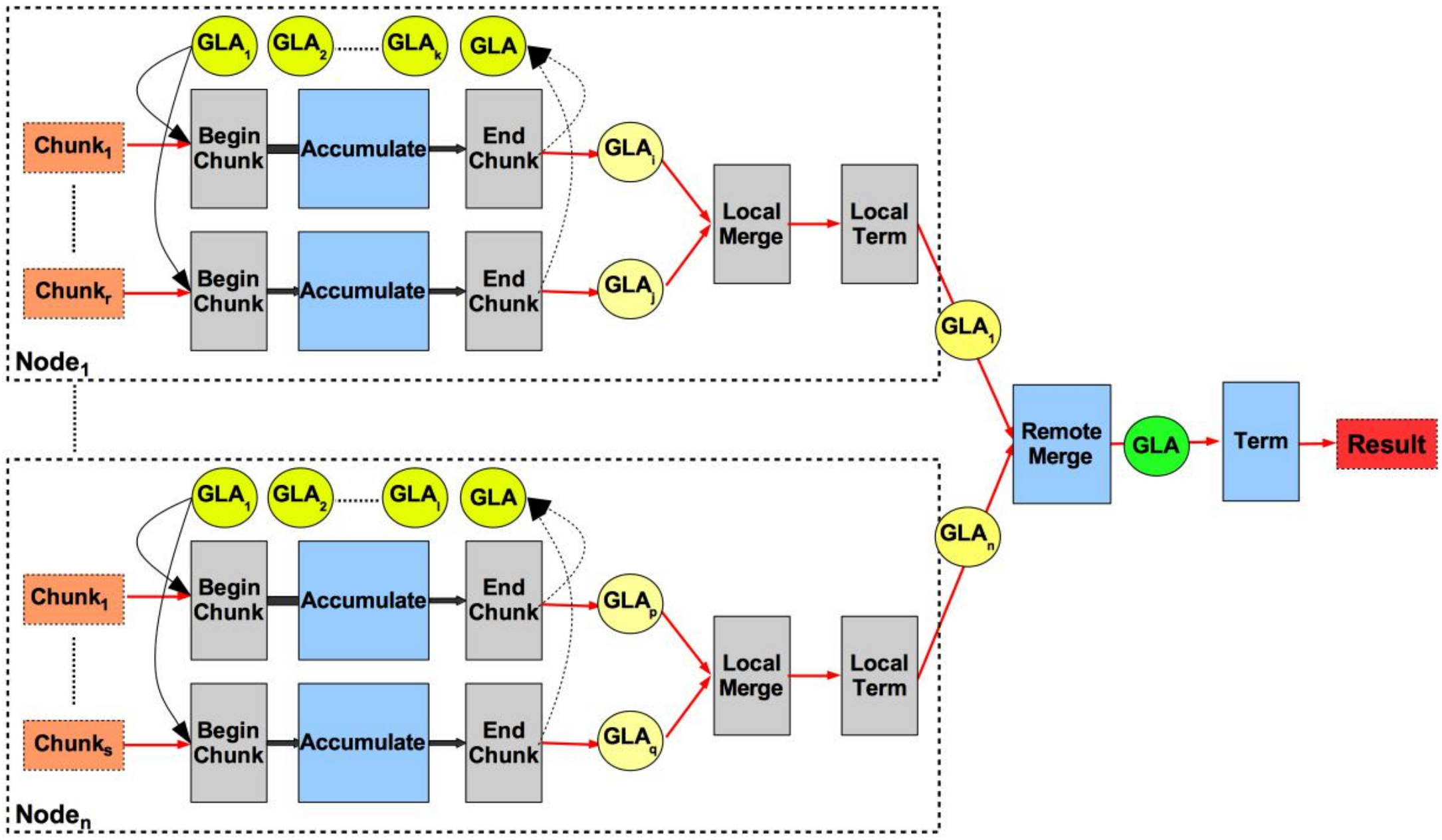

**Figure 4.2**: GLA interface.

## 4.2 GLA

In this section, we will provide some details about the GLA, and present how the quantile algorithms are adapted in the GLA.

The GLA is short for Generalized Linear Aggregate [9]. It is a user interface of *GLADE* system for user-defined aggregate functions. Figure 4.2 shows how the user-defined functions works in GLAs. We will describe some of the GLA functions in details.

**BeginChunk.** This function will be executed before the system starts to process a chunk of data.

**AddItem.** This function will be a call for each tuple in the chunk. A tuple can contain multiple attributes, which is defined by the user while initially loaded to the database system. The *GLADE* will pass a tuple to AddItem as a parameter, and the user can process the data in this function.

**EndChunk.** Our system will execute EndChunk when a chunk of data is finished.

**AddState.** This function will be used for merging GLAs in one node. Note that, a GLA will be created and executed when there is a thread available; the working GLA will keep running until no more chunks are to be processed. If there are two GLAs finished in one node, *GLADE* will merges these two GLAs by executing AddState function.

**LocalFinalize.** After finishing all chunks of data in a node and merging all GLAs, the system will call LocalFinalized to finalize the work of this node.

**Serializer and deserializer.** Serializer and deserializer functions are used in the communication between nodes. User defines the serializer function to tell the system which variables need to be serialized and how to serialize them. When a node finishes its job, it will serialize the GLA by calling the serializer function and sending the serialized data to its parent node or master node. When a node receives the data, it will execute the deserializer function to retrieve the GLA. The user should define the deserializer function in the proper ways, such that the system can recover the information.

**AddGlobalState.** When a node receives a GLA from another node, it will execute AddGlobalState to merge that GLA with its own and send the result to its parent or master node. Typically, a node will wait until all its child nodes finish their jobs and get their GLAs. A node will merge all GLAs, including its own and the GLAs received from its children.

**Finalize.** This is the final process. The finalize function will be executed in the master node. After the master node receives the final GLA from other nodes, it will call the finalize function to allow tthe user to do some necessary processes before the program ends.

## 4.3 Implementations in GLADE

In this section, we will present how we integrate the quantile computation algorithms (GK, Sampling Based, Q-Digest, FastQDigest, and Random Mergeable Summaries) into *GLADE*. We first implemented the algorithms in a normal computer with C++ and test all functionality and correctness; then we implemented the *GLADE* version.

### 4.3.1 Normal Version

We implemented all five algorithms as described above in Chapter 3. Each of the algorithms keeps their own data structures as they need and contains the following common functions, which have been declared in Chapter 3.

- Build: insert items into the synopses
- Merge: merge two synopses
- Estimation: answer a quantile query

We implemented a driver for each algorithm, which partitions the data into several parts, and reads the data one-by-one in one partition then calls the Build function. After processing all the data, it calls the Merge function (for the Sampling-Based algorithm, it is the Improved Merge for Tree Model) to get a final synopses, and then it queries this synopses with designed quantiles.

In this version, data is stored in plain text files. The details about our experimental data will be discussed in Chapter 5.

### 4.3.2 GLADE Version

After we finished and checked the correctness of the normal version quantile algorithms, we integrate them into *GLADE*. We implemented one GLA for each of the five algorithms, and in each GLA we will call the Build, Merge and Estimation functions. In the constructor of GLA, we initialize the data structure and set the $\varepsilon$. We will give the details implementation of each quantile computation algorithm, using the Build, Merge (or Improved Merge for Tree Model), and Estimation functions we declared in Chapter 3.

Algorithm 10 shows the **GLADE** version GK algorithm. We call the Build function in AddItem, and the Build function will determine when to compress the synopses. Two synopses are merged when two GLAs are merged in AddState and AddGlobalState. In Finalize, we call the Estimation function to get the $\phi$-quantile.

---

**Algorithm 10** GLADE_VERSION_GK

**Input:** pre-loaded data chunks, $\varepsilon$, $\phi$
**Output:** estimation of $\phi$-quantile

**procedure** AddItem($v$)
call Build($v$)
**end procedure**

**procedure** AddState()
Merge two GLAs by calling Merge() of GK algorithm to merge two synopses.
**end procedure**

**procedure** AddGlobalState()
Merge two GLAs from different nodes by calling Merge() of GK algorithm to merge two synopses.
**end procedure**

**procedure** Finalize()
Call Estimation($\phi$) to answer queries.
**end procedure**

---

Algorithm 11 shows the **GLADE** version Sampling-Based algorithm. We call the Build function in AddItem to sample data and store the selected items in synopses. Before merging or estimating, a synopses must compute its local rank first, and this computation is done only once. Two synopses are merged when two GLAs are merged in AddState and AddGlobalState. In Finalize, we call the Estimation function to get the $\phi$-quantile.

**Algorithm 11** GLADE_VERSION_Sampling-Based

**Input:** pre-loaded data chunks, $\varepsilon$, $\phi$
**Output:** estimation of $\phi$-quantile

**procedure** AddItem($v$)
call Build($v$)
**end procedure**

**procedure** AddState()
Compute local rank for each synopses if it is not done before.
Merge two GLAs by calling Improved Merging for Tree Model of Sampling-Based algorithm to merge two synopses.
**end procedure**

**procedure** LocalFinalize()
compute local rank for each item in synopses if it is not done before.
**end procedure**

**procedure** AddGlobalState()
Compute local rank for each synopses if it is not done before.
Merge two GLAs from different nodes by calling Improved Merging for Tree Model of Sampling-Based algorithm to merge two synopses.
**end procedure**

**procedure** Finalize()
Call Estimation($\phi$) to answer queries.
**end procedure**

Algorithm 12 shows the **GLADE** version q-digest algorithm. We call the Build function to build the binary tree and compress the tree in EndChunk. Two binary trees are merged when two GLAs are merged in AddState and AddGlobalState. In Finalize, we call the Estimation function to get the $\phi$-quantile.

---

**Algorithm 12** GLADE_VERSION_QDigest

**Input:** pre-loaded data chunks, $\varepsilon$, $\phi$
**Output:** estimation of $\phi$-quantile

**procedure** AddItem($v$)
call Build($v$) to add item $v$ to the binary tree.
**end procedure**

**procedure** EndChunk()
Compress the binary tree.
**end procedure**

**procedure** AddState()
Merge two GLAs by calling Merge() of q-digest algorithm to merge two synopses.
**end procedure**

**procedure** AddGlobalState()
Merge two GLAs from different nodes by calling Merge() of q-digest algorithm to merge two synopses.
**end procedure**

**procedure** Finalize()
Call Estimation($\phi$) to answer queries.
**end procedure**

---

Algorithm 13 shows the **GLADE** version of the Random Mergeable Summaries algorithm. We call the Build function in AddItem to randomly pick items and put them in a hierarchy. Two synopses are merged when two GLAs are merged in AddState and AddGlobalState. In Finalize, we call the Estimation function to get the $\phi$-quantile.

**Algorithm 13** GLADE_VERSION_Random_Mergeable_Summaries

**Input:** pre-loaded data chunks, $\varepsilon$, $\phi$

**Output:** estimation of $\phi$-quantile

```
procedure AddItem(v)
call Build(v).
end procedure

procedure AddState()
Merge two GLAs by calling Merge() of Random Mergeable Summaries algorithm to
merge two synopses.
end procedure

procedure AddGlobalState()
Merge two GLAs from different nodes by calling Merge() of Random Mergeable
Summaries algorithm to merge two synopses.
end procedure

procedure Finalize()
Call Estimation(φ) to answer queries.
end procedure
```

The data is loaded into the database system prior to executing the experiments. We loaded the data from the files we used in the normal version. The data will be partitioned into several chunks, and the number of chunks is configurable. For example, if the total number of items is 100 and we set 10 chunks, then the first 10 items will be loaded into one chunk, the next 10 items will be loaded into another chunk, and so on. If there is only one thread, the GLA will process the chunks one by one; however, when there are multiple threads, the *GLADE* will determine which GLA processes which chunks.

# Chapter 5

# Experiment

The purpose of this experiment is to evaluate and compare the five algorithms, GK [1], q-digest [4], Sampling-Based [5], and Random Mergeable Summaries [6], and to investigate the accuracy, space usage, and execution time across different datasets.

**Implementation.** We implemented GK [1], q-digest [4], Sampling-Based [5], and Random Mergeable Summaries [6] algorithms in C++, and the random variables generation are based on Sketch-Based Estimations [7, 8].

**System.** We execute the experiment in *GLADE* on a standard server with 2 AMD Opteron 6128 series 8-core processors – a total of 16 cores – 40 GB of memory, and four 2 TB 7200 RPM SAS hard-drives configured RAID-0 in software. Each processor has 12 MB of L3 cache, while each core has 128 KB L1 and 512 KB L2 local caches. The storage system supports 240, 436, and 1600 MB/second minimum, average, and maximum read rates, respectively—based on the Ubuntu disk utility. The cached and buffered read rates are 3 GB/second and 565 MB/second, respectively. Ubuntu 14.04.2 SMP 64-bit with Linux kernel 3.13.0-43 is the operating system. We use the *GLADE* framework to execute our experiment, and the details about *GLADE* and how we integrate the algorithms to *GLADE* have been discussed in Chapter 4. We load the data into 1,024 chunks, and we use only one node. We will have various numbers of threads in our experiment.

## 5.1 Setup

**DataSet.** In order to evaluate how the algorithms perform in space, time and accuracy for different configuration parameters, $\varepsilon$ and number of threads, we generated a set of data and designed several experiments. For the space, we calculate the maximum space usage and total space usage. The maximum space usage is the size of largest summaries of transmission between nodes; and the total space usage is the total size of the summaries transmission between nodes. The dataset we use is randomly generated in different Zifian distributions. For different distributions of data, we want to see if the distribution of the data would affect the performance to different algorithms. The data is totally 1 billion integers ranging from 0 to 1 million. And the dataset is partitioned into 1,024 parts. There are three Zifian parameters: 0, 0.5, and 1; basically 0 means an uniform distribution, while 1 means that there are a lot of smaller values and fewer larger values. In addition, the data are in a sorted- and random-order so that we can see the different performance of the five algorithms. The sorted data is in increasing order.

**Configurations.** Except Zifian and order, we give the algorithms different $\varepsilon$: 0.1, 0.01, 0.001, and 0.0001. In addition, we run them in different numbers of threads: 1, 2, 4, 8, and 16. We will query 19 $\phi$-quantiles, 0.05, 0.1, 0.15, ..., 0.9, 0.95.

**Measurements.** There are three measurements we care about: space, time, and accuracy. Since communication cost is the most important factor in quantiles problem, we will measure the total size and maximum size of summaries before merging. The size is measured in bytes of all the variables that need to be serialized. For the time, we measure the total time spent for the whole process in seconds. We also show the ratio of running time against the number of threads, and we define the ratio-time to be $\frac{Time_1}{Time_i}$ where the $Time_i$ means the running time for $i$ threads. We will measure the error as the average rank error for the 19 $\phi$-quantiles query. The rank error is how the minimum rank ($r_{min}$) and maximum rank ($r_{max}$) of the output (since the value may duplicated) compared to actual rank ($r$). If $r$ is between $r_{min}$ and $r_{max}$, rank error is 0; if $r$ is less than $r_{min}$, rank error is $\frac{r_{min}-r}{n}$ ($n$ is size of data); and if $r$ is greater than $r_{max}$, rank error is $\frac{r-r_{max}}{n}$.

## 5.2 Results and Comparisons

In this subsection, we will show the results of the five algorithms and effects of the factors then compare these algorithms in certain configurations. I won't show all configurations in this paper, I just choose some to show and discuss here.

### 5.2.1 GK

As mentioned in Section 3.1, we use GKMixed here for the experiment. We have implemented GK, GKAdaptive (another variant [2]), and GKMixed algorithms, we see that GKMixed has similar accuracy but faster than GK, in the mean time, it uses much smaller space and has better accuracy than GKAdaptive, which is the fastest.

Figure 5.1 shows the relationship between $\varepsilon$ and accuracy for the GK algorithm with 8 threads. Figure 5.1a is 0 Zifian distributed data, while Figure 5.1b is 0.5 Zifian distributed data. The error increases linearly when $\varepsilon$ increases. We can probably say based on Figure 5.1 that the two curves are slightly different from each other, which can be negligible when $\varepsilon$ is greater than 0.01; however, when $\varepsilon$ is smaller than 0.01, we can see a conspicuous difference between two curves. Since the GKMixed algorithm removes the tuple if it is removable immediately when we insert it, if the data is sorted, we always insert into the last position of the list and it is not removable. Moreover, we compress the summary when its size doubles, so we compress more when the data is sorted. This is why the sorted data has high error.

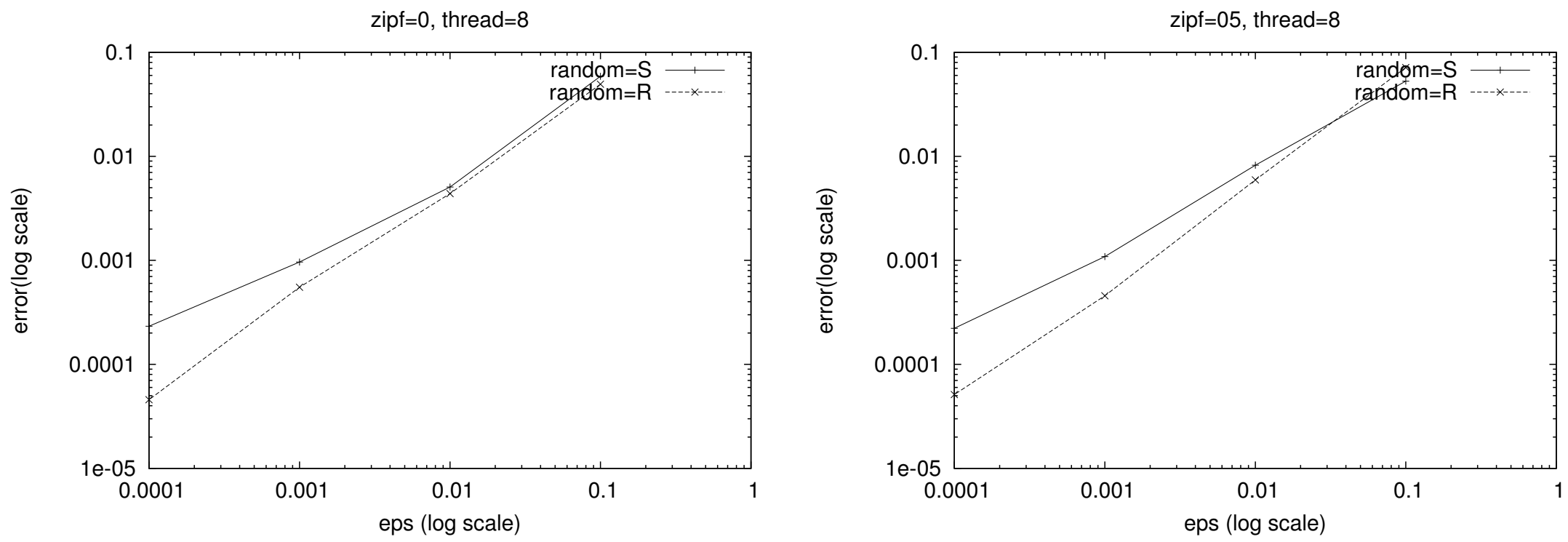

(a) zipf=0 (b) zipf=0.5

**Figure 5.1**: GK, ε-error for zipf 0 and 0.5, with 8 threads.

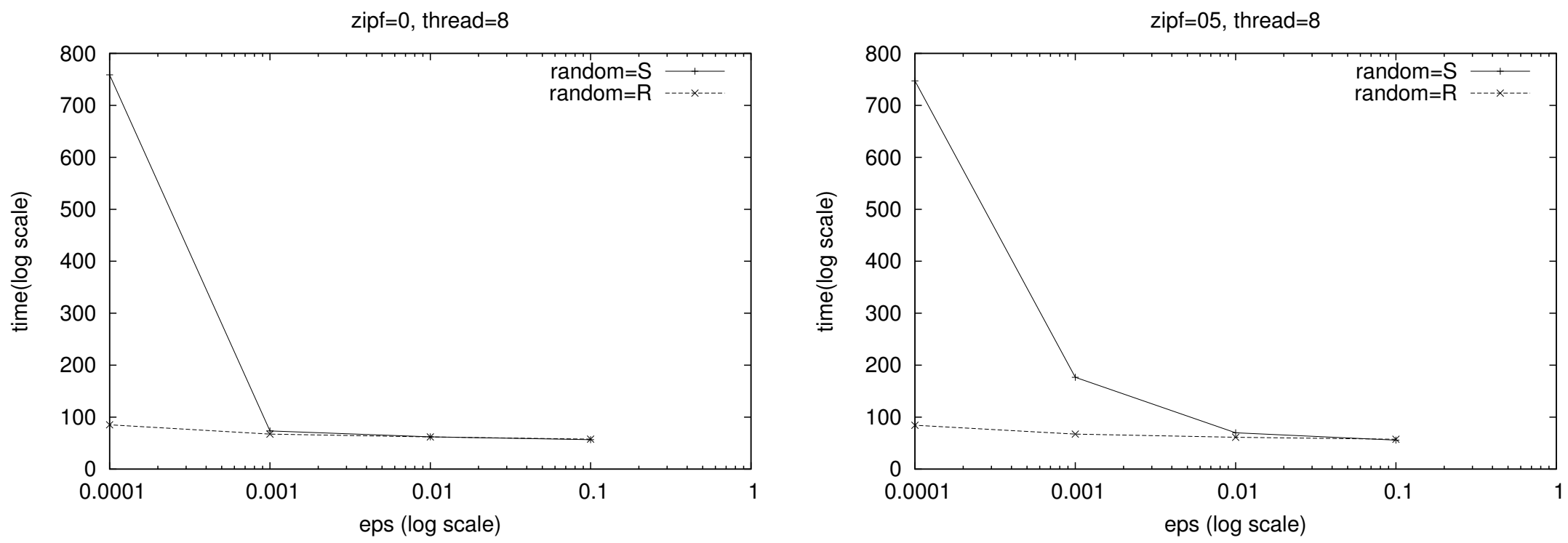

(a) zipf=0 (b) zipf=0.5

**Figure 5.2**: GK, ε-time for zipf 0, 0.5, with 8 threads.

Figure 5.2 shows the relationship between ε and running time for GK algorithm with 8 threads. Figure 5.2a is 0 Zifian-distributed data, while Figure 5.2b is 0.5 Zifian-distributed data. In these figures, we can see that the running time decreases slightly when ε increases for random data; however, for sorted data, the running time decreases sharply when ε increases from 0.0001 to 0.001. Each time when an item arrives, we need to do a binary search to decide where to put this item. At the same time, when ε is very small, the algorithm keeps a very large summary, which certainly increases the running

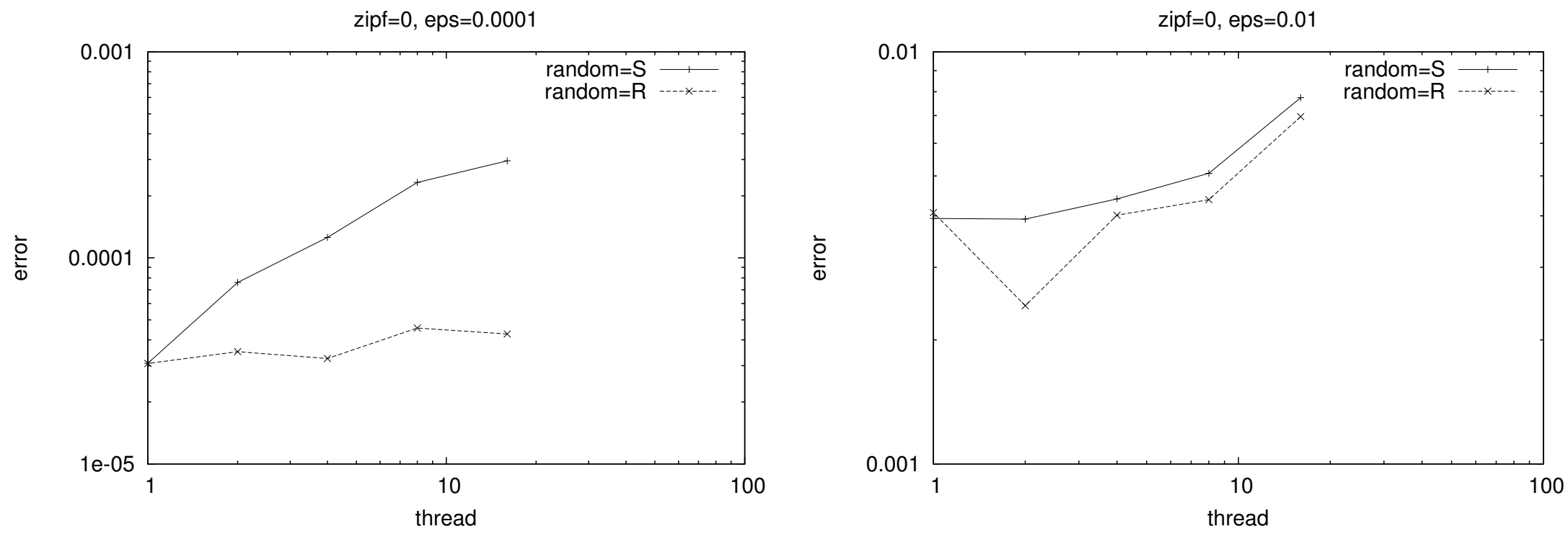

(a) $\varepsilon = 0.0001$ (b) $\varepsilon = 0.01$

**Figure 5.3**: GK, threads-error for ε 0.01, 0.0001, with 0 zipf.

time. In addition, the sorted data is the worst case for the binary search.

Figure 5.3 shows the relationship between the number of threads and errors. As shown in Figure 5.3a, we can see that the curve for the sorted data increased with the number of threads. When there is more than one thread, the assignment of chunks to the GLAs is determined by the system, and the GLA can get a chunk of data that has a very large gap with the previous ones. This causes the curve of sorted data to go up when the number of threads increases. When the number of threads is one, the data is continuous; however the more we have the more gap between two chunk of data. The number of threads does not affect the curve for the random data since the data is in random-order.

In Figure 5.4, we could see the relationship between number of threads and the space; Figures 5.4a and 5.4b show the total space used by all the threads; 5.4c and 5.4d show the maximum space used by each thread. It is obvious that the total space used increases sharply when number of threads increased, especially for random-ordered data. Basically, the number of threads means the number of merging, and more merging of course uses more space. We could see in Figure 5.4 that with more threads the sorted-ordered data uses much more space than the random. Since for the sorted data, items are in increasing order, and every incoming item will be inserted in the last position.

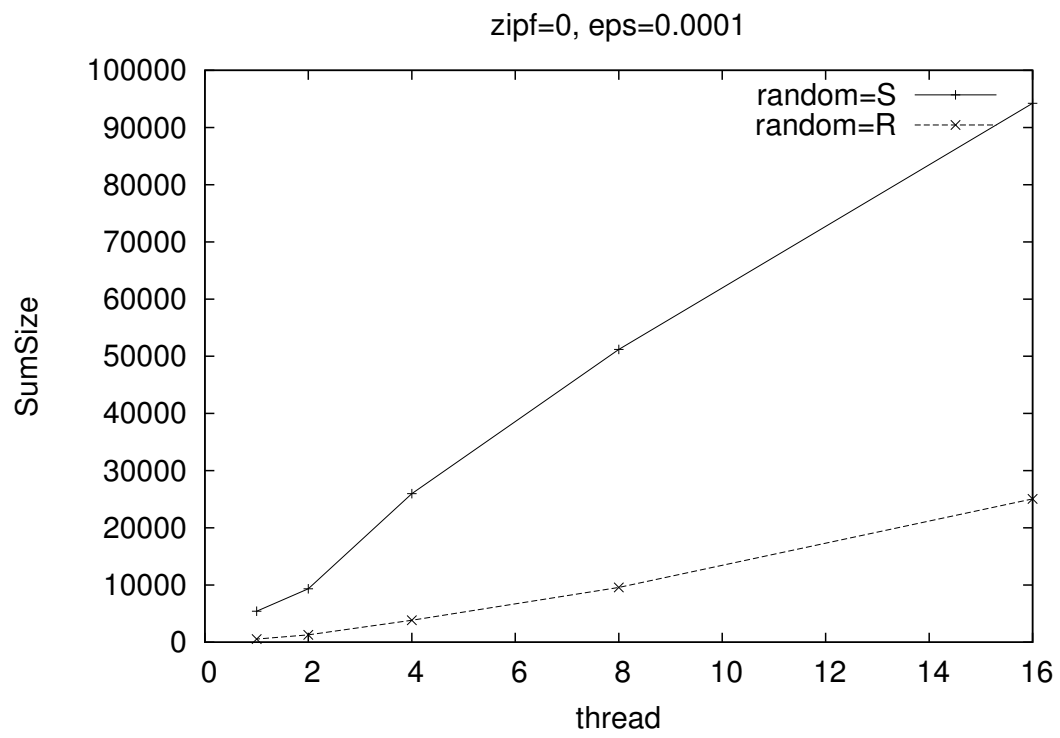

(a) zipf=0, total space usage

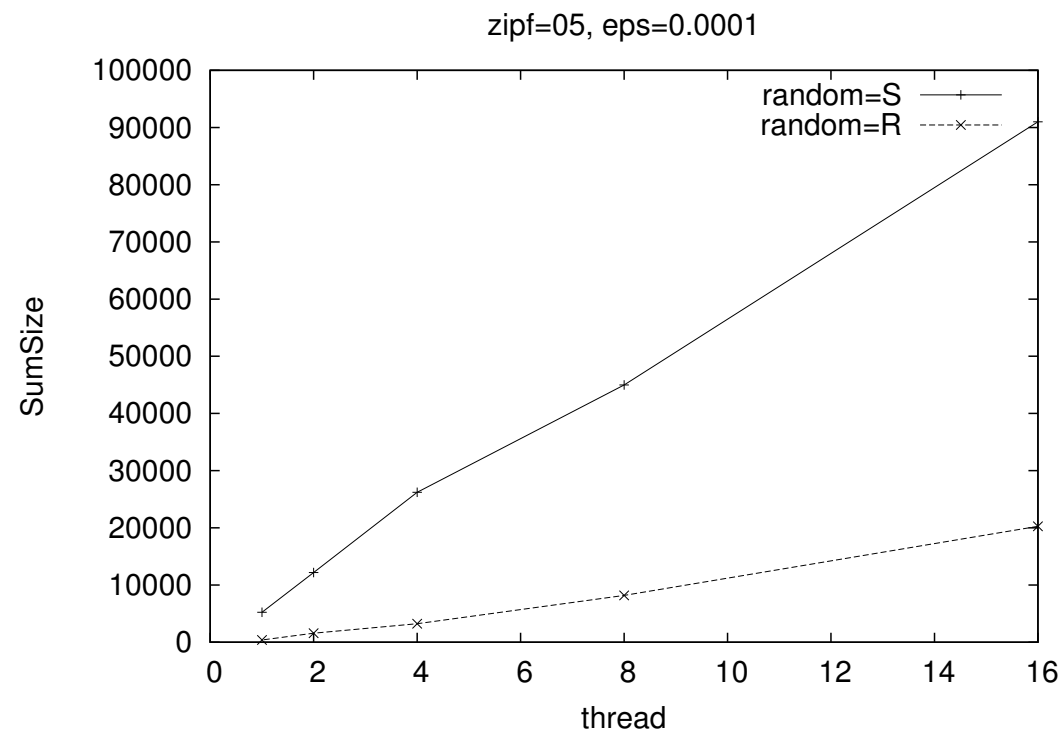

(b) zipf=0.5, total space usage

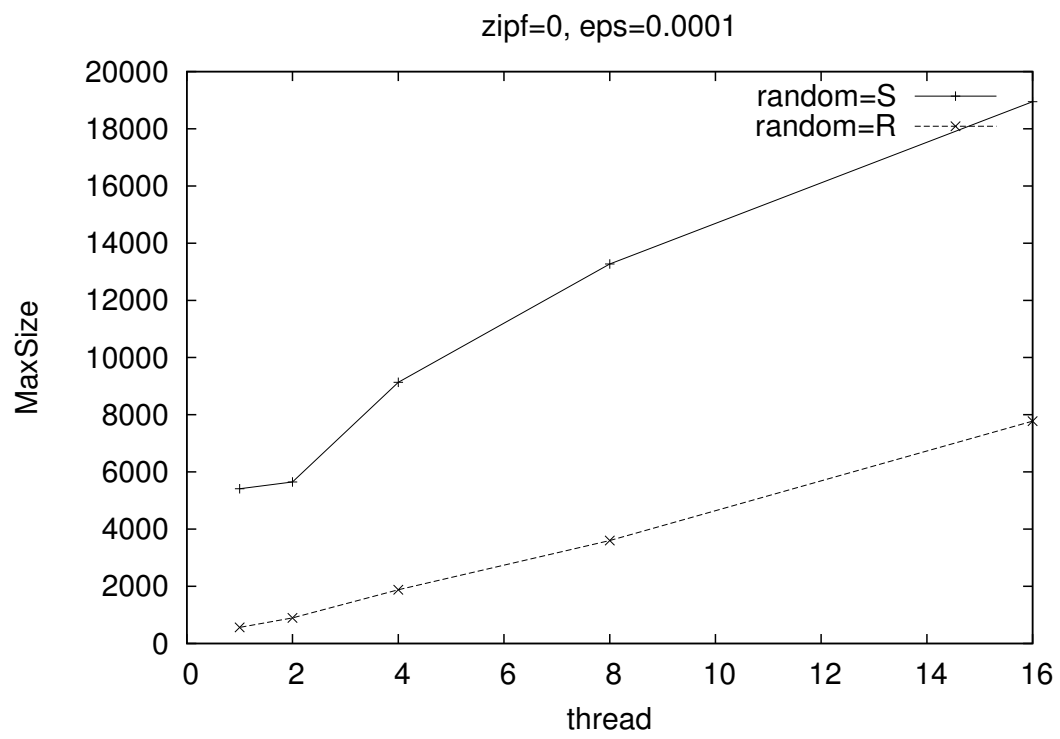

(c) zipf=0, maximum space usage

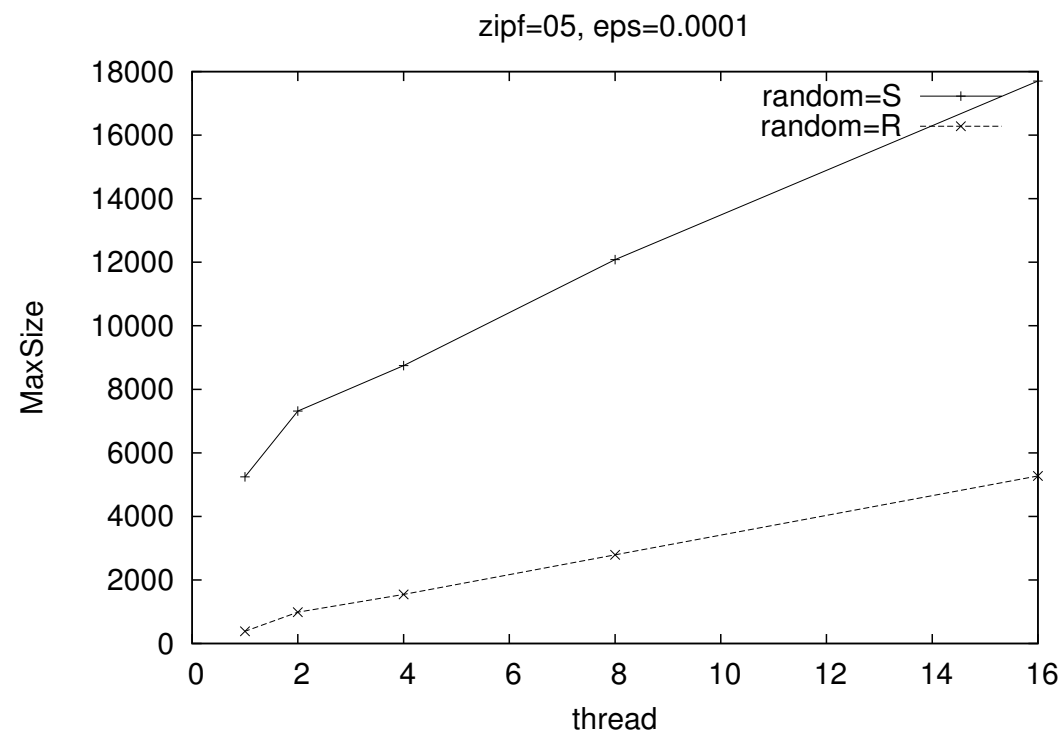

(d) zipf=0.5, maximum space usage

**Figure 5.4**: GK, threads-size for zipf 0, 0.5, with 0.0001 ε.

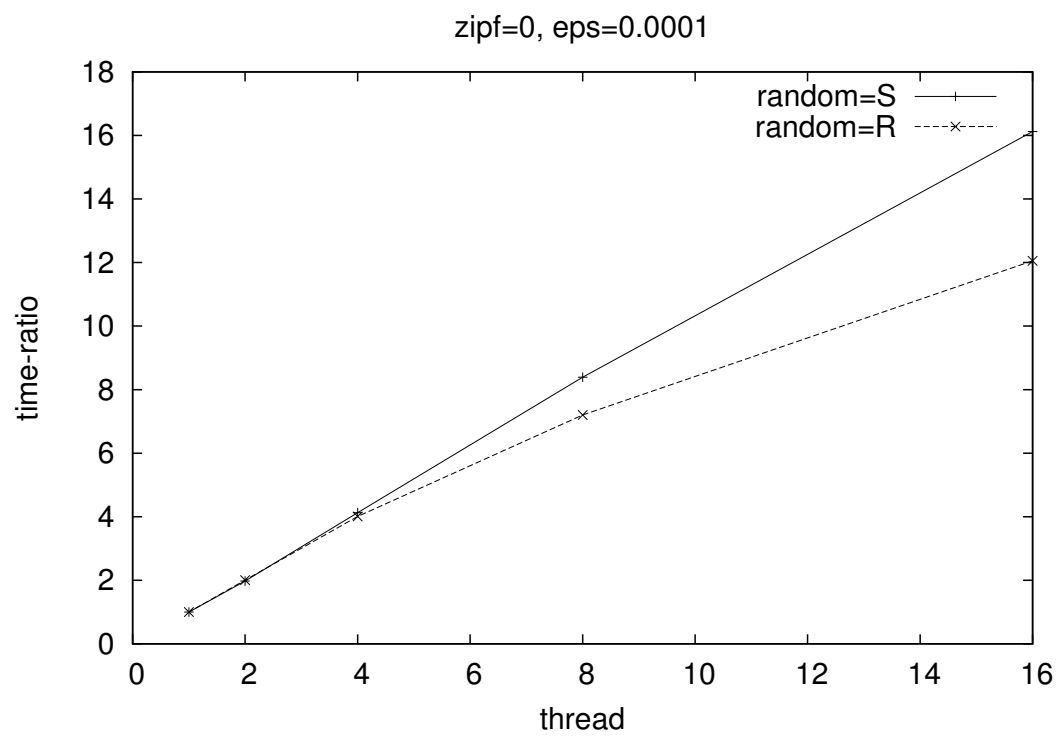

(a) zipf=0

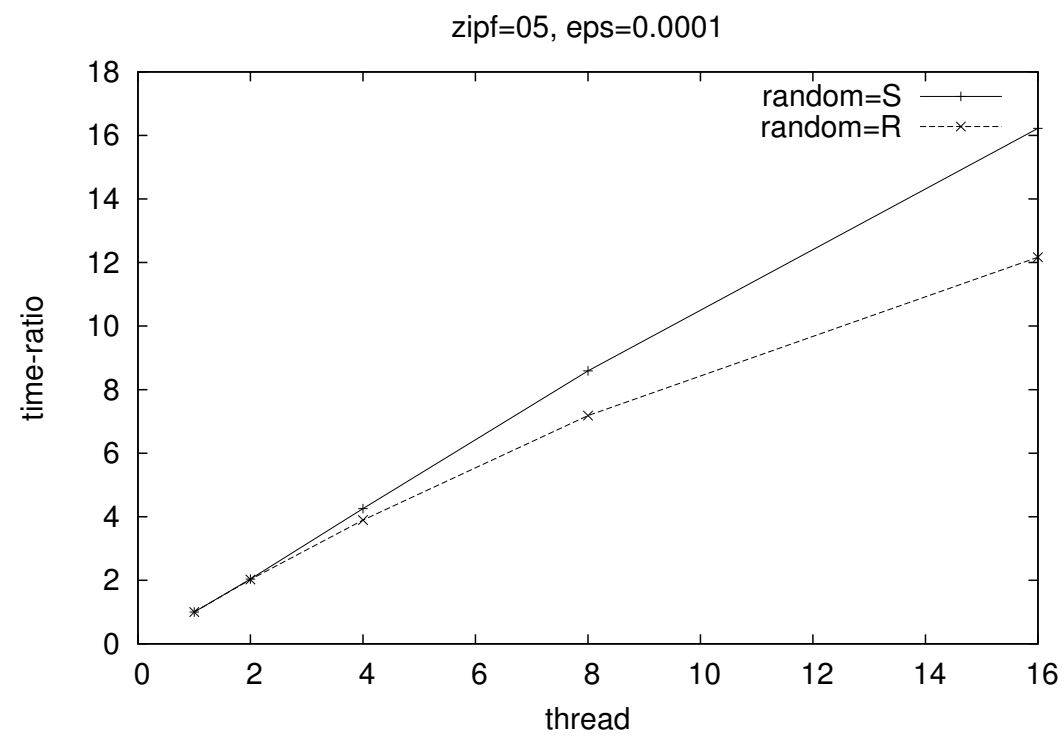

(b) zipf=0.5

**Figure 5.5**: GK, threads-time in ratio ($\frac{Time_1}{Time_n}$) for zipf 0, 0.5, with 0.0001 ε.

As described in Section 3.1, tuple $(v, 1, 0)$ will be inserted if $v$ is the smallest or largest in the list. Thus if the data is sorted, the tuples in the summary would always be $(v_i, 1, 0)$ before compressing. In addition, since the δ are 0, the *capacity* are almost same, very few tuples will be removed.

Figure 5.5 shows how the number of threads improve the running time. We can see that for the sorted data, 16 threads is 16 times faster than single thread; and for random data, it is almost 12 times faster. This indicates that when we have very large data we can use multiple threads or machine to improve the performance.

### 5.2.2 Sampling-Based

Since this algorithm is based on sampling with certain probabilities, some weird thing might happen accidentally.

Figure 5.6 shows the relationship between ε and accuracy for Sampling-Based algorithm with 8 threads. The Figure 5.6a is 0 Zifian distributed data, while Figure 5.6b is 0.5 Zifian distributed data. The error increases linearly when ε increases. We can probably say based on Figure 5.6 that the two curves have same tendency; the only conspicuous difference between the two curves is when ε is 0.0001 in 5.6a, which can be seen as contingency.

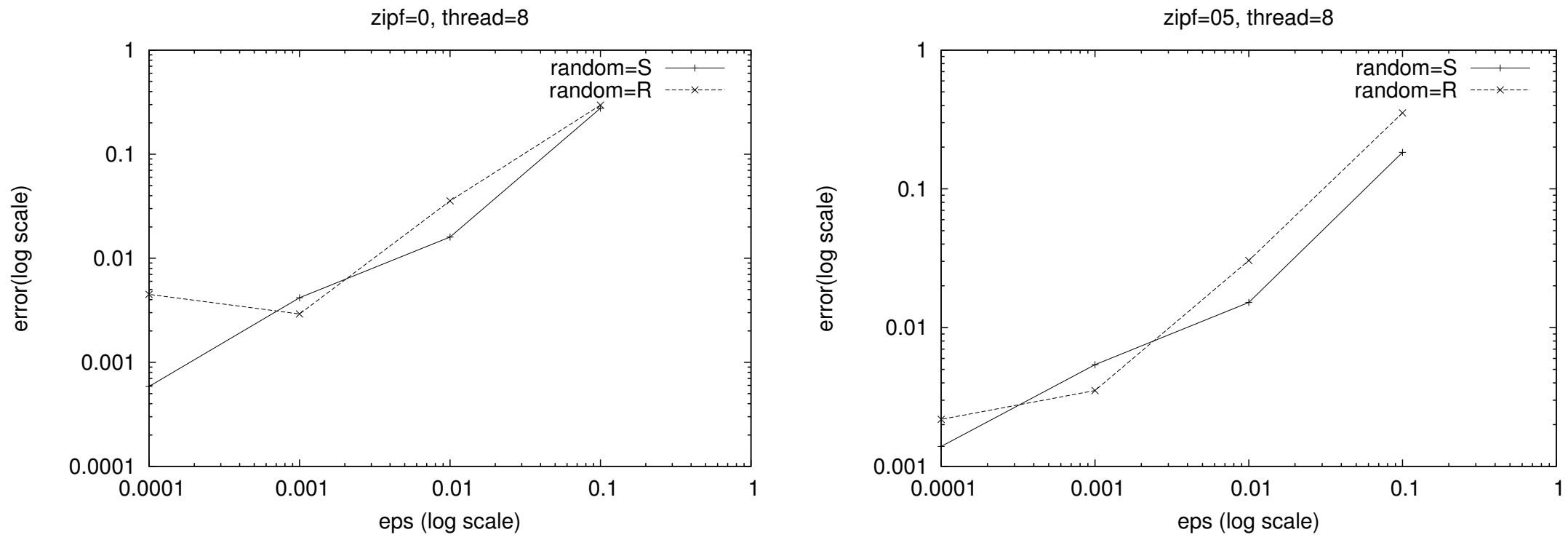

(a) zipf=0 (b) zipf=0.5

**Figure 5.6**: Sampling-Based, ε-error for zipf 0 and 0.5, with 8 threads.

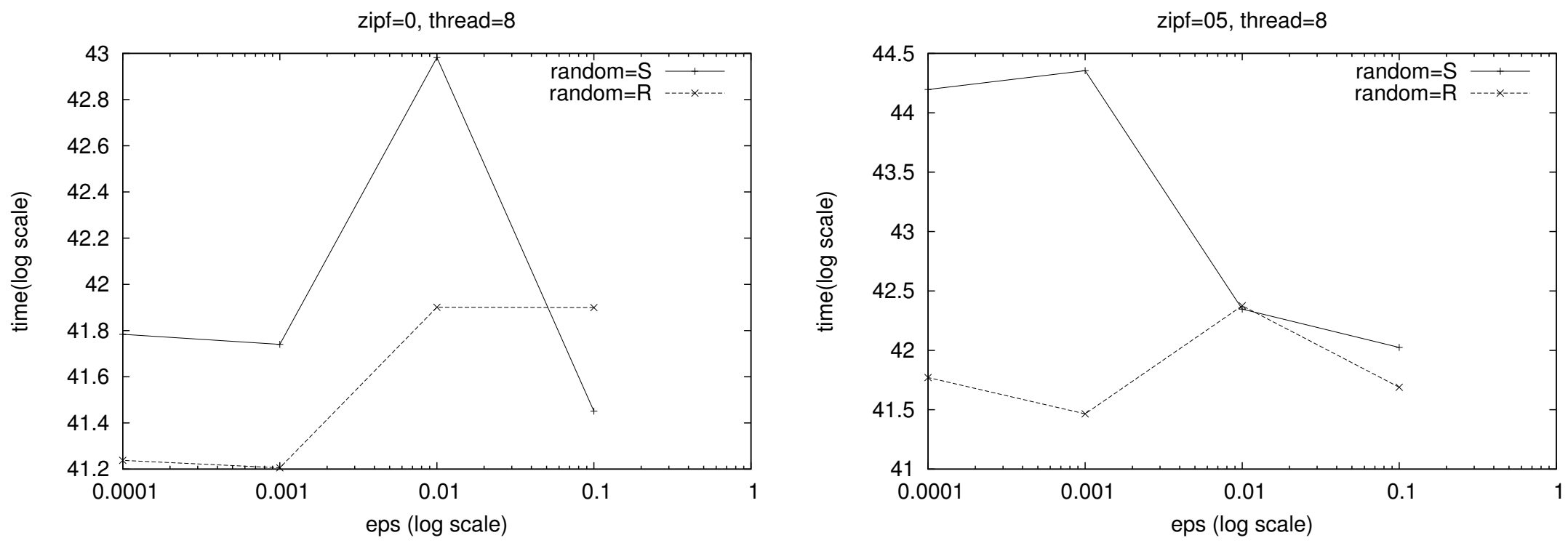

(a) zipf=0 (b) zipf=0.5

**Figure 5.7**: Sampling-Based, ε-time for zipf 0, 0.5, with 8 threads.

Figure 5.7 shows the relationship between ε and running time for q-digest algorithm with 8 threads. The Figure 5.7a is 0 Zifian-distributed data, while Figure 5.7b is 0.5 Zifian-distributed data. In these figures, we can see that the running time keeps almost monotonous for both curves. The running time varies from 41 to 45 seconds, which is very slight change. Since the algorithm is based on sampling, the increase of ε only affects the probabilities.

Figure 5.8 shows the relationship between the number of threads and errors. As shown in Figure 5.8, we can see the curves are ruleless. In addition, since the range of errors is small, especially for Figure 5.8a, there isn't distinct difference between random- and sorted-ordered data, which we already see in Figure 5.6.

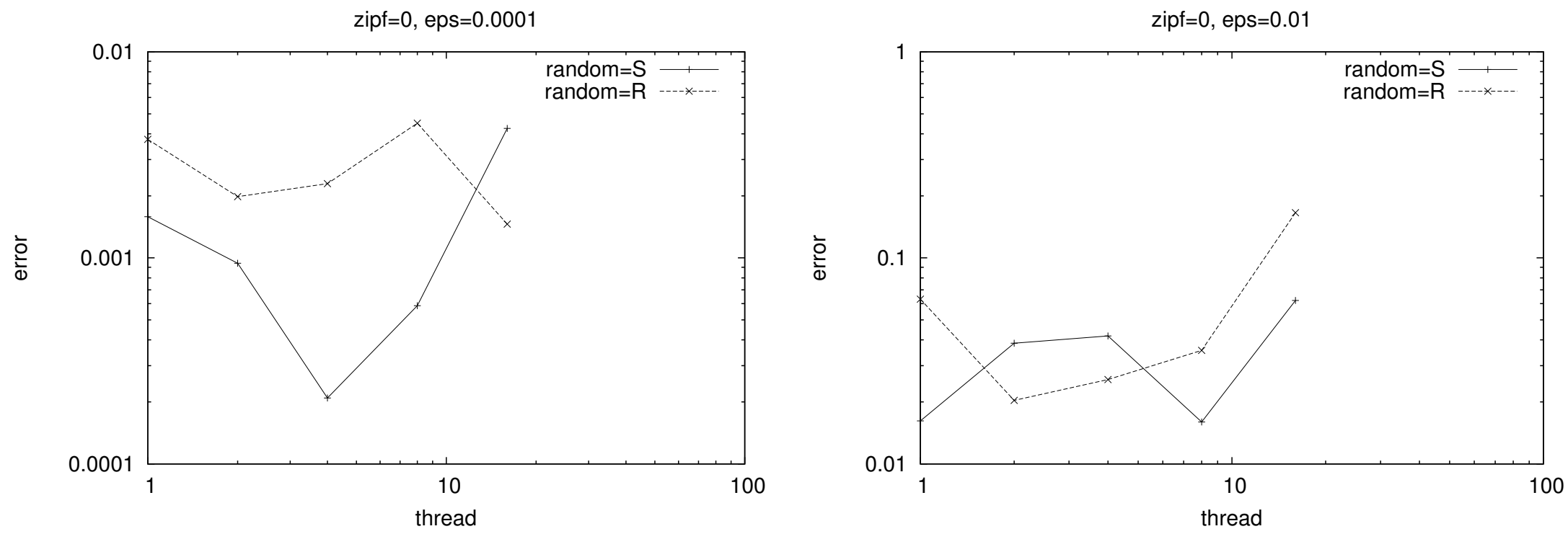

(a) $\varepsilon = 0.0001$ (b) $\varepsilon = 0.01$

**Figure 5.8**: Sampling-Based, threads-error for ε 0.0001 and 0.01, with 0 zipf.

In Figure 5.9, we could see the relationship between number of threads and the space; Figures 5.9a and 5.9b show the total space used by all thread; 5.9c and 5.9d shows the maximum space usage by each threads. Basically, when the number of threads increases, the occupied space will increase because we need to do more merging processes. We can see that there is very slight difference between random and sorted data in Figure 5.9. This is because the algorithm is a Sampling-Based algorithm, it samples the data in certain probabilities, and the order of data does not affect the size of summaries. Both total space and maximum space increase linearly when the number of threads increases.

Figure 5.10 shows how the number of threads improve the running time. We can see that 16 threads is 9 times faster than single thread for both sorted data and random data. We can see that the curves slow down when the number of threads increase from 8 to 16. Even though, we still could expect that when we have very large data we can use multiple threads or machines to improve the performance.

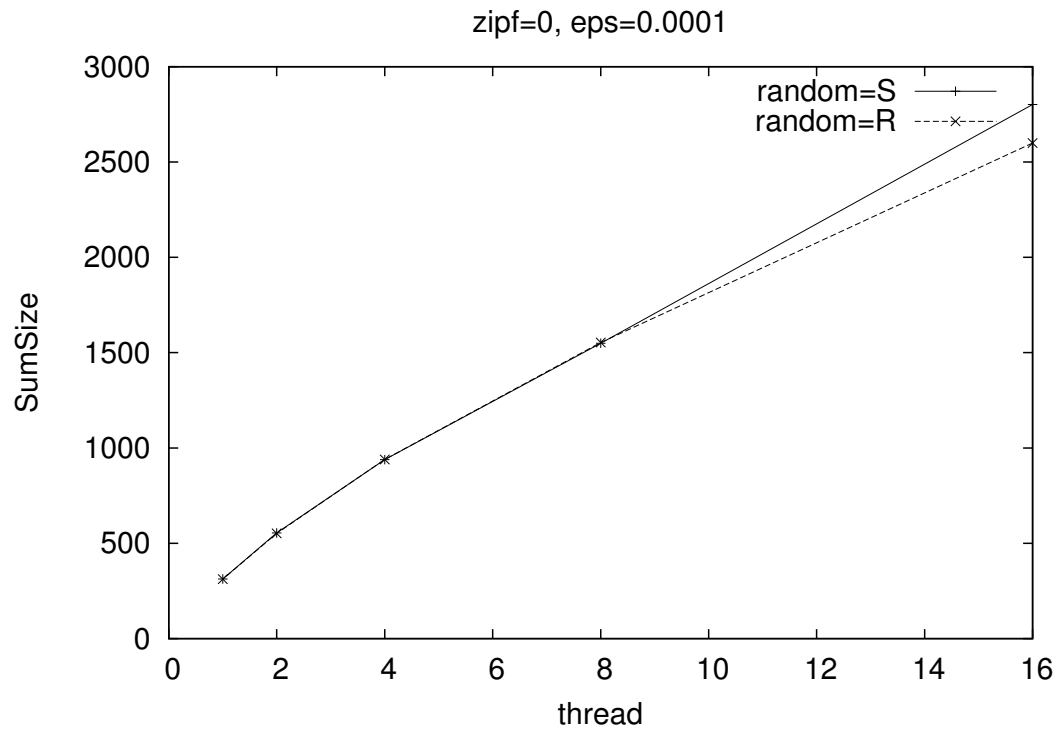

(a) zipf=0, total space usage

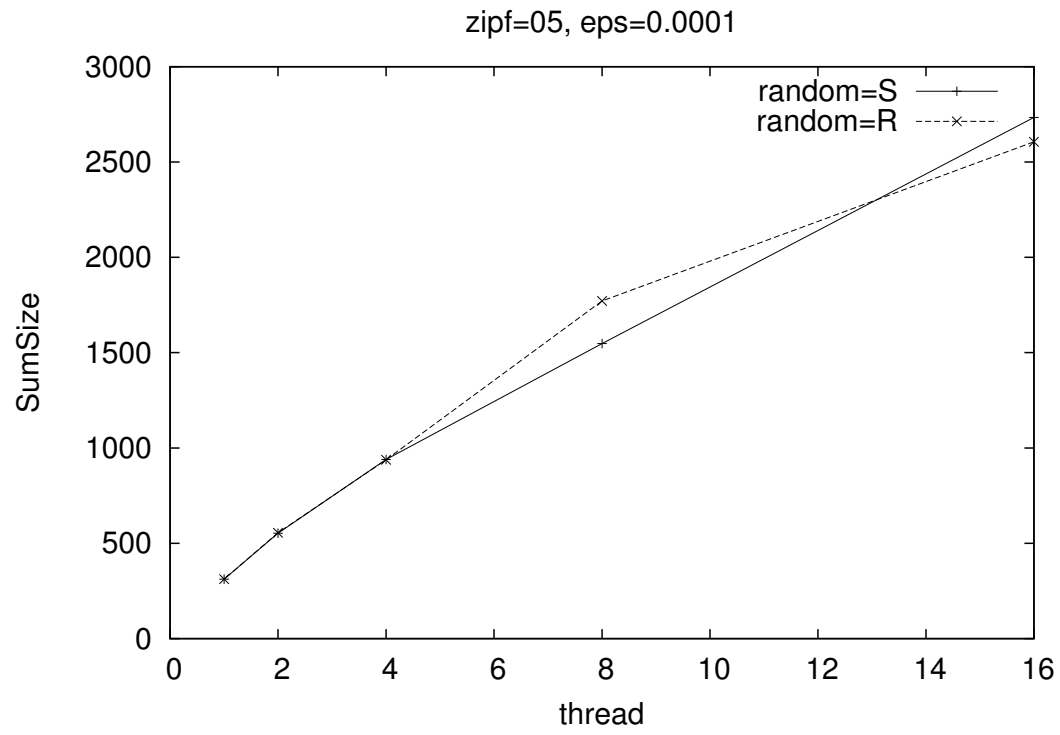

(b) zipf=0.5, total space usage

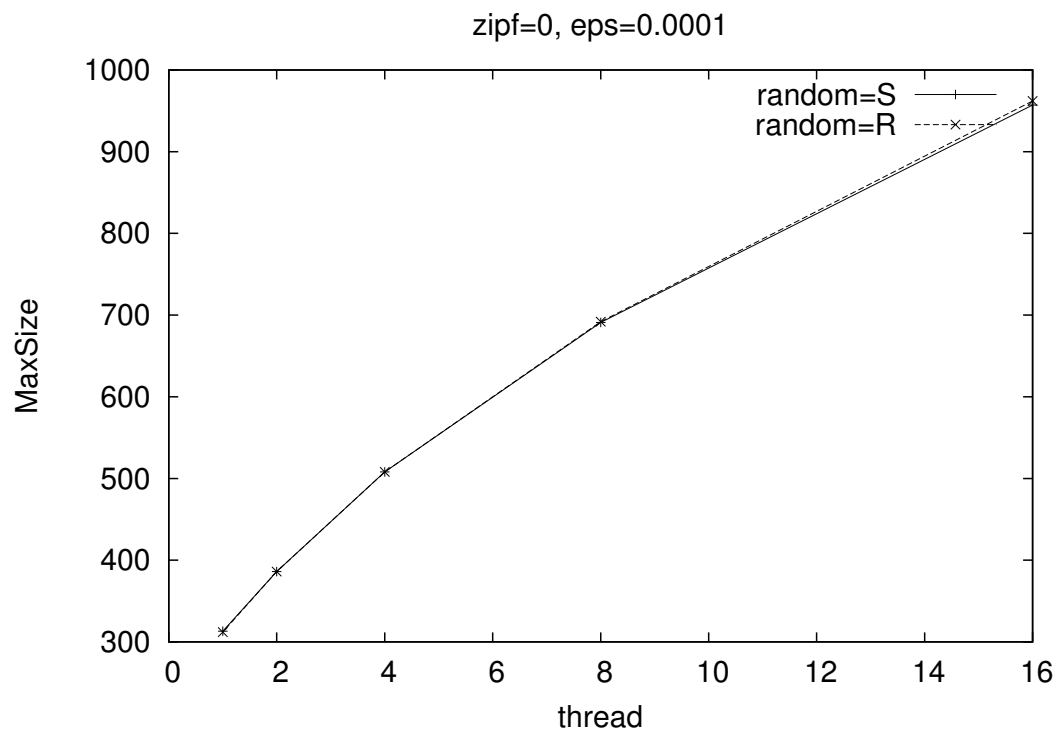

(c) zipf=0, maximum space usage

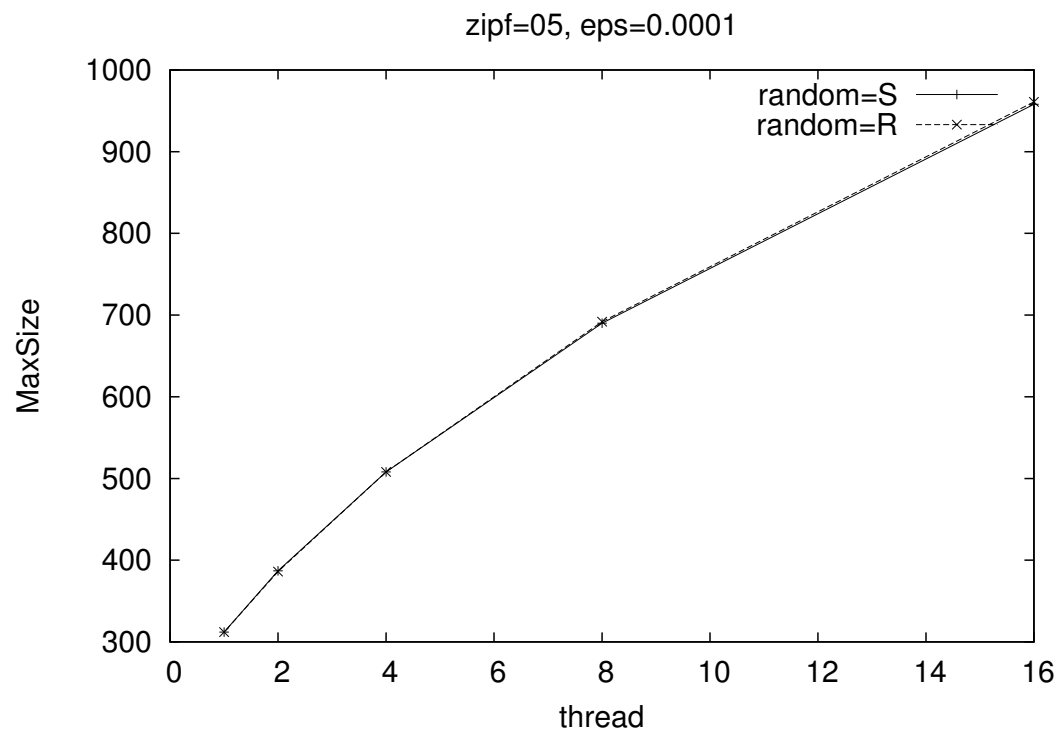

(d) zipf=0.5, maximum space usage

**Figure 5.9**: Sampling-Based, threads-size for zipf 0, 0.5, with 0.0001 ε.

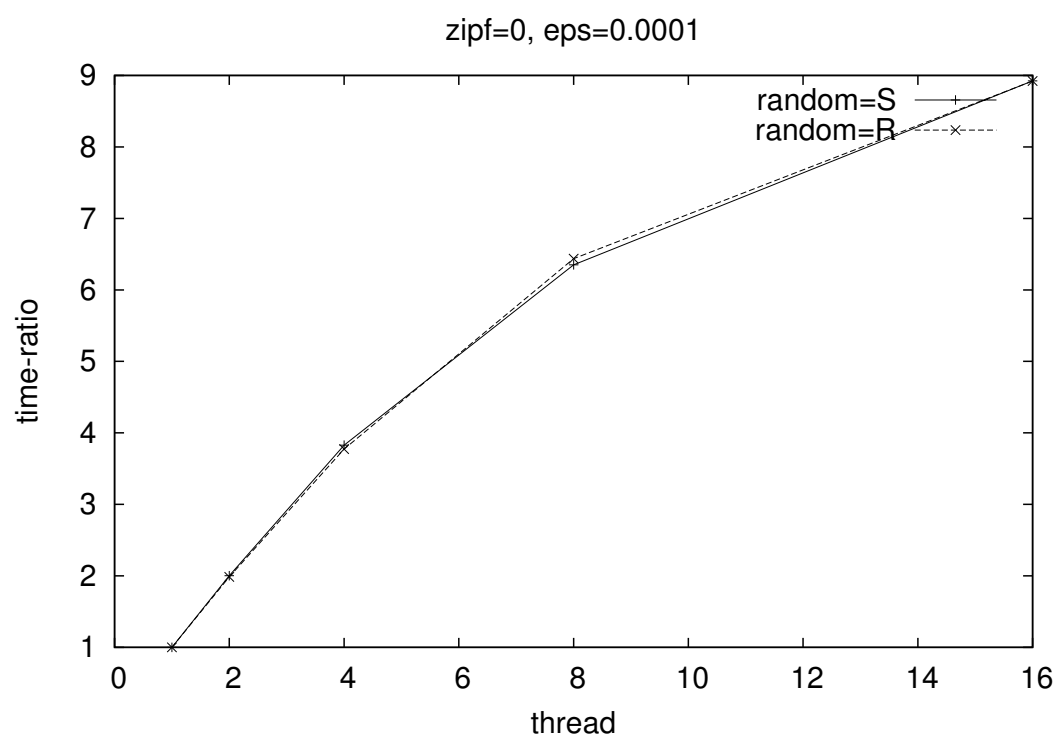

(a) zipf=0

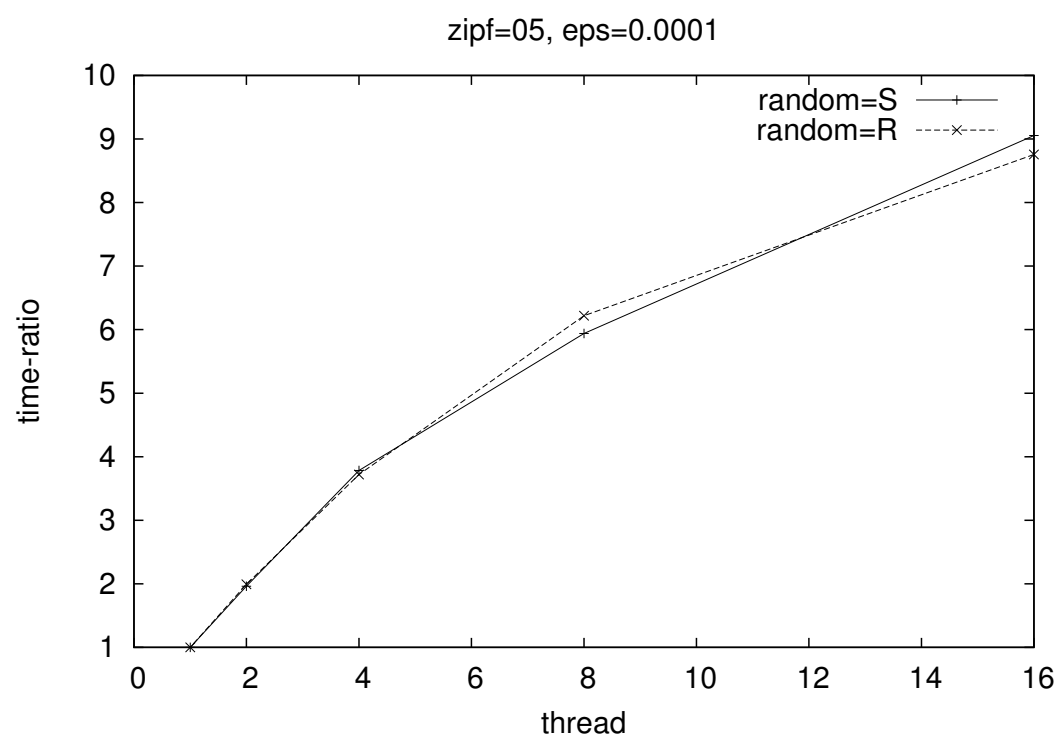

(b) zipf=0.5

**Figure 5.10**: Sampling-Based, threads-time in ratio ($\frac{Time_1}{Time_n}$) for zipf 0, 0.5, with 0.0001 ε.

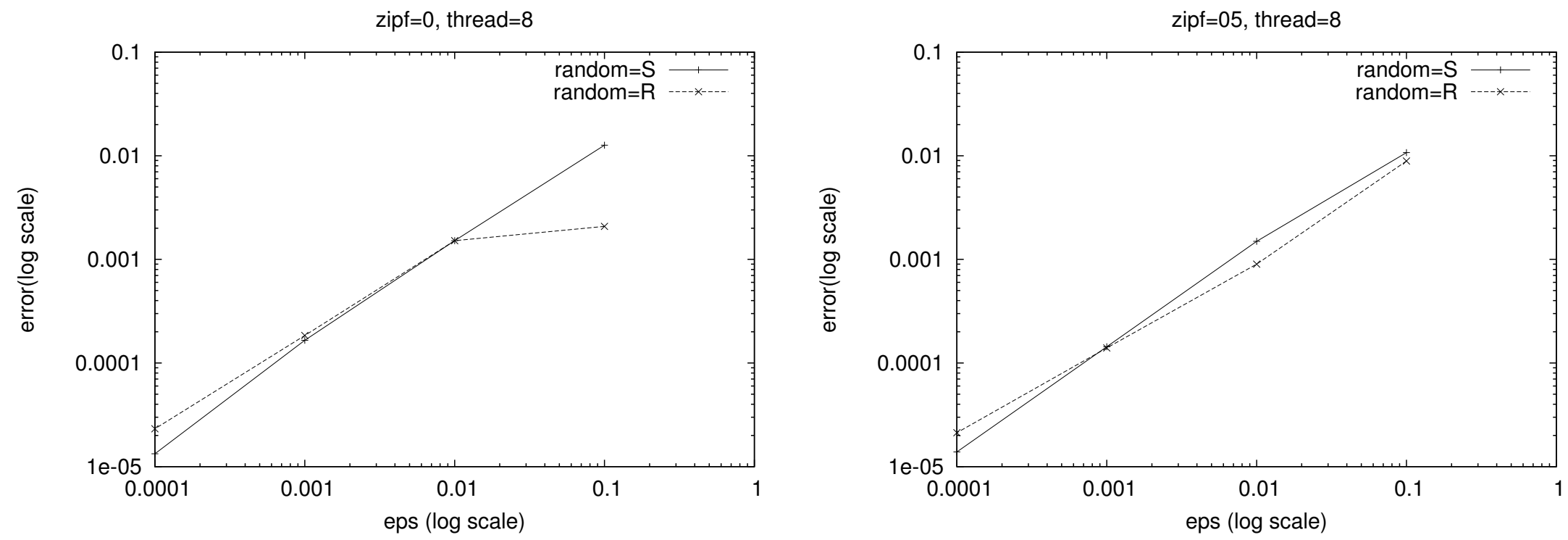

(a) zipf=0 (b) zipf=0.5

**Figure 5.11**: Q-Digest, ε-error for zipf 0 and 0.5, with 8 threads.

### 5.2.3 Q-Digest

As we described above, the q-digest algorithm has few limitations. It must have a fixed universe, and it only supports the integers. Moreover, it has to read all the data first and get the histogram, so it will need much more memories at the beginning to gather the histogram. In this subsection, we will discuss more details about its performance based on its experiment.

Figure 5.11 shows the relationship between ε and accuracy for the q-digest algorithm with 8 threads. The Figure 5.11a is 0 Zifian-distributed data, while Figure 5.11b is 0.5 Zifian-distributed data. The error increases linearly when ε increases. We can probably say based on Figure 5.11 that the two curves are slightly different from each other, which can be negligible, this is because the q-digest algorithm reads all data first, thus no matter if the data is in sorted- or random-order, the compressing and querying processes are the same. The weird thing in Figure 5.11a is the point of random-ordered data when ε is 0.1, the error is much less than the sorted data. However, we cannot find a similar situation in Figure 5.11b and other related experimental figures that are not listed in this paper. Thus, we would like to say such phenomenon is a coincidence. In addition, for the same reason as above, the distribution of the data does not cause any difference

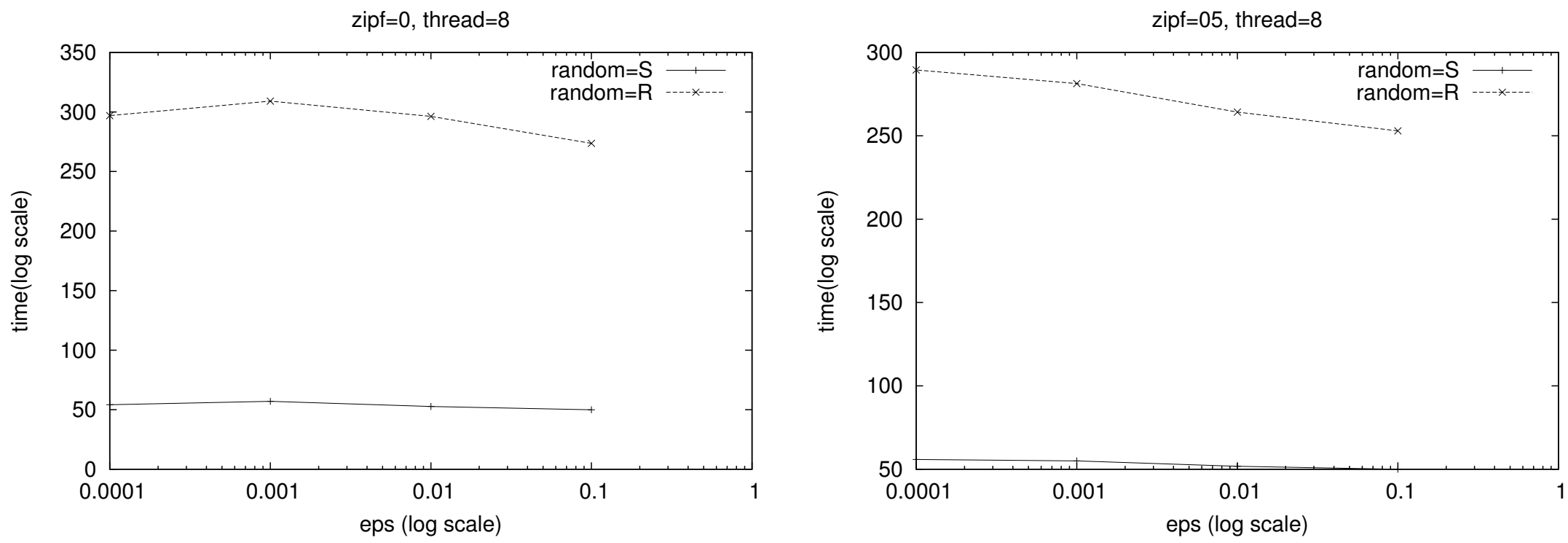

(a) zipf=0 (b) zipf=0.5

**Figure 5.12**: Q-Digest, ε-time for zipf 0, 0.5, with 8 threads.

for the accuracy (Figures 5.11a and 5.11b).

Figure 5.12 shows the relationship between ε and running time for the q-digest algorithm with 8 threads. The Figure 5.12a is 0 Zifian-distributed data, while Figure 5.12b is 0.5 Zifian-distributed data. In these figures, we can see that the running time decreases slightly when ε increases. In conclusion the ε and the distribution of the data (Figures 5.12a and 5.12b) do not affect the running time too much because no matter in sorted or random situation, the algorithm goes through every node in the tree and check wether it violate the condition (mentioned in Section 3.3). However, in both 5.12a and 5.12b, we can see a big difference between sorted- and random-ordered data. For the sorted data, each time an item comes we only need to decide to put it either in the current bucket or the next bucket ($O(n)$); however, for the random-ordered data, each time an item comes, we need to do a binary search and then decide which bucket the item should be put in($O(nlogn)$).

Figure 5.13 shows the relationship between the number of threads and error. As shown in Figure 5.13, we found that the number of threads does not affect the accuracy for q-digest algorithm. The curves are almost flat with slight fluctuation. In Figure 5.13a, considering the range of error is $10^{-5} \sim 1^{-4}$, the change of error is inconspicuous when the number of threads increases, besides the different order of data only has a very slight difference.

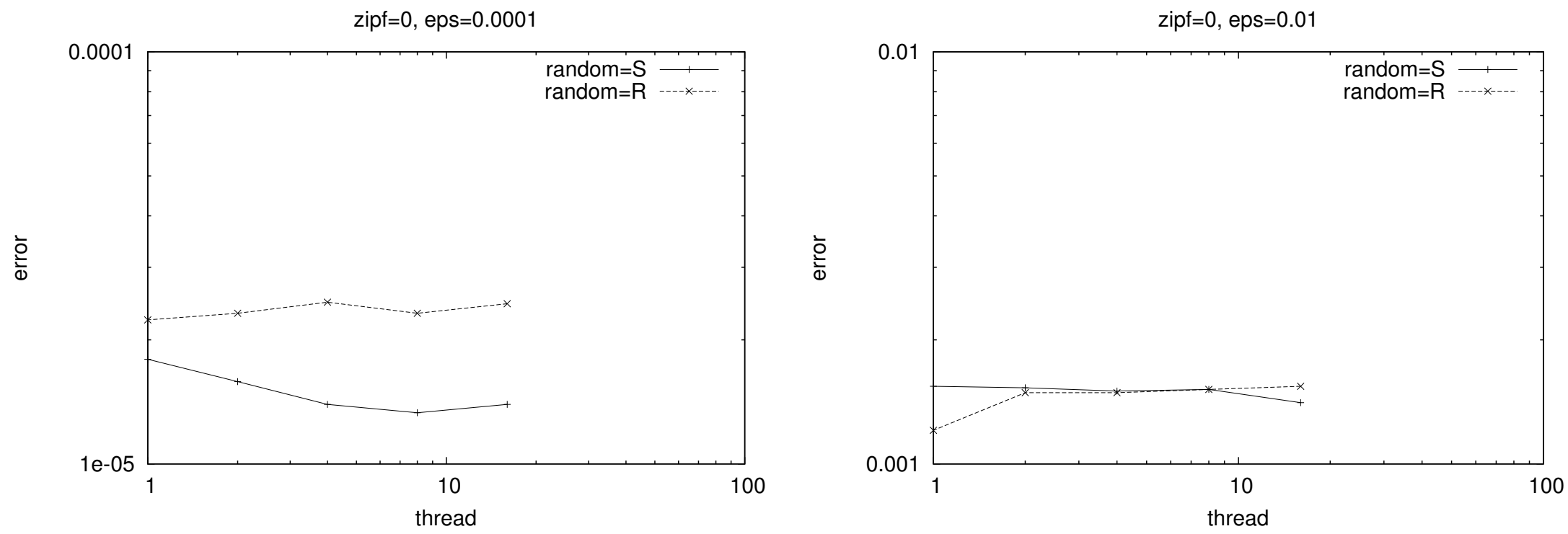

(a) ε=0.0001 (b) ε=0.01

**Figure 5.13**: Q-Digest, threads-error for ε 0.0001 and 0.01, with 0 zipf.

In Figure 5.14, we could see the relationship between the number of threads and the space; Figures 5.14a and 5.14b show the total space used by all the threads; 5.14c and 5.14d shows the maximum space used by each threads. It is obvious that the total space used increases sharply when the number of threads increases, especially for random-ordered data. Basically, the number of threads is the number of merging, and more merging of course uses more space. Figures 5.14a and 5.14b show that with more threads the random-ordered data use much more space than that of the sorted. Since we generate the data first in a sorted- or random-order and then partition them into 1,024 chunks, if the data is sorted, then each chunk has a small range of data, but if the data is random, then each chunk has the full range of data. Thus, it is clear that it uses much less space when data are sorted since the data in each chunk are in a relatively small range. In Figures 5.14c and 5.14d, we could see that there is a huge difference between sorted- and random-ordered data for the maximum space usage in a different number of threads. For the sorted data, the maximum space usage keep same when the number of threads increases. For the random data, the maximum space increases sharply with more threads.

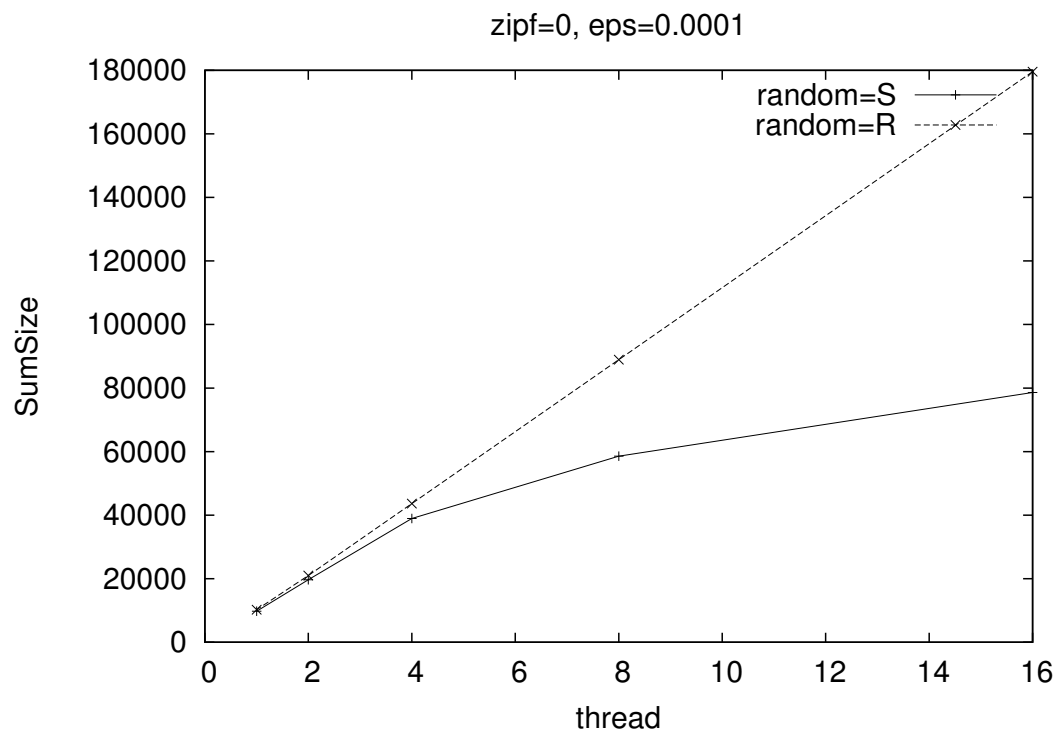

(a) zipf=0, total space usage

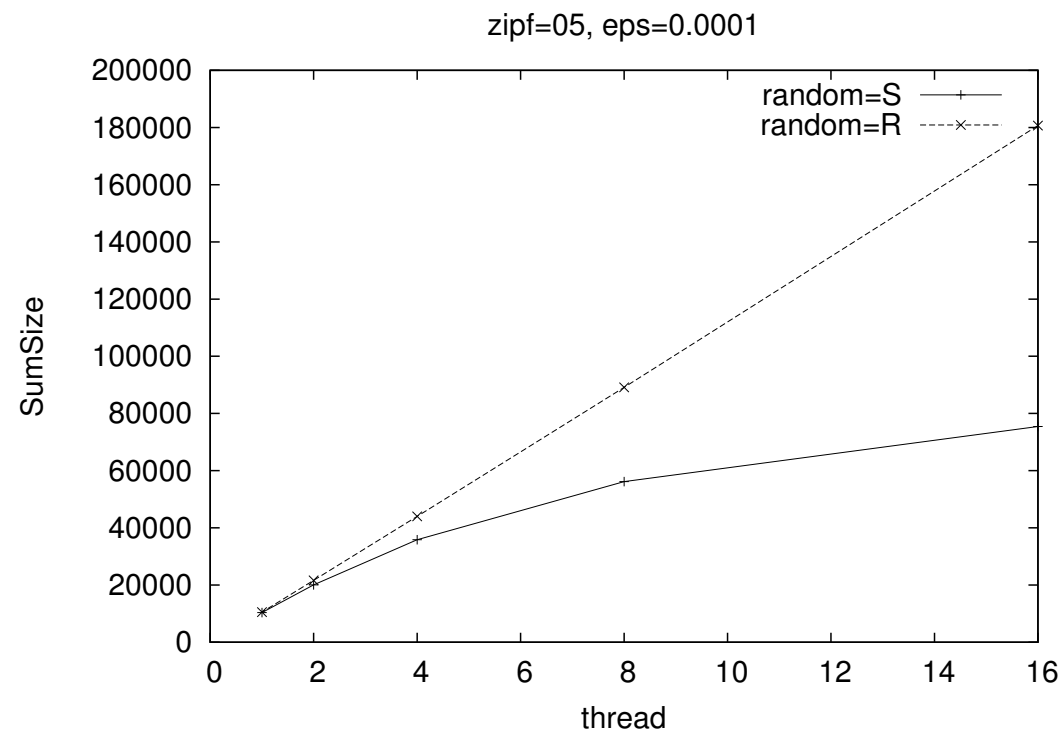

(b) zipf=0.5, total space usage

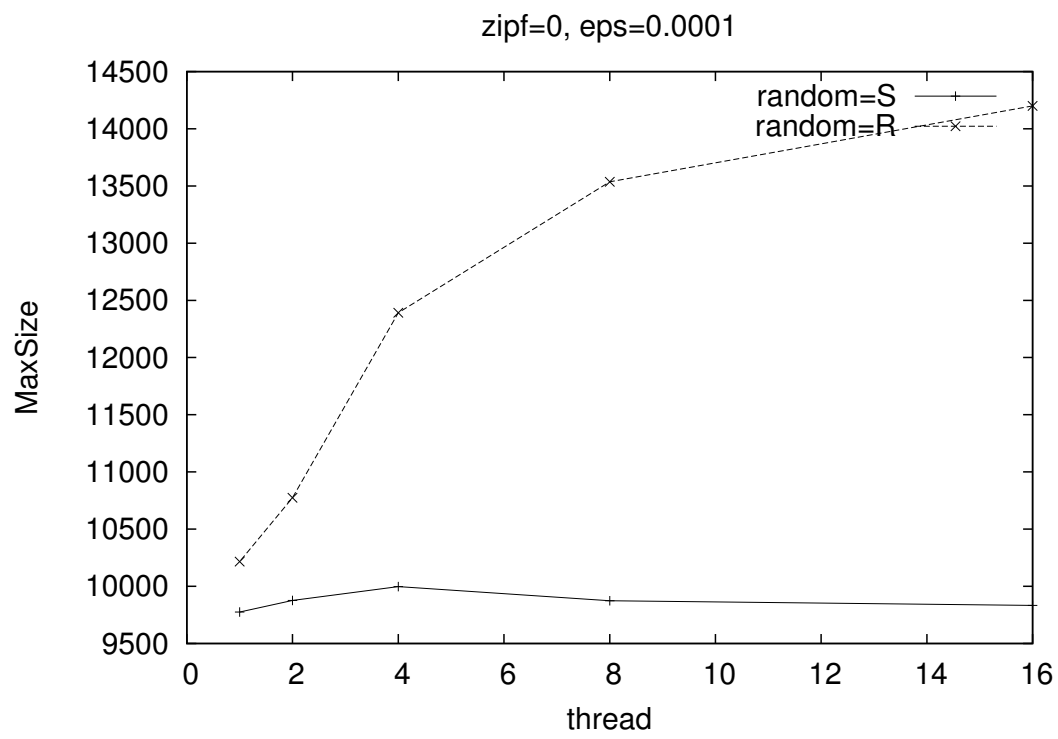

(c) zipf=0, maximum space usage

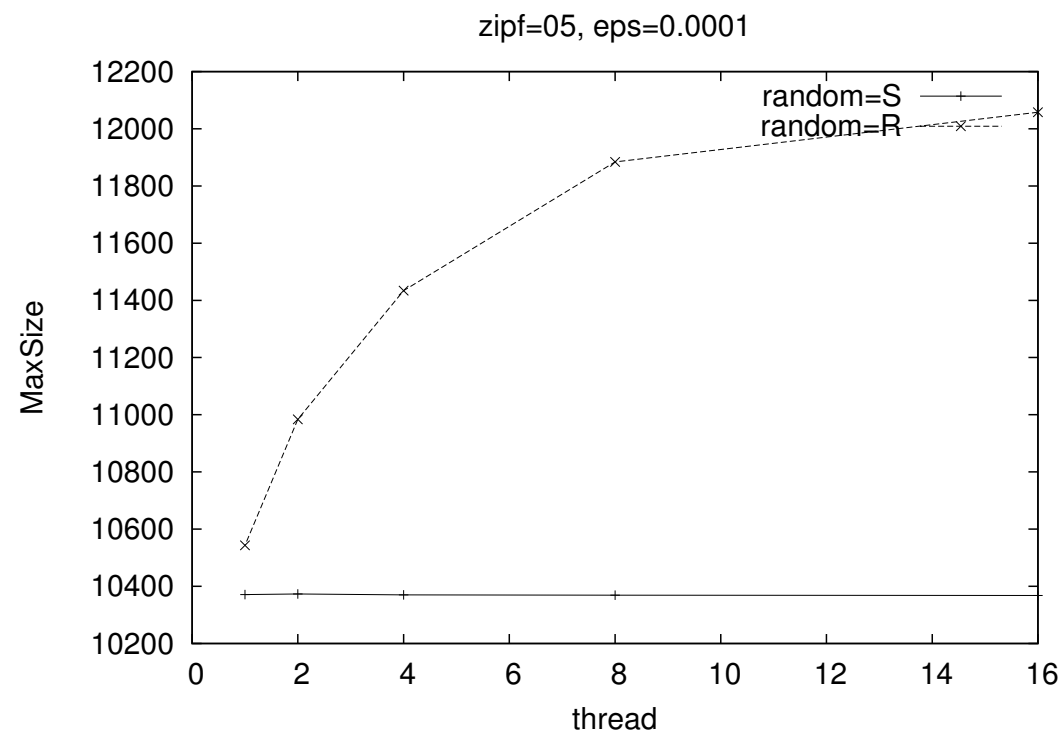

(d) zipf=0.5, maximum space usage

**Figure 5.14**: Q-Digest, threads-size for zipf 0, 0.5, with 0.0001 ε.

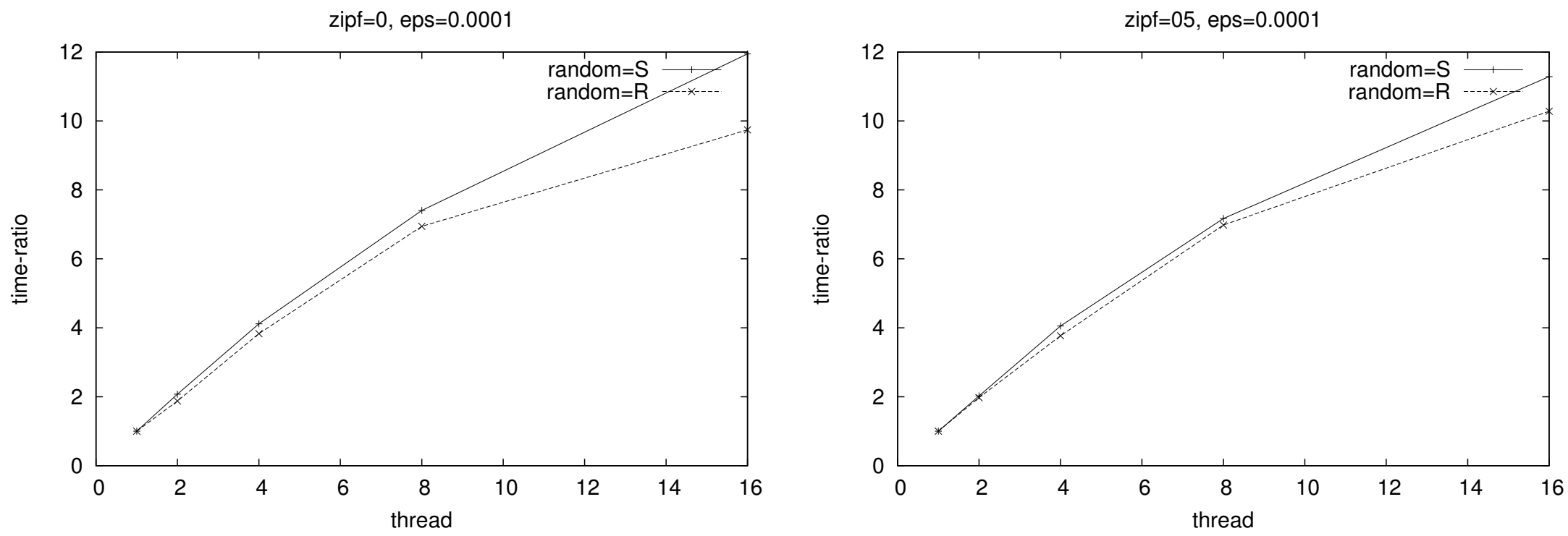

(a) zipf=0 (b) zipf=0.5

**Figure 5.15**: Q-Digest, threads-time in ratio ($\frac{Time_1}{Time_n}$) for zipf 0, 0.5, with 0.0001 ε.

If the data is random, the number of unique data is much larger, and the more threads we have, the more number of unique values we will have.

Figure 5.15 shows how the number of threads improves the running time. We can see that for the sorted data, 16 threads is 12 times faster than single thread; and for random data, it is almost 10 times faster. This indicates that when we have very large data we can use multiple threads or machines to improve the performance.

### 5.2.4 FASTQDigest

As mentioned above, the q-digest algorithm has several limitations; this variant solves these limitations. In the FastQDigest algorithm, when an item comes, we either add a counter in an existing node in the tree or create a new node based on the same conditions with q-digest algorithm (Section 3.3). Unlike the q-digest algorithm, FastQDigest processes the item when it comes, thus the performance of FastQDigest will be affected by the order and the distribution of the data. The reason is that if there are several similar items that have been processed for the incoming item, it is more likely that we just need to increase the counter of existing nodes instead of creating a new node, and this probably will cause fewer errors and shorter running time.

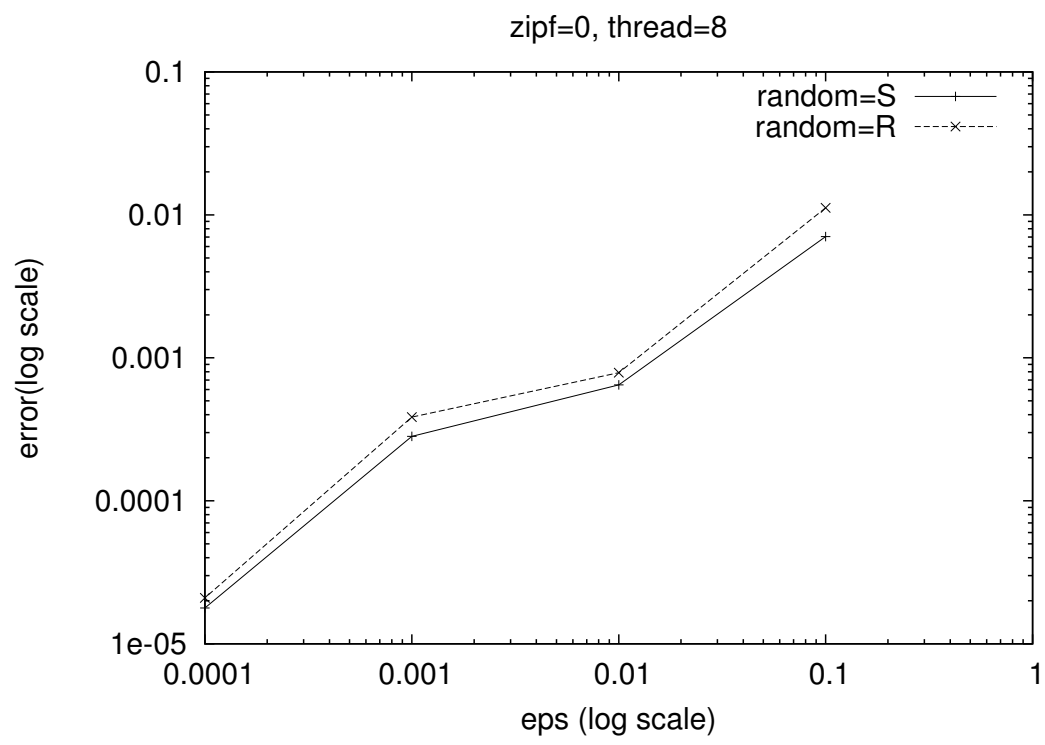

(a) zipf=0

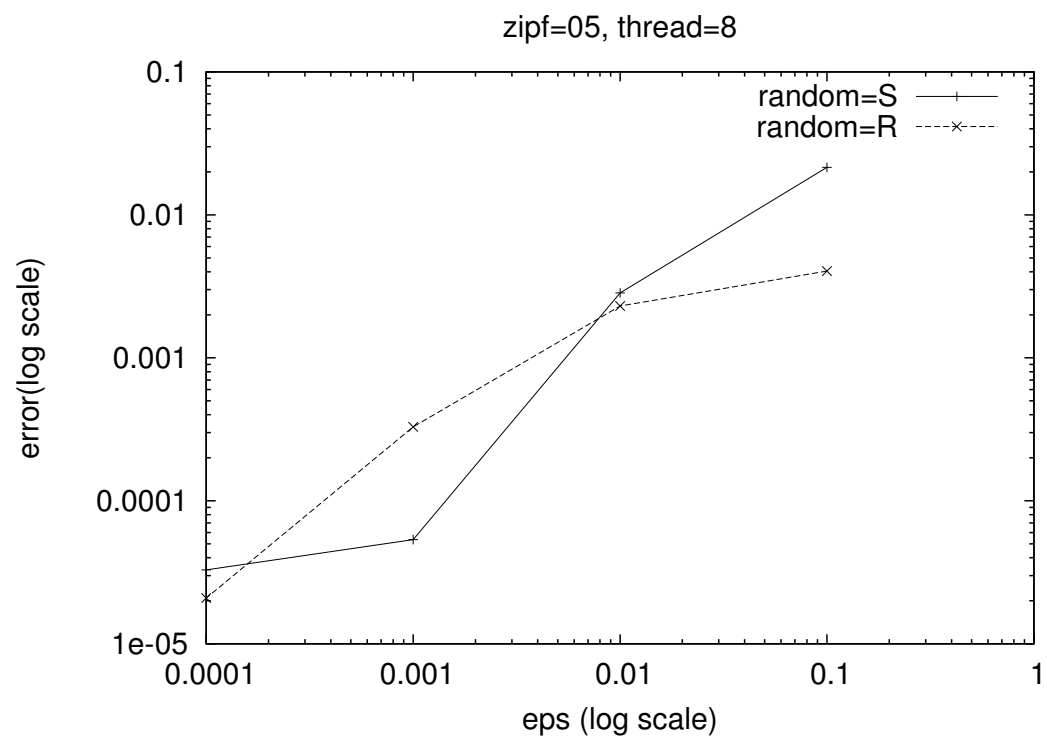

(b) zipf=0.5

**Figure 5.16**: FASTqdigest, ε-error for zipf 0 and 0.5, with 8 threads.

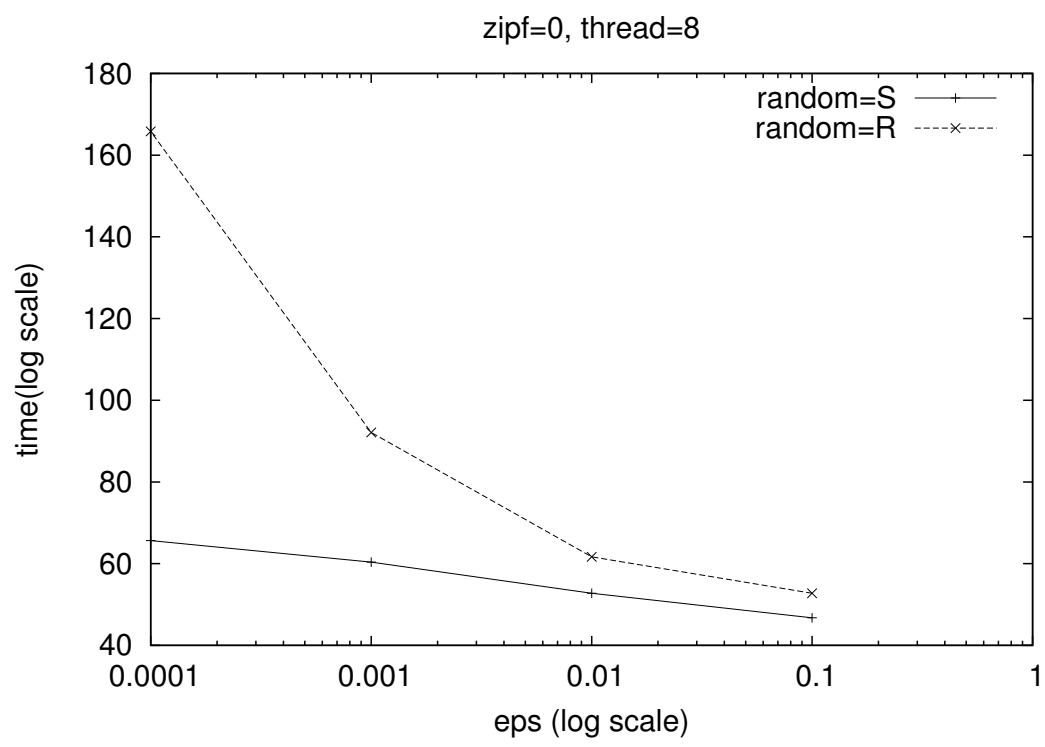

(a) zipf=0

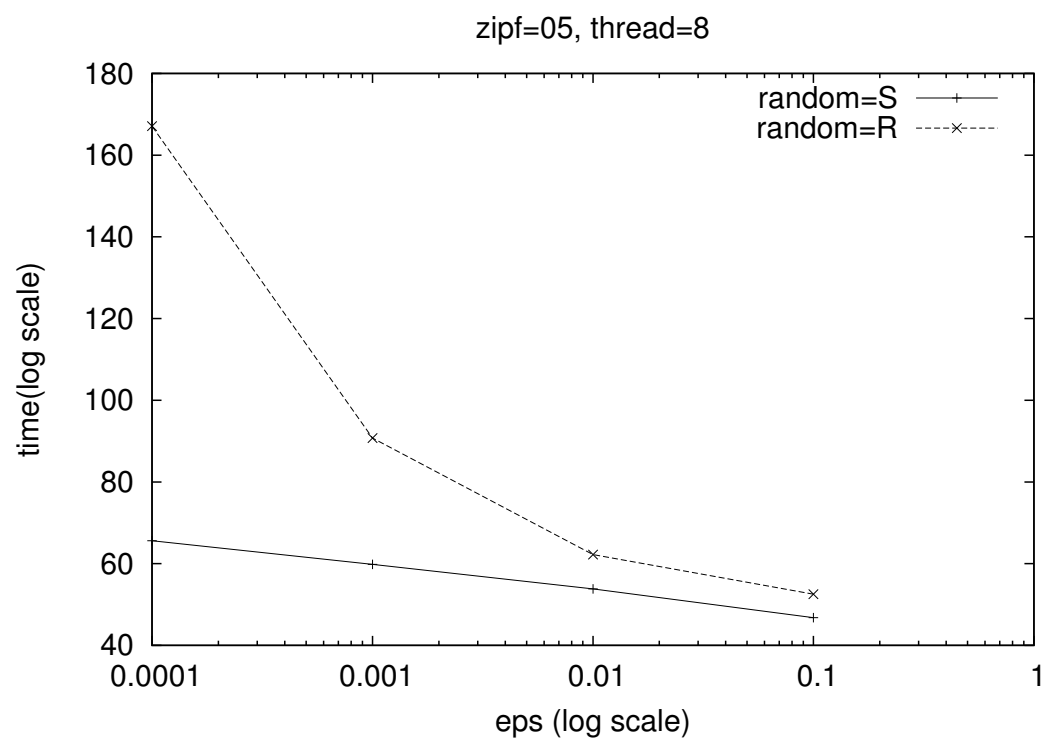

(b) zipf=0.5

**Figure 5.17**: FASTqdigest, ε-time for zipf 0, 0.5, with 8 threads.

In Figure 5.16, we could see the relationship between ε and accuracy. In Figure 5.16a, the curves of error for sorted and random data are very similar to each other; while in figure 5.16b, random data has more errors when ε is 0.001, but it has fewer errors when ε is 0.1. When Zifian becomes larger, the dataset contains more smaller numbers. For sorted-ordered data, some chunks will contain large amount of smaller data, but some chunks do not. For the random-ordered data, however, it is really unpredictable. Thus, the accuracy of random data may be better or worse than the sorted.

Figure 5.17 shows the relationship between ε and running time for FastQDigest algorithm with 8 threads. Figure 5.17a is 0 Zifian-distributed data, while Figure 5.17b is 0.5 Zifian-distributed data. In these figures, we can see that the running time decreases sharply for random-ordered data when ε increases. When ε is 0.0001, the running time for random data is much larger than sorted data; however, when ε is 0.1, the difference is very slight. For sorted data, numbers in chunks are sequential; for random data, numbers in chunks are irrelevant. Thus, the sorted data uses a shorter running time than random data, especially when ε is small, since the conditions (Section 3.3) are more restrictive.

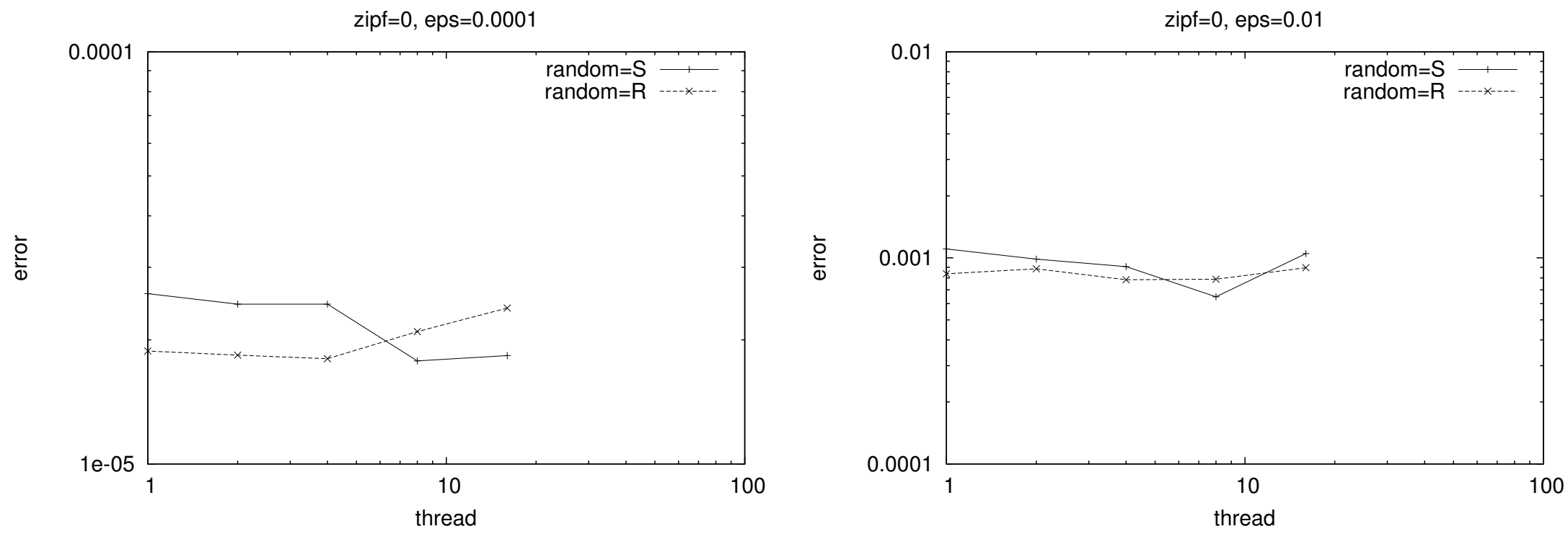

(a) ε=0.0001 (b) ε=0.01

**Figure 5.18**: FASTqdigest, threads-error for ε 0.0001 and 0.01, with 0 zipf.

Figure 5.18 shows the relationship between the number of threads and errors. As shown in Figure 5.18, we found that the number of threads does not affect the accuracy of the FastQDigest algorithm. The curves are almost flat with slight fluctuations. In Figure 5.18a, considering that the range of error is $10^{-5} \sim 1^{-4}$, the change of error is inconspicuous when the number of threads increases, besides the different order of data only has a very slight difference.

In Figure 5.19, we could see the relationship between the number of threads and the space; Figures 5.19a and 5.19b show the total space usage by all threads; 5.19c and 5.19d show the maximum space used by each thread. Basically, when the number of threads increases, the occupied space will increase because we need to do more merging processes. However, we can see that the sorted data use a very large space when the number of threads is 1, and when number of threads is 2, it decreases sharply. This is because the sorted data become random when we have more than 1 thread. While data in each chunk is sorted, a thread will not see consecutive chunks, it will see the sorted data with gaps.

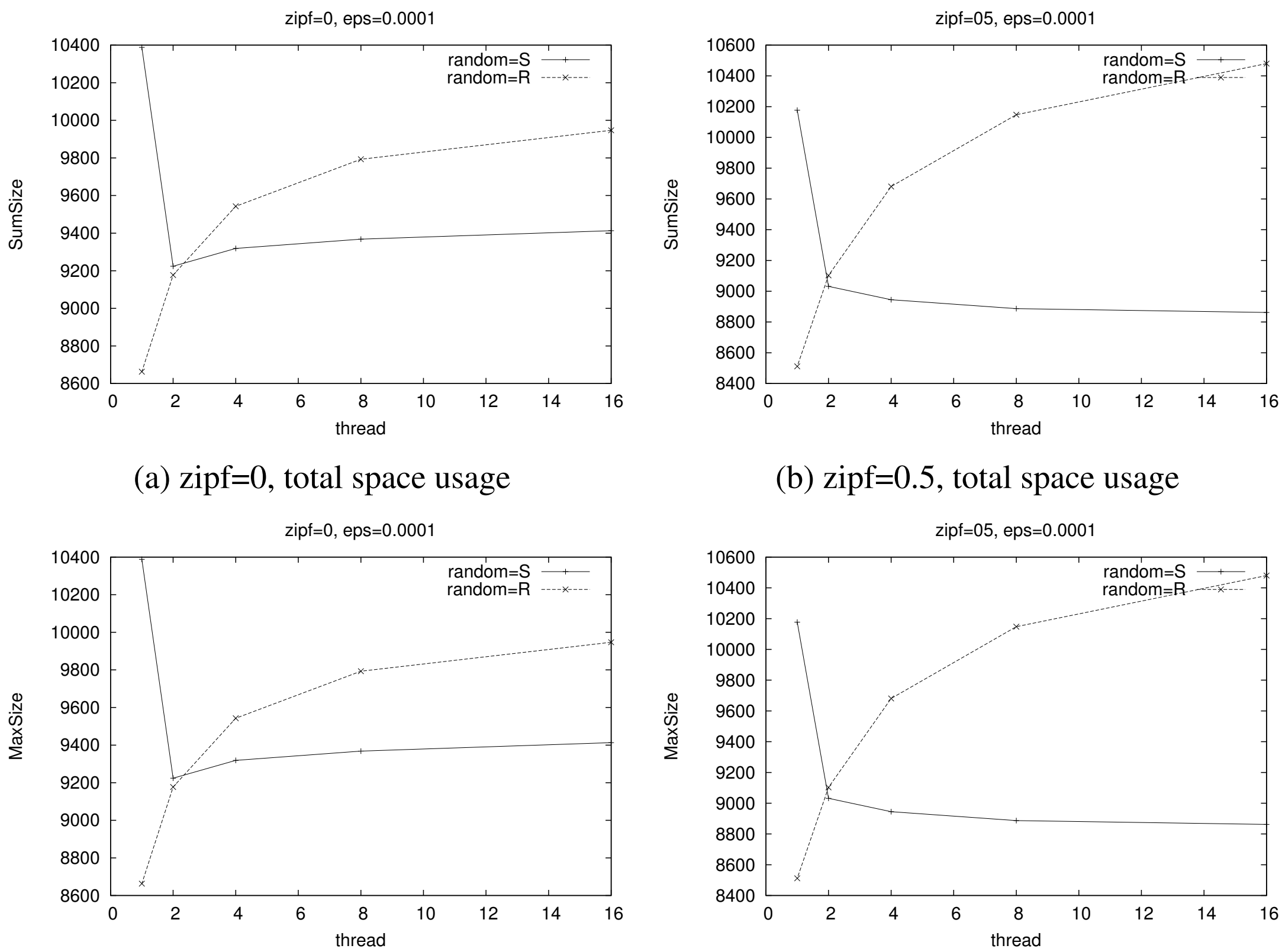

(a) zipf=0, total space usage

(b) zipf=0.5, total space usage

(c) zipf=0, maximum space usage

(d) zipf=0.5, maximum space usage

**Figure 5.19**: FASTqdigest, threads-size for zipf 0, 0.5, with 0.0001 ε.

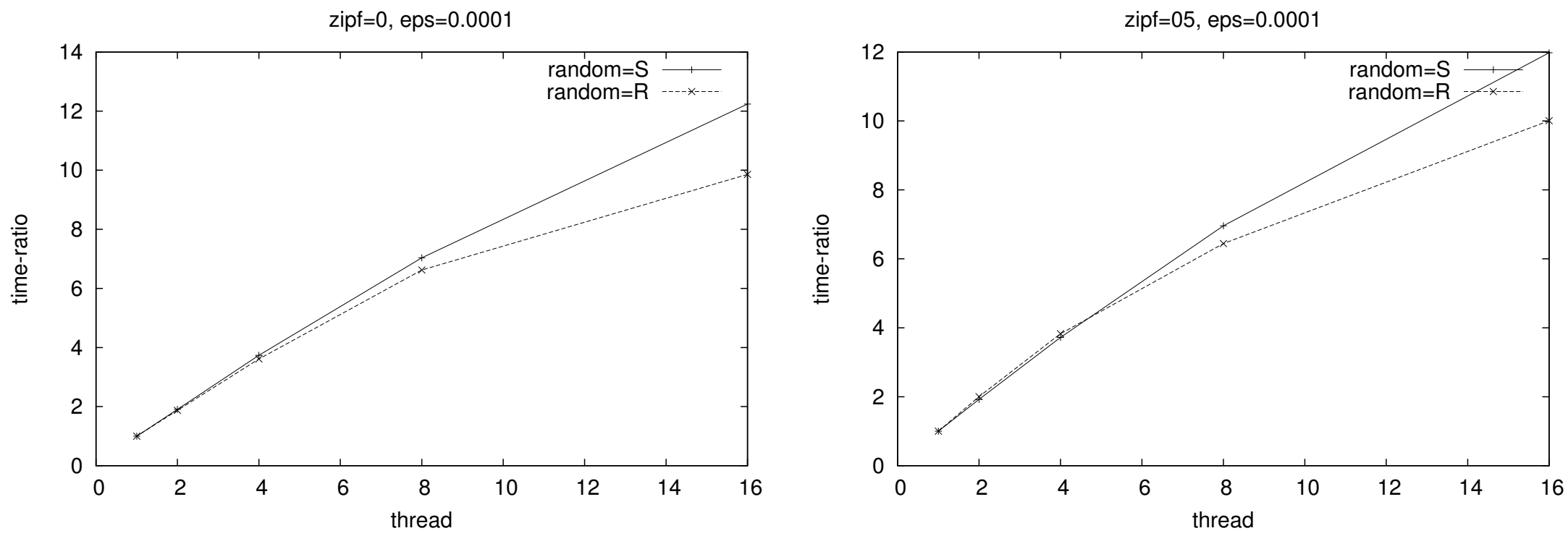

(a) zipf=0 (b) zipf=0.5

**Figure 5.20**: FASTqdigest, threads-time in ratio ($\frac{Time_1}{Time_n}$) for zipf 0, 0.5, with 0.0001 ε.

Figure 5.20 shows how the number of threads improve the running time. Similar with q-digest, FastQDigest is also scalable. We can see that for the sorted data 16 threads is 12 times faster than a single thread; and for random data it is almost 10 times faster. This indicates that when we have very large amounts of data we can use multiple threads or machines to improve the performance.

### 5.2.5 Random Mergeable Summaries

The Random Mergeable Summaries algorithm is another non-deterministic quantile algorithm besides the Sampling-Based algorithm in this paper. Similar to Sampling-Based algorithm, we may also see some unpredictable phenomenon in our experiment.

In Figure 5.21, we could see the relationship between ε and accuracy. Unlike the other four algorithms, the error doesn't increase when ε increase. This algorithm samples the data totally random (not a factor of ε). The ε affects the number of buffers, the size of the buffer and the *constLevel*. Based on Figure 5.21, we could say that these factors do not affect the accuracy. We can also see that the tendency of the curves are very similar, so the distribution of data doesn't influence the accuracy as well.

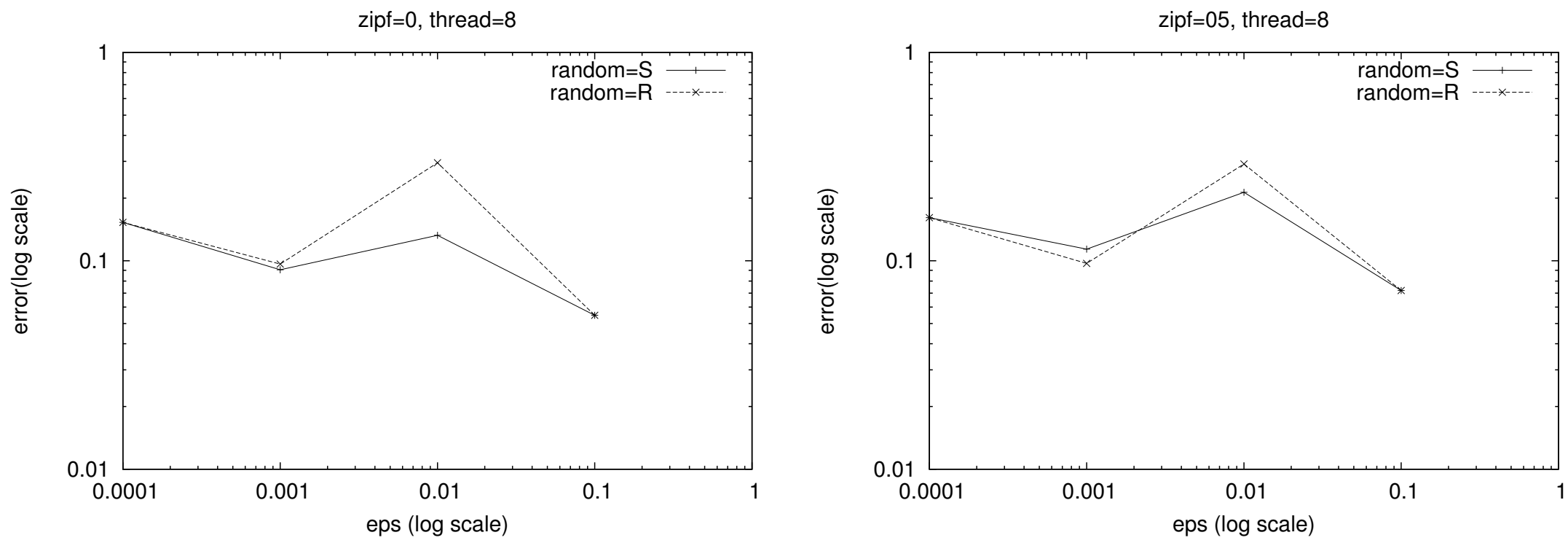

(a) zipf=0 (b) zipf=0.5

**Figure 5.21**: Random Mergeable Summaries, ε-error for zipf 0 and 0.5, with 8 threads.

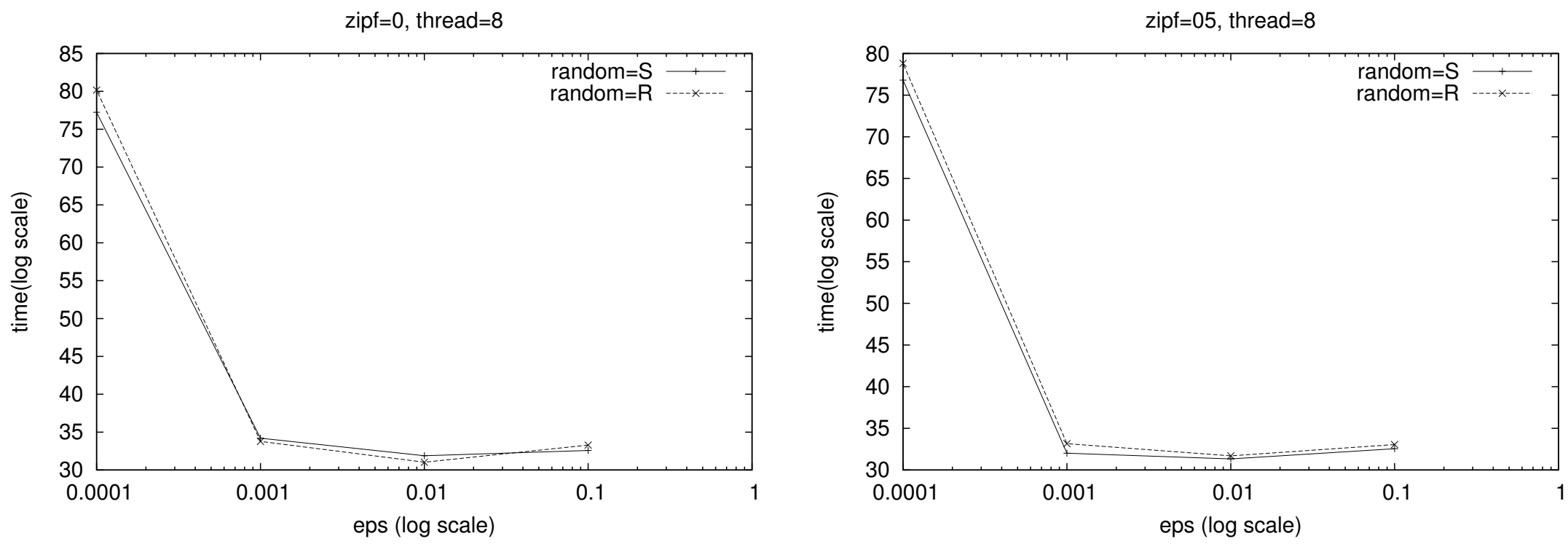

(a) zipf=0 (b) zipf=0.5

**Figure 5.22**: Random Mergeable Summaries, ε-time for zipf 0, 0.5, with 8 threads.

As we know, that the smaller ε is, the larger size of buffer we will have. Since the larger size of buffer will consume more time on sorting and merging, the running time should be much longer when ε is small. Figure 5.22 certifies that when ε is 0.0001, the running time is 2 to 3 times longer than the running time when ε is larger. However, we can also see that the curves are almost flat when ε is equal or greater than 0.001.

Figure 5.23 shows the relationship between the number of threads and errors.

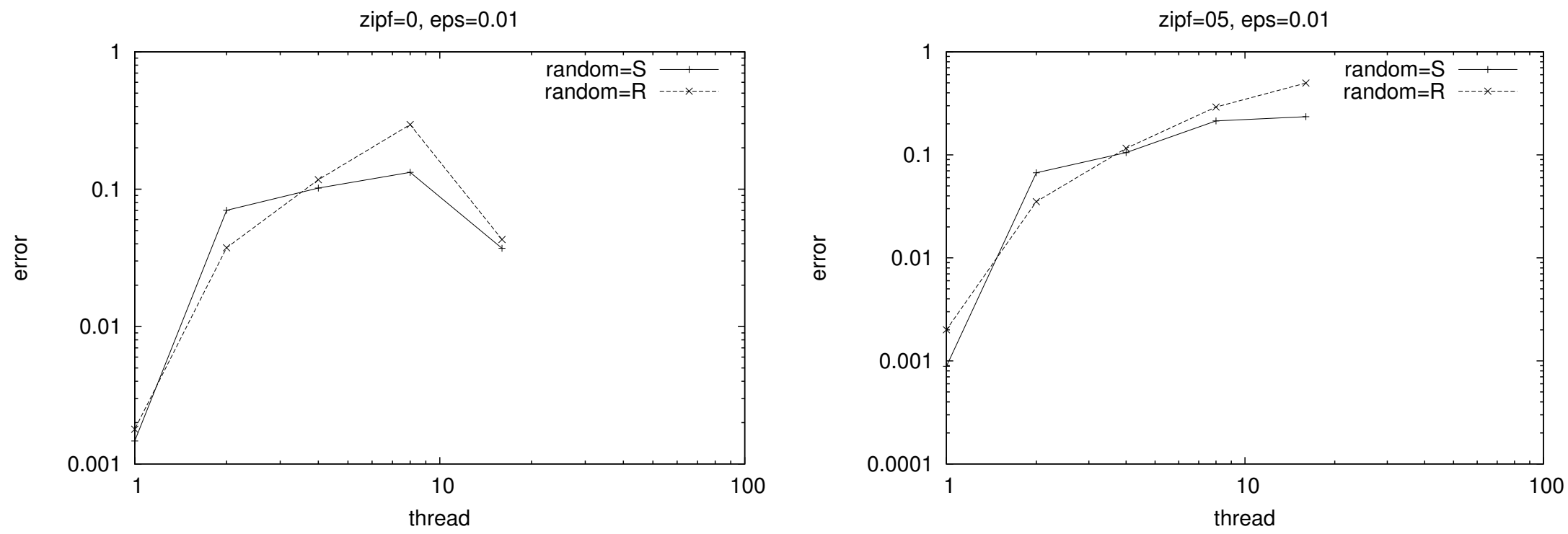

(a) zipf=0 (b) zipf=0.5

**Figure 5.23**: Random Mergeable Summaries, threads-error for zipf 0, 0.5, with 0.0001 ε.

As shown in Figure 5.23, we found that the error increases with the number of threads. The process of merging is the main factor that causes the increase of errors. The errors increase sharply when the number of threads increases from 1 to 2, and the curves become gradual after 2 threads. The merging process introduces a sampling progress on the synopses.

In Figure 5.24, we could see the relationship between number of threads and the space; Figures 5.24a and 5.24b show the total space used by all the threads; 5.24c and 5.24d show the maximum space used by each threads. Basically, when number of threads increases, the occupied space will increase because we need to do more merging processes. Thus, we can see that the total space used increases linearly when the number of threads increase. However, the maximum space usage seems unpredictable with the number of threads. The program uses fewer maximum space when the number of threads is 2 or 4, but it uses more space when the number of threads is 1, 8, or 16. Based on other related figures that haven't been presented in this paper, we could say that the maximum space usage is indeterminate with the number of threads.

Figure 5.25 shows how the number of threads improve the running time. As we can see in Figure 5.25, the Random Mergeable Summary algorithm is scalable. The 16

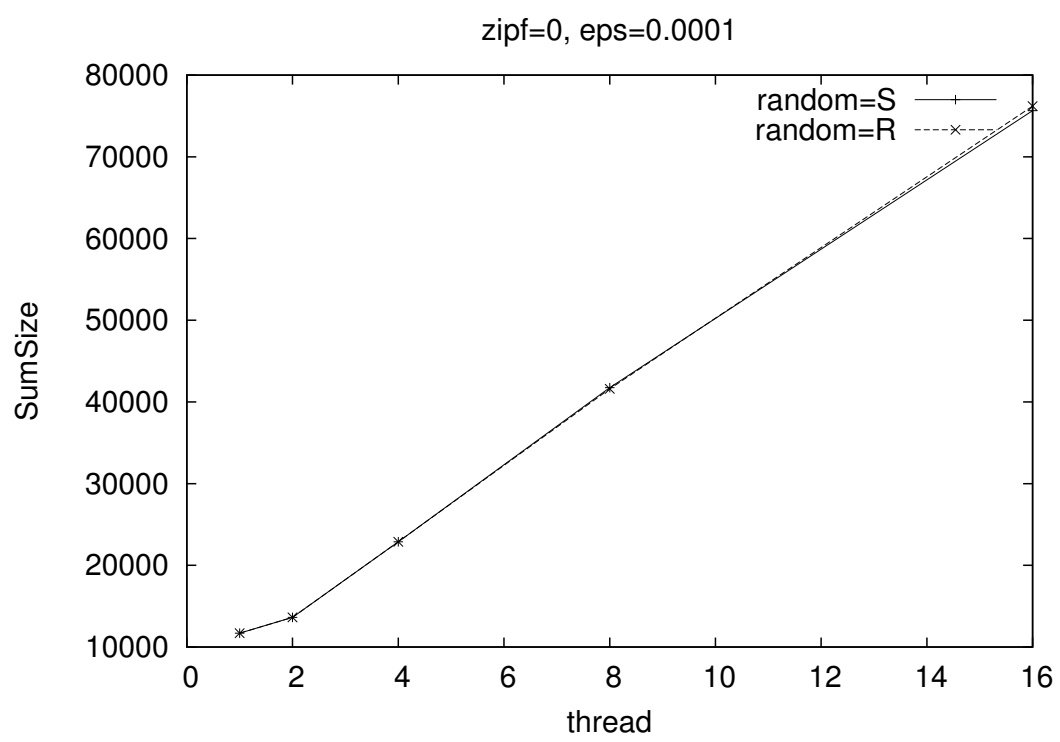

(a) zipf=0, total space usage

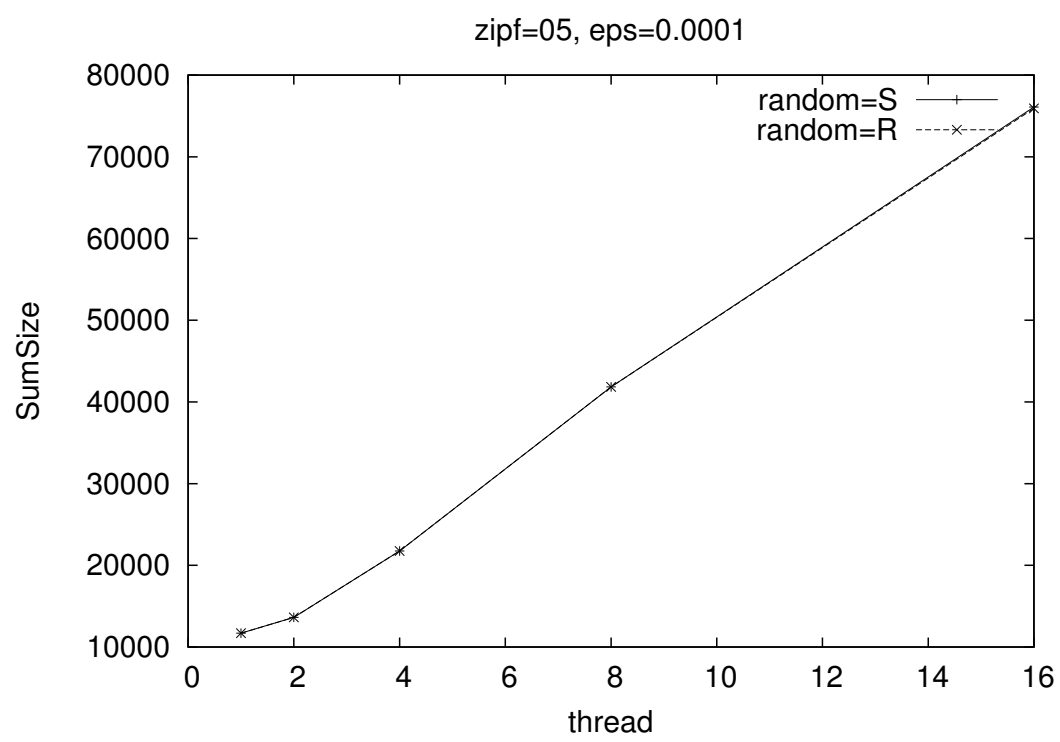

(b) zipf=0.5, total space usage

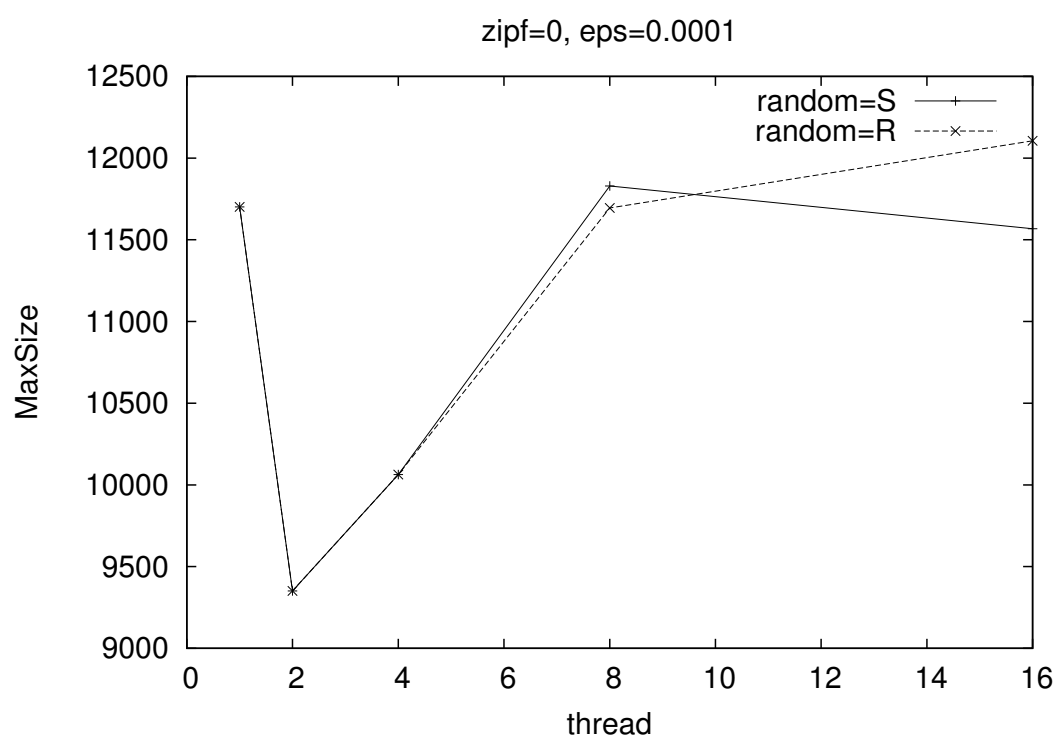

(c) zipf=0, maximum space usage

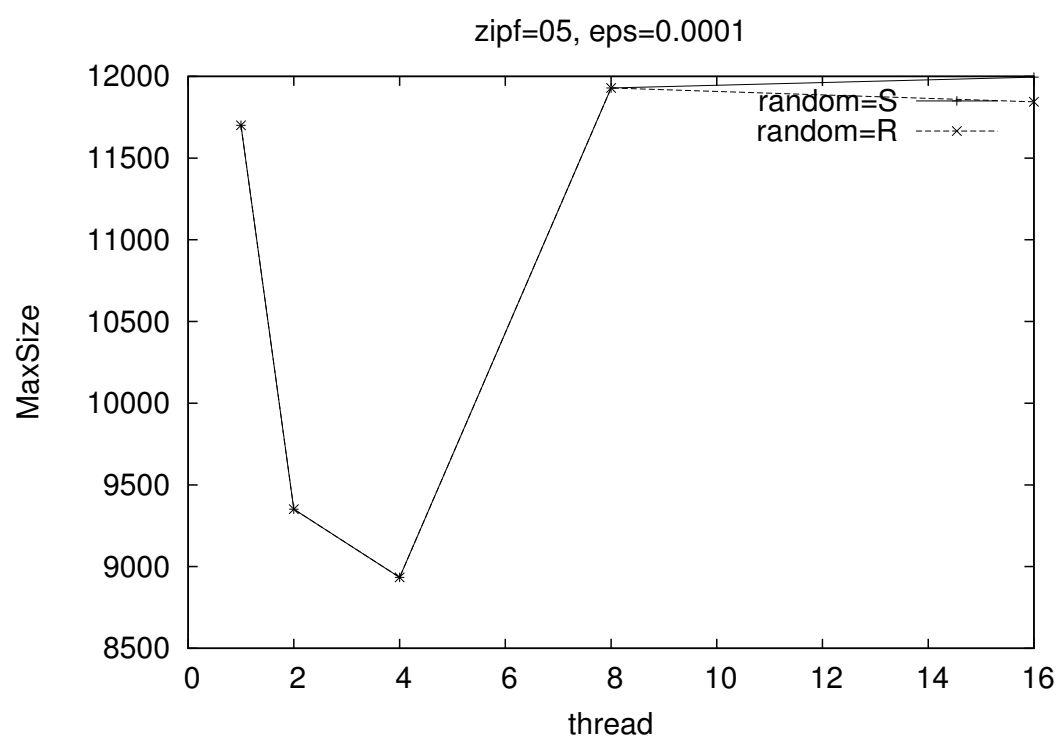

(d) zipf=0.5, maximum space usage

**Figure 5.24**: Random Mergeable Summaries, threads-size for zipf 0, 0.5, with 0.0001 ε.

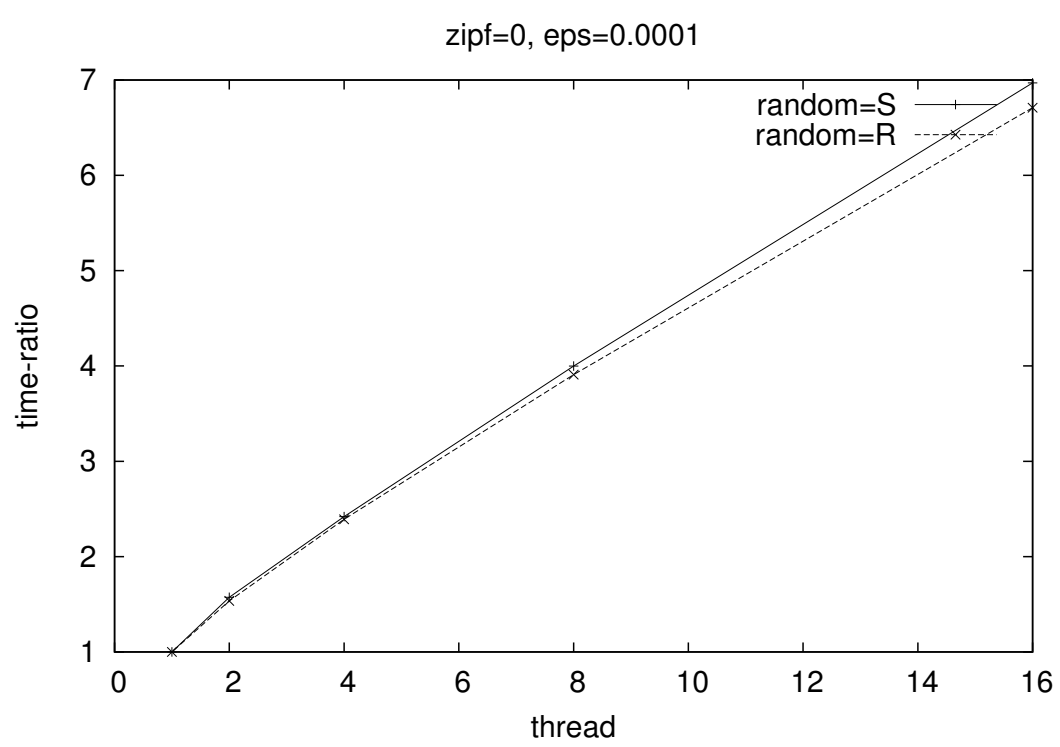

(a) zipf=0

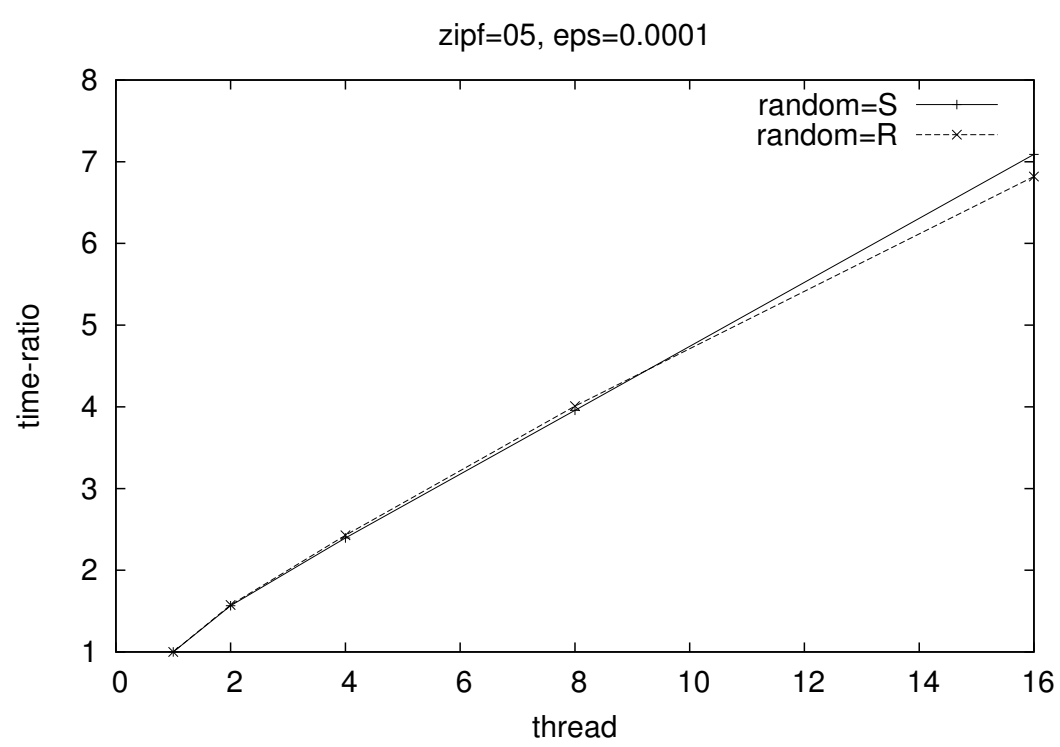

(b) zipf=0.5

**Figure 5.25**: Random Mergeable Summaries, threads-time in ratio ($\frac{Time_1}{Time_n}$) for zipf 0, 0.5, with 0.0001 ε.

threads are 7 times faster than the single thread. This indicates that when we have very large data we can use multiple threads or machines to improve the performance.

In conclusion for above figures and analysis, we could see that the error will increase and the running time will decrease when $\varepsilon$ increase. Figures 5.1, 5.6, 5.11, 5.16, and 5.21 show the relationship between $\varepsilon$ and error for different algorithms. In almost all the algorithms except the Random Mergeable Summaries algorithm, we can see a very clear increase of errors when $\varepsilon$ increases; Figure 5.21 shows that the errors keep stable and decrease a little bit when $\varepsilon$ increase. Figures 5.2, 5.7, 5.12, 5.17, and 5.22 show the relationship between $\varepsilon$ and running time. For algorithm GK, Random Mergeable Summaries and FastQDigest, we can see a decrease in time when $\varepsilon$ increases; however, $\varepsilon$ seems not to affect the running time for the q-digest and Sampling-Based algorithm.

For the large dataset, the scalability of an algorithm is very important. Thus, we test all five algorithms with up to 16 threads and see the affects to error, space, and time. To better see how much faster it runs with more threads, we show the ratio of the running time compared to the running time of single thread ($\frac{Time_1}{Time_n}$). Figures 5.3, 5.8, 5.13, 5.18, and 5.23 show the relationship between the number of threads and errors. More threads means more merging, so the error would increase, but we can see that q-digest (Figure 5.13) and FastQDigest (Figure 5.18) is almost flat, which means their merging algorithm does not affect accuracy. Figures 5.4, 5.9, 5.14, and 5.24 show that total space usage increases with more threads for algorithms GK, Sampling-Based, q-digest, and Random Mergeable Summaries; however, Figure 5.19 shows that when the FastQDigest only has 1 thread and the data is sorted, it uses a lot of space, than if it has more than 1 thread. In Figures 5.5, 5.15, and 5.20, we can see that multi-thread mostly accelerates linearly for sorted data, but random-ordered data decelerates in 16 threads. Sampling-Based (Figure 5.10), and Random Mergeable Summaries (Figure 5.25) are not affected by the order of the data, but the Sampling-Based algorithm seems to slow down with 16 threads.

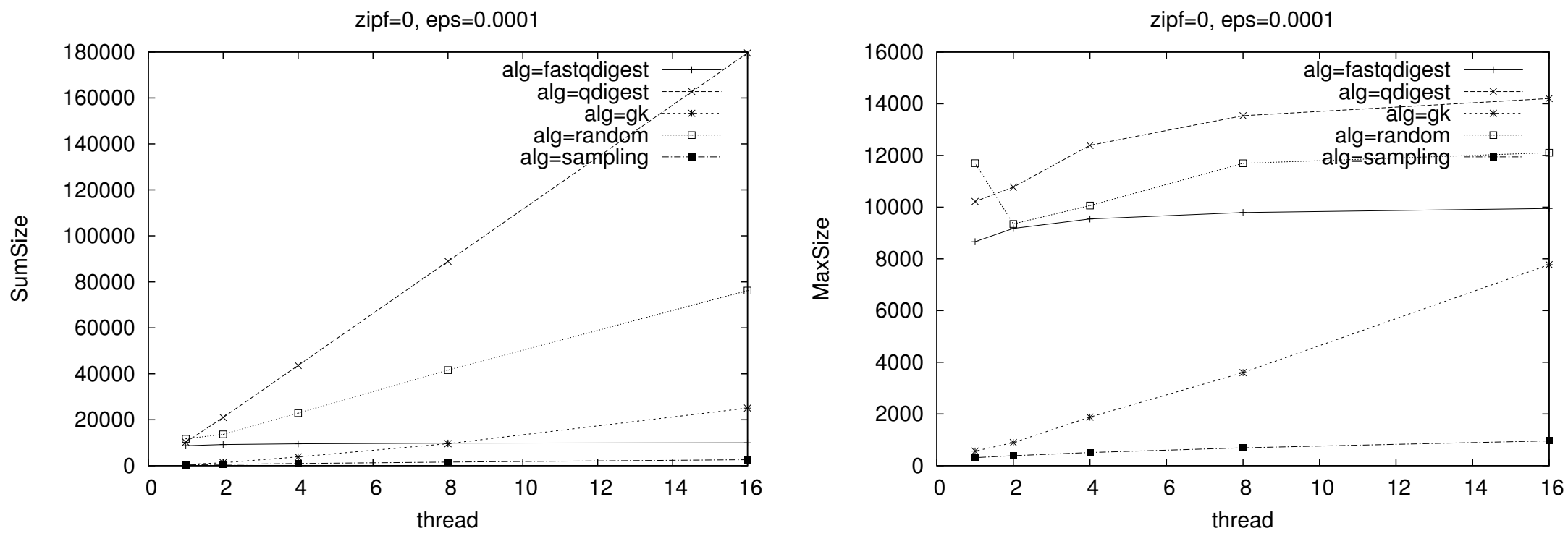

(a) total space usage (b) maximum space usage

**Figure 5.26**: threads-size for zipf 0, with 0.0001 ε and random-order data.

### 5.2.6 Comparison of all algorithms

In this subsection, we will compare all five algorithms with several experimental results. All experiments in this subsection use random-ordered data.

Figure 5.26 shows the relationship between number of threads and the space for all five algorithms; Figure 5.26a shows the total space used by all threads; 5.26b shows the maximum space usage by each threads. In Figure 5.26a, we can see that the q-digest uses the most space and the Random Mergeable Summaries algorithm is the second. However, we can find that there is a very slight difference of the total space when number of threads is 1, and a very large difference when number of threads is 16. Since the q-digest algorithm keeps a binary tree for each thread, it uses much more space than other algorithms with more threads. On the other hand, the variant of q-digest, FastQDigest, seems to be constant when number of threads increased. The Random Mergeable Summaries keep fixed-size buffer in each thread, so its total space used increases linearly as well with the number of threads. In Figures 5.26a and 5.26b, we could see that the total and maximum space of GK algorithm both increase linearly with the number of threads. This is because that the GK algorithm compresses the synopses after a number of insertion, and when there are more threads, each thread has fewer data, then the algorithm compresses the synopses less.

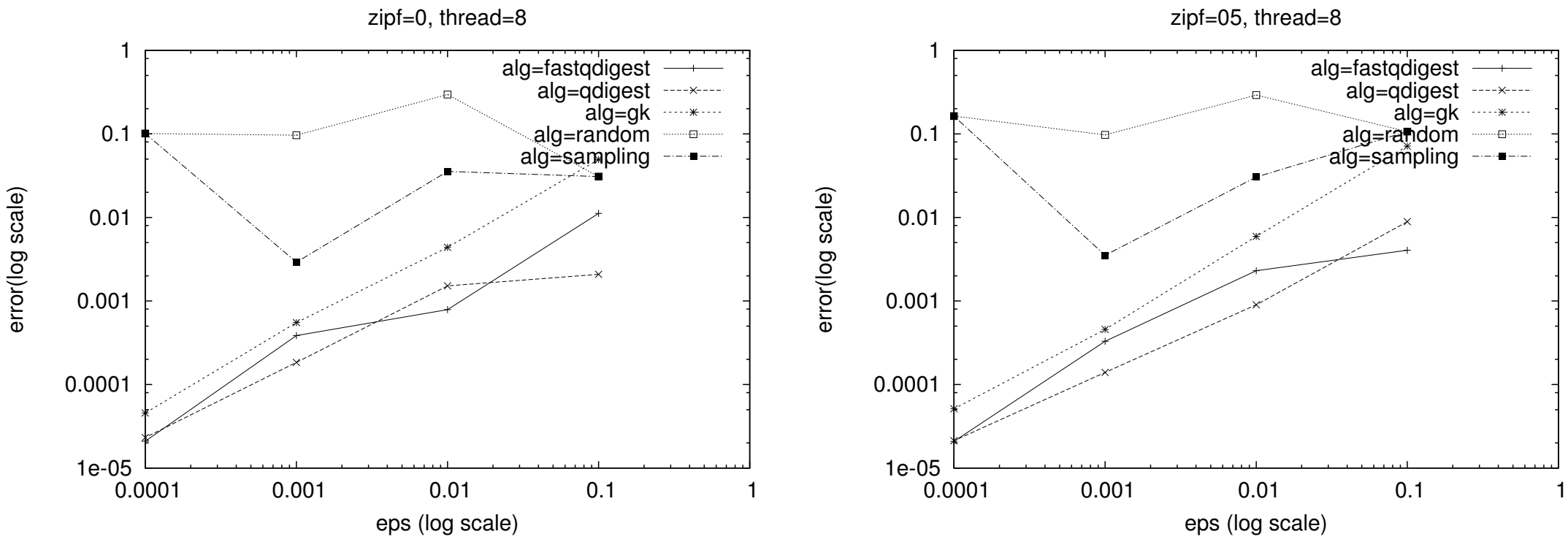

(a) zipf=0 (b) zipf=0.5

**Figure 5.27**: ε-error for zipf 0, 0.5, with 8 threads.

Figure 5.27 shows the relationship between ε and accuracy for all five algorithms. We could find that the error of the deterministic algorithms, GK, q-digest and FastQDigest, increase linearly with ε; in the other hand the error of the non-deterministic algorithms, Sampling-Based and Random Mergeable Summaries, keep flat with fluctuation as ε increase. From figure 5.27 we can see that the error of the deterministic algorithms is always under the ε, but the error of the non-deterministic algorithms does not.

Figure 5.28 show the relationship between ε and the running time for all five algorithms. We can find that the q-digest algorithm takes much longer time than other algorithms; other algorithms do not have much difference but we still can see that the non-deterministic algorithms, Sampling-Based and Random Mergeable Summaries, are fastest among all algorithms. We can also find that except for FastQDigest, there is only slight running time change when ε increase. This is because, in FastQDigest, if we have fewer critical constrain, we can avoid some nodes creating.

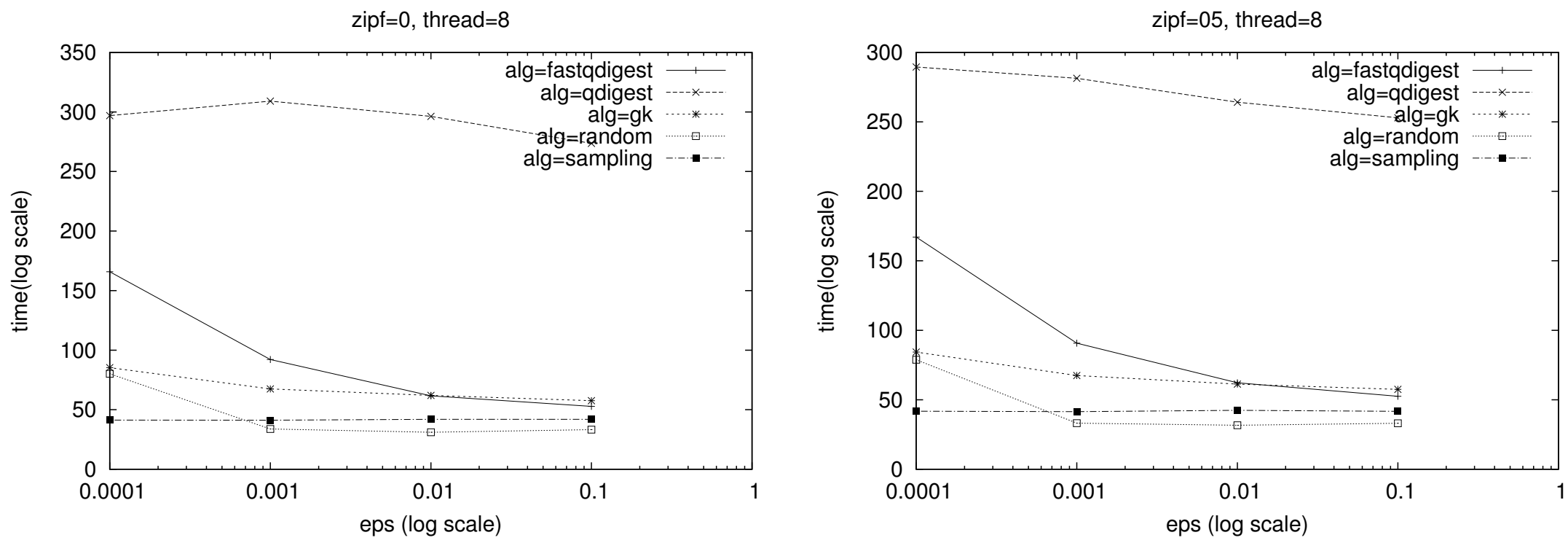

(a) zipf=0 (b) zipf=0.5

**Figure 5.28**: ε-time for zipf 0, 0.5, with 8 threads.

## 5.3 Discussion

### 5.3.1 GK

Figures 5.1 to 5.5 show our experimental results for GK algorithm. Through the figures, we can see that the order of source data plays a big role for GK algorithm, especially for the space usage. In addition, the GK algorithm seems to have better performance in accuracy, space usage, and running time with the random-ordered data than the sorted.

### 5.3.2 Sampling-Based

Figures 5.6 to 5.10 show our experimental results for Sampling-Based algorithm. Through the figures, we can see that the algorithm doesn't perform a distinctly difference based on the order of source data. Across Figures 5.6 and 5.8, the accuracy of the algorithm is not good enough and the error is larger than the ε.

### 5.3.3 Q-Digest & FASTQDigest

Figures 5.11 to 5.15 show our experimental results for q-digest algorithm. Figures 5.16 to 5.20 show our experimental results for FASTQDigest algorithm. For both algorithms, the performance of space usage and running time have a big gap for different data order. As I mentioned, the FASTQDigest is a variant of Q-Digest algorithm, it does not require a fixed universe of the data. As the results show, FASTQDigest performs even better than Q-Digest.

### 5.3.4 Random

Figures 5.21 to 5.25 show our experimental results for Random Mergeable Summaries algorithm. Across Figures 5.21 and 5.23, the $\varepsilon$ and number of threads does not affect accuracy for this algorithm. The major advantage of this algorithm is the speed; we can find it in Figure 5.22.

### 5.3.5 Comparison

Figures 5.26 to 5.28 show our experimental results for all five quantile computation algorithms. Through the figures, we can see that the non-deterministic algorithms (Sampling-Based and Random Mergeable Summaries) have fewer accuracy but also fewer running time than deterministic algorithms. Especially, the q-digest algorithm has the best accuracy but the longest running time among all algorithms. For the space usage, q-digest and Random Mergeable Summaries use much more space than other algorithms. The GK algorithm has the best overall performance across accuracy, running time and space usage.

# Chapter 6

# Conclusion

Quantile is a very helpful statistics information for large data sets analysis. In this paper we introduced and compared four quantile computation algorithms, GK [1], q-digest [4], Sampling Based [5] and Random Mergeable Summaries [6], and a variant FastQDigest [2] in distributed setting in shared memory multi-core database system. Different than the existing survey of L. Wang et al. [2], which is on a centralized setting, we mainly focused on the distributed setting since it is very popular and important in today's database system. In addition, we explored possible implementations for some of the algorithms on the extension to a distributed setting.

We expressed the quantile computation algorithms in a single formalism given by GLAs and gave the *GLADE* version quantile algorithms. We executed the algorithms in the *GLADE* with several $\varepsilon$s and numbers of threads, and measured them in several aspects: ranking error, running time, space usage, and ratio time. Finally, we analyzed the experimental results and compared all five quantile computation algorithms. The GK, q-digest and FastQDigest are deterministic algorithms, while the Sampling Based and Random Mergeable Summaries are non-deterministic, for which the algorithms diverge: deterministic algorithms have better accuracy while non-deterministic algorithms are faster. The GK algorithm has the best overall performance across accuracy, running time, and space usage.